%% file: MinSpace_v04_FINAL.tex
\documentclass{article}
\usepackage{amsmath,amsfonts,amsthm}
\usepackage{amstext}
\usepackage{amsgen}
\usepackage{amsbsy}
\usepackage{amsopn}
\usepackage{amssymb}
\usepackage{graphicx}
\usepackage{epsfig}
\theoremstyle{definition}

\theoremstyle{remark}

\numberwithin{equation}{section}

\input{tcilatex}
\include{definitions}

\begin{document}

\title{Minimum search space and efficient methods for structural cluster optimization}
\author{Carlos Barr\'{o}n-Romero \\
cbarron@cimat.mx\\
Centro de Investigaci\'{o}n en Matem\'{a}ticas, A.C. \\ \\
CIMAT, Dept. of Computer Science\\
Jalisco S/N Mineral de Valenciana, Apdo. Postal 402 36000 \\
 Guanajuato, Gto., CP 36240, MEXICO\\
http://www.cimat.mx/reportes/enlinea/I-05-06.pdf}


\date{April 8, 2005}%

\maketitle

\begin{abstract}
A novel unification for the problem of search of optimal clusters under a
well pair potential function is presented. My formulation introduces appropriate
sets and lattices from where efficient methods can address this problem.
First, as results of my propositions a discrete set is depicted such that
the solution of a continuous
and discrete search of an optimal cluster is the same. Then, this discrete set
is approximated by a special lattice IF. IF stands for a lattice that
combines lattices IC and FC together.
In fact, two lattices IF with 9483 and 1739 particles are obtained
with the property that they include all putative optimal clusters from 2 trough
1000 particles, even the difficult optimal Lennard-Jones clusters,
$C^{\ast}_{38}$, $C^{\ast}_{98}$, and the Ino's decahedrons. $C^{\ast}_{98}$ is
the only cluster where its initial configuration has a different geometry than
the putative optimal cluster in term of the adjacency matrix stated by Hoare.
My paper is not a benchmark, I develop a theory and a numerical experiment for
the state of the art of the optimal Lennard-Jones clusters and even I found new
optimal Lennard-Jones clusters with a greedy search method called Modified
Peeling Method. The paper includes all the necessary data  to allow the researchers
reproduce the state of the art of the optimal Lennard-Jones clusters at April
8, 2005. This novel formulation unifies the geometrical motifs of the optimal
Lennard-Jones clusters and gives new insight towards the understanding of the
complexity of the NP problems.

\end{abstract}
\section{Introduction}
Many methods have been proposed for the problem of search of optimal
clusters (SOC)~\cite{aml:Barron1999, jcics:Cai2002, T:Cai2002, prl:Deaven1995,
cpl:Deaven1996, jcp:Doye1998, jcc:Hartke1999, ap:Hoare1975, ap:Hoare1983,
jgo:Leary1997, ps:Maier1992, jgo:Maranas1994, jcp:Northby1987,
jgo:Pardalos1994, cpc:Pullan1997, cpc:Romero1999A, jcics:Shao2004,
ape:Solovyov2003, jpca:Wales1997, jpc:Wolf1998, jpca:Xiang2004A, jgo:Xue1994}.
It takes a while until a novel method is able to validity its performance and
found new putative optimal Lennard-Jones (LJ) clusters. Nowadays, Shao {\it et
al}.~are pushing the frontier of the size of the putative optimal LJ clusters
over 309 particles~\cite{jcics:Cai2002, T:Cai2002, jpca:Jiang2003,
jpca:Xiang2004A, jcics:Shao2004, cp:Shao2004, pc:Shao2005}. The author has kept
contact with this group for collaboration. Huang {\it et
al}.~\cite{jgo:Huang2002}  give equivalent formulations for LJ Potential.
Xue~\cite{jgo:Xue1997} presents several properties of the LJ Potential
formulation $\frac{1}{x^{12}}-\frac{2}{r^6}$. Pardalos {\it et
al}.~\cite{jgo:Pardalos1994} describe the conditions of a well pair potential
function and present several optimization methods for SOC.

Maranas and Floudas~\cite{jgo:Maranas1994} present a method of global
optimization for molecules:
\begin{quote}
 "Given the connectivity of the atoms in a
molecule and the force field according to which they interact, find the
molecular conformation(s) in the three-dimensional Euclidian space involving
the global minimum potential energy".
\end{quote}
This method uses the connectivity of atoms in a molecule to partitioning in
several sets based on the distance of pairs of atoms. Several properties are
presented and a global optimization algorithm is presented. The complexity of
this algorithm is exponential over the number of variables.

Some Authors resist adding to much knowledge and heuristics for the design of
an algorithm or  method for SOC. However, successful methods based on
molecular dynamics, molecular chemistry or physics~\cite{ap:Hoare1983,
arcp:Wille1999, pre:Leary1999, jgo:Xue1994, jcp:Northby1987,
ps:Maier1992,jpca:Wales1997, jcp:Doye1998, jcc:Hartke1999, jpca:Xiang2004A}
reduce the Hoare complexity
$\exp(-2.5176+0.3572n+0.028n^2)$~\cite{ap:Hoare1983} to a polynomial time
$(0.05 \pm 0.02)n^{2.8 \pm 0.1}$~\cite{jcc:Hartke1999},
$0.02n^{2.9}$~\cite{jcics:Cai2002}, and other polynomial times bigger than the
previous ones (some can be found in~\cite{arcp:Wille1999}).
However even
with the help of previous knowledge, the complexity of a discrete SOC is the same
of the NP class of problems, and this is the challenging impulse for the
creation of novel methods. Here, I do not included an extent review of methods
for SOC but some reviews are~\cite{arcp:Wille1999, jgo:Leary1997,
jgo:Pardalos1994}. In addition, for the limitation of time and space is not
possible to review or mention all previous methods or classify them, instead
the article focus in the closed related methods under my perspective. My
apologies, if I omit a relevant method but if this happens, it is without
prejudice.

It is probable that methods with a good background on the knowledge of the
problem and using adaptive search, simulated annealing, lattices, basin
hopping, funnels, phenotypes, fusion, evolutionary, and genetic operations have
advantages over other methods because they are exploring the discrete search
spaces of my lattice IF (hereafter only IF, see
Figs.~\ref{fig:latticeicfc9483}, and~\ref{fig:min_lattice1739}). It was stated
by Northby~\cite{jcp:Northby1987}:
\begin{quote}
 "The complexity of the problem lies in the fact that while it is always possible with a
 computer to allow a particular initial configuration to relax to the adjacent minimum of
the potential energy surface, unless the starting configuration has been chosen
to lie in the proper valley, or "catchment basin, the resulting configuration
will not correspond to the absolute minimum."
\end{quote}
Considering this remark, the mentioned methods can relax efficiently but the
global minimum could escape from the initial selection in a particular lattice,
making necessary the
exhaustive creation of good initial clusters from different lattices
or the transformation in good ones
before the relaxation. Some successful methods use random selection of clusters and particles
in a random way from the well know lattices type IC, FC, ID, TO and so on. Here, my
propositions close the gap stated by Northby between the initial configuration
and the global minimum cluster, in the sense that exist a lattice or a set as an appropriate
search space from where it is possible to repeat all the
putative optimal LJ clusters reported in The Cambridge Cluster Database
(CCD)~\cite{http:CCD}. As an example, a lattice and a set are
presented, IF9483 and MIF1739 respectively, such that they contain particular clusters that
match with the putative optimal LJ clusters from $n=2,\ldots,100$ in one
relaxation (minimization procedure). The complexity of this type of telephone
directory method on IF is at most $O(n^3)$ (the complexity of the relaxation
multiplying by the complexity of the evaluation of a pair potential function
for $n$ particles). However, this is not a lower bound for the complexity of
discrete methods of SOC using IF. There are cases where it is possible to reduce
the numbers of operations in the evaluation of the potential of a cluster by the symmetry
inherited from IF. $C_{13}^\ast$ is an example where the cost of computing the potential
is
4 instead of 91 operations (section~\ref{ssc:simmandPot} depicts this).
IF allows to have automatic classification of clusters, a measurement in term
of the number of adjustment for solving SOC in discrete fashion, and let to
study NP complexity.

IF (as a discrete search space) coincides with the well know result from Quantum
Mechanic that the particles interact in discrete fashion.
Moreover, the existence of particles forming an IF can be seeing as particles
in a hot temperature where the positions IC and FC could be occupied with equal
likelihood. What it makes difficult to predict the geometric shape of small
clusters is the mobility of the potential energy surface (PES).  PES changes
from small cluster to larger ones in the sense that
the displacement of a particle in the outer shell from its lattice's position has
more free in the small clusters
than in large ones in order to reduce the total cluster energy. On the other hand, for
large clusters the transition to stable structures corresponds to a change of
geometrical structure from IC to a decahedral lattice~\cite{jgo:Leary1997,
pc:Shao2005} where the PES has less freedom. Section~\ref{sc:results} has
Figures where the normalized gradient is depicted. From these figures the PES's
mobility for the particles in the outer shells of a small cluster can be
explained.

The notation and some conventions used in this report are given in
Section~\ref{sc:notation}. Section~\ref{sc:propertiesPotential} describes the
properties of the potential where this methodology can be applied.
Sections~\ref{sc:Unified lattice} and~\ref{sc:methods} describe the special
IF9483,  MIF1739 and methods for them. Section~\ref{sc:results} presents MIF1739.
It contains all the putative optimal LJ
clusters, tables~\ref{tb:On_off_0}-\ref{tb:On_off_32} give in an efficient and
short notation all the indices to build the initial clusters,
tables~\ref{tb:Changes_0}-\ref{tb:Changes_9} present the geometrical type, the
initial and minimum LJ potential, and a measure of the adjustment necessary to
transform from $C_{n+1}$ to $C_n$, and
tables~\ref{tb:idMinLattice_0}-\ref{tb:idMinLattice_15} give the coordinates of
MIF1739 in order to reproduce the numerical results presented here. In
addition, this section includes novel figures of the difficult clusters inside
of IF9483, and novel descriptions of some clusters. Finally,
Section~\ref{sc:conclusions and future work} presents my conclusions and future
work.

\section{Notation}~\label{sc:notation}
$\Natural$ is the set of the natural numbers, $\Rational$ is the set of the rational
numbers, and $\Real$ is the set of the real numbers.

A lattice, $\Omega =\left\{ p_{i}\right\} _{i\in \text{I}}$,
$p_{i}\in \Real^{3}\forall i\in $I where I is a set of indexes
(I=$\Real$ or I=$\Natural)$ is a set of points in a regular pattern in $\Real^3$.

A cluster of size $n$ is
$C_{n}=\left\{p_{i_{_{1}}},p_{i_{2}},...,p_{_{i_{n}}}\right\}$,
 $p_{i_{j}}\in \Omega $, $\forall j=1,\ldots,n$.

$\overrightarrow{\cdot }:L\rightarrow \Real^{3\cdot }$ means
$\overrightarrow{\cdot}\left( C_{n}\right) =\overrightarrow{C}_{n}$ $=$
$\left(p_{i_{_{1}}},p_{i_{2}},...,p_{_{i_{N}}}\right) \in \Real^{3N}$, it is a
vector representation of a cluster where $p_{i_{l}}=(x_{i_{l}}, y_{i_{l}},
z_{i_{l}})$ is mapped into cylindrical coordinates $(\rho_{i_{l}},
\alpha_{i_{l}}, \beta_{i_{l}})$, $\rho_{i_{l}}\in\Real^+$,
$\alpha_{i_{l}}\in[0,\pi]$ is zero on the semi-axes $Y^+$ and the
$\theta_{i_{l}}\in[0,2\pi]$. Then coordinates $p_{i_{_{l}}}\leq p_{i_{m}}$,
$l\leq m$ if $\rho_{i_{l}} \leq \rho_{i_{n}}$; or if $\rho_{i_{l}} =
\rho_{i_{n}}$ and $\alpha_{i_{l}} \leq \alpha_{i_{m}}$; or if
 $\rho_{i_{l}} = \rho_{i_{n}}$, $\alpha_{i_{l}} = \alpha_{i_{m}}$, and
 $\beta_{i_{l}} \leq \beta_{i_{m}}$.

$\overrightarrow{C}_{n} $ is used as an element of
 the metric space $\Real^{3n}$.

Let $p_{j}$, $p_{k}\in \Real^{3}$ and $r_{j,k}$ = $\sqrt{\left( x_j - x_k
\right)^{2}+
       \left( y_j - y_k \right)^{2}+
       \left( z_j - z_k \right)^{2}}$. Some potential functions \cite{jgo:Maranas1994} are

Buckingham Potential (BU):

\begin{equation*}
\hbox{VBU}_{i,j}=\alpha _{i,,j}e^{\beta _{i,,j}r_{i,,j}}+\frac{\gamma _{i,,j}}{%
r_{i,,j}^{6}}
\end{equation*}
where $\alpha _{i,j}$, $\beta _{i,j}$, and $\gamma _{i,j}$ are parameters
for the type of particles.

Kihara Potential (KI):

\begin{equation*}
\hbox{VKI}_{i,j}=4\epsilon _{0}\left[ \left( \frac{1-\gamma }{r_{i,,j}/\sigma
-\gamma }\right) ^{12}+\left( \frac{1-\gamma }{r_{i,,j}/\sigma -\gamma }%
\right) ^{6}\right]
\end{equation*}
where $\epsilon _{0}$, $\sigma $, and $\gamma $ are parameters for the type
of particles.

Lennard-Jones potential (LJ):
\begin{equation*}
\hbox{VLJ}_{i,j}=V_{i,j}=4\epsilon _{0}\left[ \left( \frac{\sigma _{i,j}}{r_{i,j}}%
\right) ^{12}-\left( \frac{\sigma _{i,j}}{r_{i,j}}\right) ^{6}\right]
\end{equation*}
where $\epsilon _{0}$ and $\sigma _{i,j}$ are parameters for the type
of particles. For the examples
in this paper, $\epsilon _{0}=\sigma _{i,j}=1$.

Morse Potential (MO) \cite{ap:Hoare1983} is

\begin{equation*}
\hbox{VMO}_{i,j}=\left( 1-e^{-\alpha \left( 1-r_{i,j}\right) }\right) ^{2}-1
\end{equation*}
where $\alpha $ is a parameter.

A pair potential can represented by
\begin{equation*}
\hbox{VXX}_{j,k}=EXX \left( \left\{ p_{j}, p_{k} \right\} \right)
=EXX(p_{j}, p_{k}) = EXX(r_{j,k})
\end{equation*}
where XX can be BU for Buckingham, KI for Kihara,
       LJ for Lennard-Jones, and MO for Morse Potentials.

The complete potential of a cluster is
$$
\hbox{EXX} \left( C_{n} \right) =\sum_{i=1}^{n-1}\sum_{j=i+1}^{n} \hbox{VXX}_{j,k}
$$

The Lennard-Jones Potential (LJ)\ is written also as
\begin{equation*}
\hbox{E}\left( C_{n}\right) =E\left( \overrightarrow{C}_{n}\right)
=4\sum_{i=1}^{n-1}\sum_{j=i+1}^{n}(\frac{1}{r_{jk}^{12}}-\frac{1}{r_{jk}^{6}}
)
\end{equation*}
when there is not confusion with the other potentials.

The conventions follows in the paper for the problems are:
\begin{list}{}{}
    \addtolength{\topsep}{-\baselineskip}
    \addtolength{\itemsep}{-\baselineskip}
    \setlength{\parskip}{\baselineskip}
    \item SOC denotes the problem of search of optimal clusters.
    \item SOCC denotes SOC solved in a continuous search space.
    \item SOCD denotes SOC solved in a discrete search space. Here it is
    assumed that an appropriate discrete set or lattice of points in $\Real^3$
    exists.
    \item SOC($n$) denotes SOC for clusters of size $n$.
    \item SOCC($n$) denotes SOCC for clusters of size $n$.
    \item SOCD($n$) denotes SOCD for clusters of size $n$.
    \item SOCYXX denotes one of the previous problems, where Y=C or Y=D, and
    XX is BU, KI,  LJ, and MO Potentials.
    \item SOCYXX($n$) denotes one of the previous problems
    for clusters of size $n$.
 \end{list}

For $n\in \Natural$, $n\geq 2$ SOCCXX($n$) can be stated:

Given $A \subset \Real^3$,
look for $\overrightarrow{C}_{n}^{\ast}=(x_{1},\ldots ,x_{n})\in
\A^n$ such that EXX$\left( \overrightarrow{C}_{n}^{\ast}\right)
$ $\leq$ EXX$\left( x\right) ,\forall x\in A^n$ where
$A^n =A \times \cdots \times A$ (n times).

In similar way SOCDXX($n$) can be stated:

Given a lattice $\Omega =$\{$p_{i}$ $|$ $p_{i}\in \Real^{3}$, $i=1,\ldots,N$ $N
\gg n$, $N \in \Natural$\}, find
$\overrightarrow{C}_{n}^{\ast}=(p_{l_{1}},\ldots ,p_{l_{n}}),$ such that
EXX$\left( \overrightarrow{C}_{n}^{\ast}\right)$ $\leq$
EXX$\left(
\overrightarrow{C}_{I^{k}}\right) $, $\forall I^{k} \subset \Natural$,
$|I^{k}|=n$ where $\left| \cdot \right| $ is the number of elements of a set.
This means
EXX$\left(\overrightarrow{C}_{n}^{\ast}\right)$ is less or equal than the potential of
any other cluster of $n$ points from $\Omega$.

The function adjust is defined as Adj:$2^{\Omega }\times 2^{\Omega
}\rightarrow \Natural$, Adj$\left( C_{n}\text{, }C_{m}\right) $ =
$\left| \left(
C_{n}\setminus C_{m}\right) \cup \left( C_{m}\setminus C_{n}\right) \right|$.

The function On is defined as On:$2^{\Omega }\times 2^{\Omega }\rightarrow \Natural$,
 On$\left( C_{n}\text{, }C_{m}\right) $=$\left| C_{m}\setminus C_{n}\right|$.

The function Off is defined as Off:$2^{\Omega }\times 2^{\Omega }\rightarrow
\Natural$,
Off$\left( C_{n}\text{, }C_{m}\right) $=$\left| C_{n}\setminus
C_{m}\right|$.

It is easy to see that Adj$\left( C_{n}\text{, }C_{m}\right)$ =
On$\left(
C_{n}\text{, }C_{m}\right) $ + Off$\left( C_{n}\text{, }C_{m}\right) $.

There are several references that explains how to build IC and
FC \cite{ps:Maier1992, jpca:Xiang2004A, ape:Solovyov2003}, in
particular Northby \cite{jcp:Northby1987} called to the combination of both
IF. Hereafter, IC$n$ represent a subset of $n$ points of type
IC, and similarly for FC and IF.

$C_{j}$ $\rightarrow C_{j}^{\ast }$ means $C_{j}^{\ast }=\min $($C_{j}$)
under some potential function and where $\min$ is a minimization procedure.
The results reported here were
computed with a version the Conjugated Gradient Method (CGM). Also, in the
text when there is not confusion $C_{j}^{\ast}$ means the putative optimal
LJ cluster.

In some figures the normalized gradient of LJ is depicted. This vector
correspond to $\nabla$ VXX $(x^{\ast})$ /
$\|$ $\nabla$ VXX ($x^{\ast }$) $\|$.
The corresponding component of the gradient is drawn as a vector in $\Real^3$
on each particle of a given cluster.

\section{Properties of the LJ Potential}~\label{sc:propertiesPotential}
Note that LJ, BU, and KI but MO Potentials share
the well potential's properties~\cite{ jgo:Pardalos1994}:
\begin{enumerate}
    \addtolength{\topsep}{-\baselineskip}
    \addtolength{\itemsep}{-\baselineskip}
    \setlength{\parskip}{\baselineskip}
\item  $\lim_{r\rightarrow r_{0}}\hbox{VXX}\left( r\right) =\infty $
\item  Each cluster under a pair potential has a basin.
\end{enumerate}

Moreover in one dimension, given
two particles, the first is fixed on $(0,0,0)$, and the second  with
coordinates $(r,0,0)$ is free to move on axes X. The following properties are
satisfied for E (similar results are given by Xue~\cite{jgo:Xue1997}):

\begin{enumerate}
    \addtolength{\topsep}{-\baselineskip}
    \addtolength{\itemsep}{-\baselineskip}
    \setlength{\parskip}{\baselineskip}
\item  E$\left( r\right) =4(\frac{1}{r^{12}}-\frac{1}{r^{6}}).$
\item  E$^{\prime}\left( r\right) =\allowbreak 24\frac{-2+r^{6}}{r^{13}}$
\item  E$^{\prime \prime }\left( r\right) =\allowbreak -24\frac{-26+7r^{6}}{%
r^{14}}$ \item  E$\left( r\right) <0$, $r>1$, $\lim_{r\rightarrow \infty}
\hbox{E}\left( r\right) =\allowbreak 0.$ \item  $r^{\ast }=\sqrt[6]{2}$, is the
global minimum of E. E$\left( r^{\ast }\right) =\allowbreak -1.0,$ E$^{\prime
}\left( r^{\ast }\right) =\allowbreak 0$ E$^{\prime \prime }\left( r^{\ast
}\right) >0$ \item  E$\left( r^{\ast }+\xi \right) \approx -1 + O\left( \xi
^{2}\right) ,0\leq \left| \xi \right| <<1.$ By a series expansion E$\left(
r^{\ast }+\xi \right) =4\left( \frac{1}{\left( \sqrt[6]{2}+\xi \right)
^{12}}-\frac{1}{\left( \sqrt[6]{2}+\xi \right) ^{6}}\right) =-1+\left( 18\left(
\sqrt[3]{2}\right) ^{2}\right) \xi ^{2}+O\left( \xi ^{3}\right) $, with $0\leq
\left| \xi \right| <<1,$ E$\left( \sqrt[6]{2}+\xi \right) \approx -1+K\xi
^{2},K=18\left( \sqrt[3]{2}\right) ^{2}.$ \item  The basin region of E is the
interval $\left(1.053666\,8,1.\,2444551\right) $. E$^{\prime \prime}\left(
r\right) >0$, $\forall r\in \left( 1.053666\,8,1.\,2444551\right)$. Therefore
E$\left( r\right) $ is convex and E$\left( r\right) <-0.786\,982\,15$, $\forall
r\in \left( 1.053666\,8,1.\,2444551\right) $.
\end{enumerate}

\begin{figure}
\centerline{
\psfig{figure=\IMAGESPATH/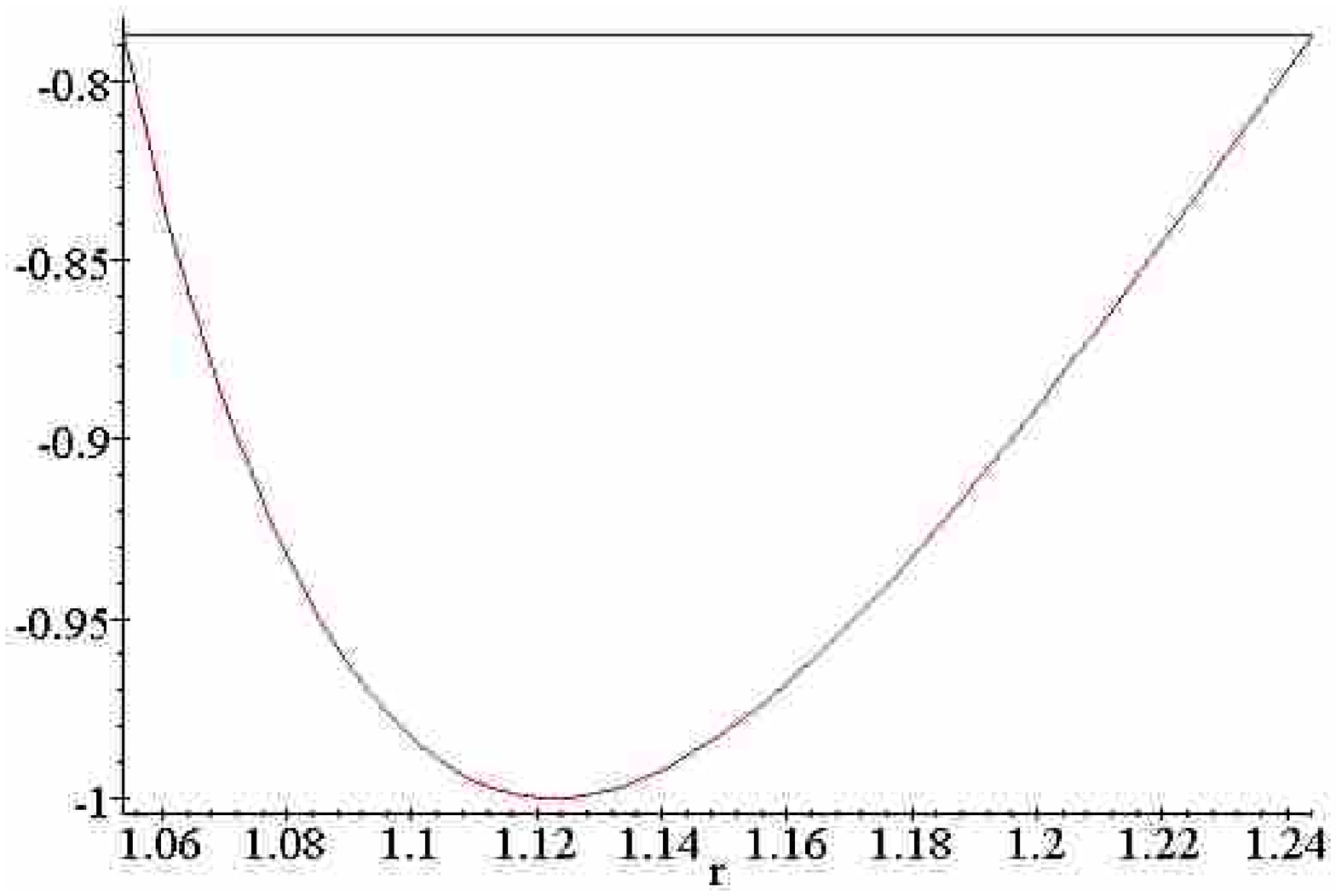,width=2.5in}
}
\centerline{ \makebox[2.5in][c]{a)}}
\centerline{
\psfig{figure=\IMAGESPATH/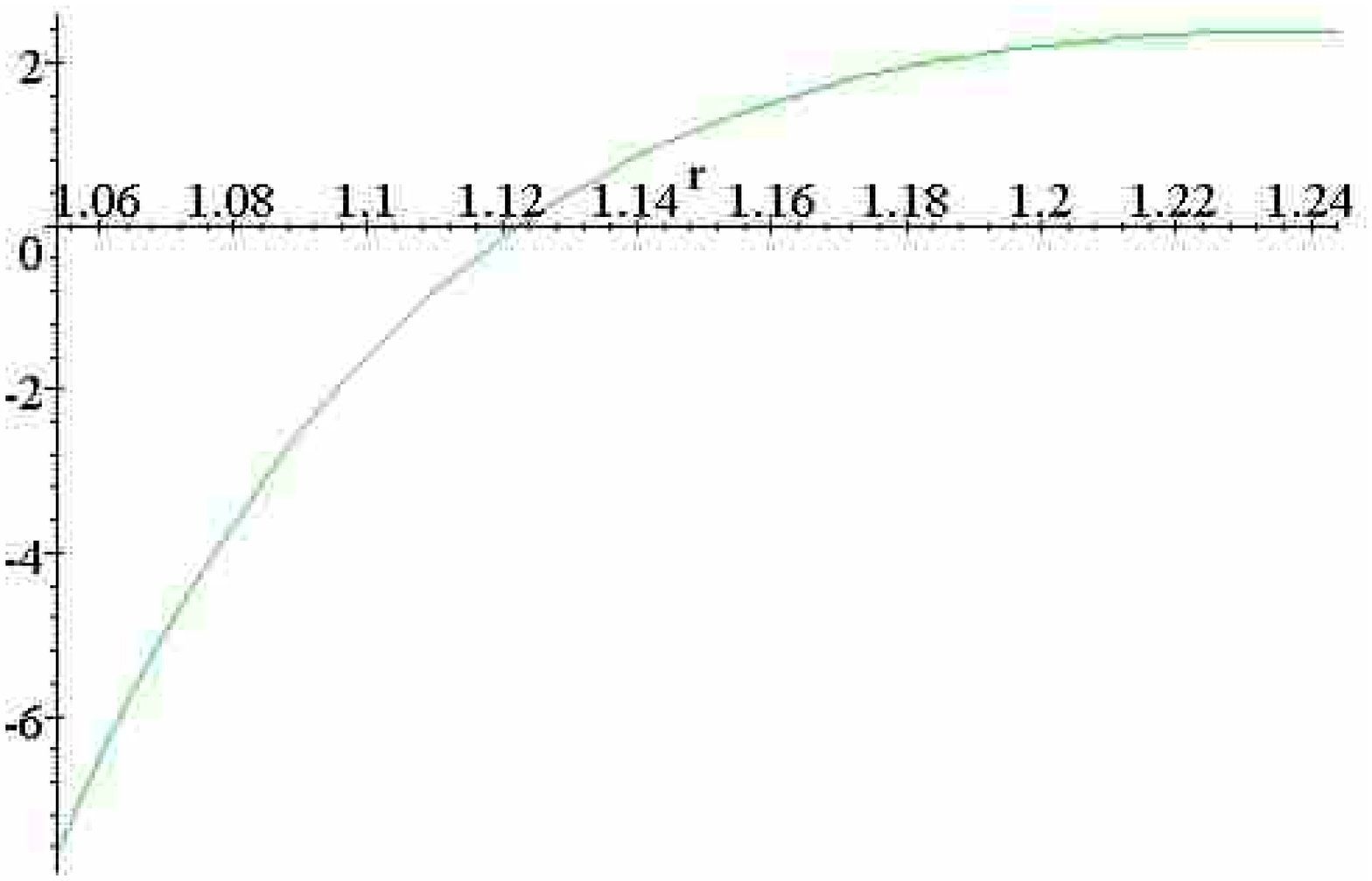,width=2.5in}
\psfig{figure=\IMAGESPATH/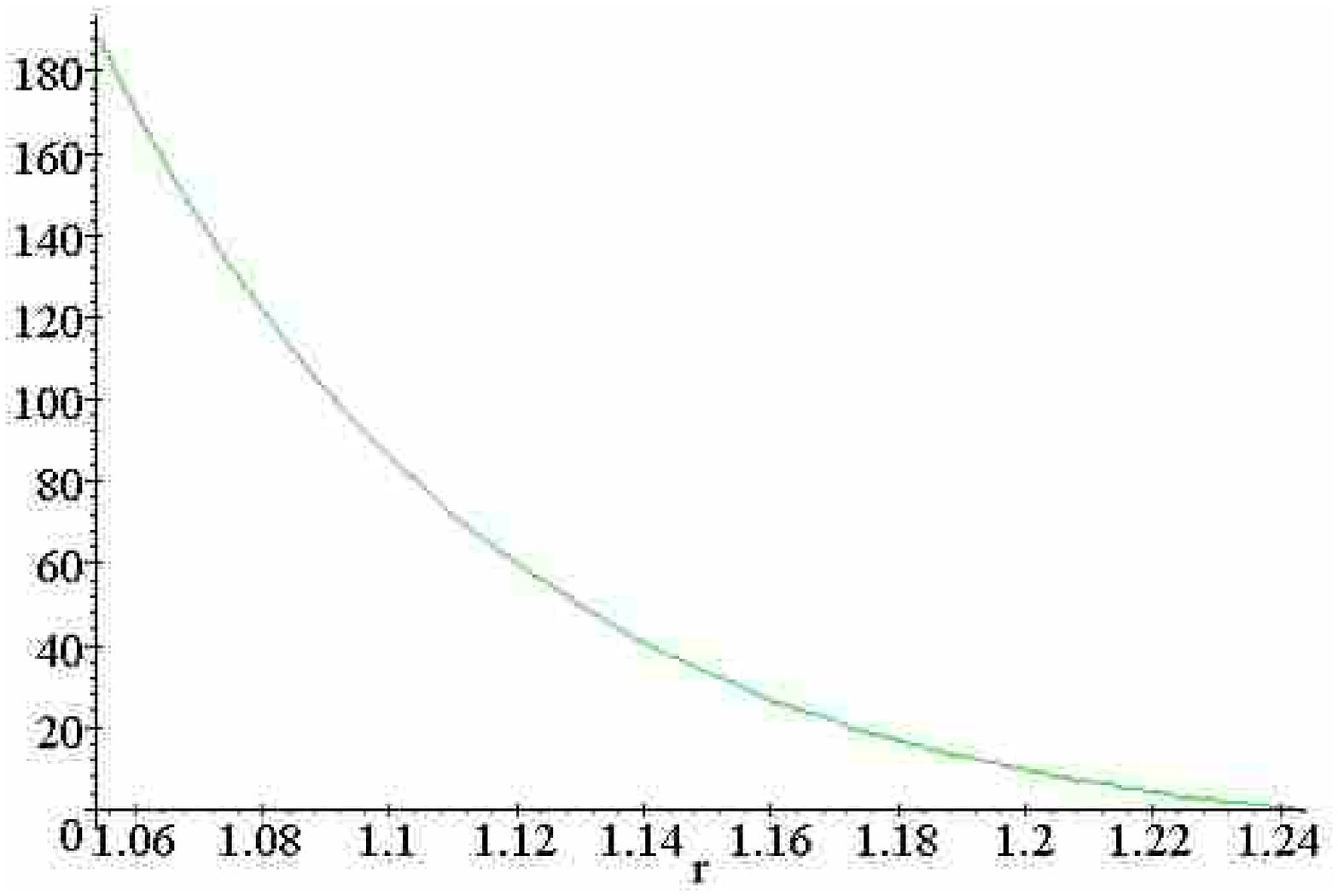,width=2.5in}
} \centerline{ \makebox[2.5in][c]{b)} \makebox[2.5in][c]{c)} }
\caption{a) E$(r)$, b) E${\prime}(r)$, and c) E${\prime \prime}(r)$.}~\label{fig:LJ_r}
\end{figure}

Figure~\ref{fig:LJ_r} depicts E$\left( r\right) $, E$^{\prime }(r),$ and
E$^{\prime \prime}\left( r\right) $ in $\left( 1.053666\,8,1.\,2444551\right)$.

\section{Unified Lattice}~\label{sc:Unified lattice}
This section explains what is the relationship between the discrete and
continuous SOC. The main proposition is:
Exist a discrete set for all optimal clusters where their potential has the
same value as in the solution of the continuous search of optimal
clusters. This type of potential function must fulfill the conditions
of a well potential~\cite{jgo:Pardalos1994}:
\begin{enumerate}
    \addtolength{\topsep}{-\baselineskip}
    \addtolength{\itemsep}{-\baselineskip}
    \setlength{\parskip}{\baselineskip}
    \item Potential function creates a infinite repulsion force when distance
between two particles goes to 0.
    \item Each cluster under this potential has a basin.
\end{enumerate}

Note that BU and LJ functions comply 1. KI function and MO function do not comply with 1.

\begin{proposition}~\label{prop:DiscreteLattice_p1}
Exist a discrete set, $\Omega $, where $\forall
j\in N$, $j\geq 2$, the potential of SOCDXX($j$) has the same optimal
value of SOCCXX($j$) for a potential function such that

\begin{enumerate}
    \addtolength{\topsep}{-\baselineskip}
    \addtolength{\itemsep}{-\baselineskip}
    \setlength{\parskip}{\baselineskip}
\item  $\lim_{r_{i,j}\rightarrow 0}\hbox{VXX}\left( r_{i,j}\right) =\infty$.
\item  $\nabla ^{2}\hbox{VXX}\left( x^{\ast }\right) $ semi-positive$,\left\|
\nabla \hbox{VXX}\left( x^{\ast }\right) \right\| \ll 1$ and $\frac{\left\| \nabla
\hbox{VXX}\left( x^{\ast }\right) \right\| }{\left| \hbox{VXX}\left( x^{\ast }\right)
\right| }<\delta _{0}$, where $0<\delta _{0}\ll 1$
\end{enumerate}
where XX is BU or LJ.
\end{proposition}

\begin{proof}
Without lost of generality, I assume a continuous search for a
cluster of size $j$ in A=$\{ p=\left( x,y,z\right)$ $ \in \Real^{3}
\ | \
\left\| p\right\| \leq r$\}, a ball of
ratio, r$\gg 0$ from where $\overrightarrow{\cdot }:$A$\rightarrow \Real^{3j}$.
The continuous search can be stated as generate $k$ random vectors $%
\left\{ \overrightarrow{C}_{j}^{l}\right\} _{l=1,\ldots ,k}$ with
coordinates in A and using a minimization routine to compute $\overrightarrow{%
C}_{j}^{\registered l}=(x_{l_{1}},x_{l_{2}},\ldots ,x_{l_{N}})$ such that
the property 2\ is fulfill.
Then select $x^{\ast }$ $=$ $\overrightarrow{C}_{j}^{\ast
}=\min_{l=1,\ldots ,k}\left\{ \overrightarrow{C}_{j}^{\registered l}\right\}
=\left( x_{1}^{\ast },\ldots ,x_{j}^{\ast }\right)$.
Repeating this procedure, A is exhaustively explored, therefore
this must provide a solution of
SOCCXX($j$). This means, EXX$\left( x^{\ast }\right)$ $\leq$ EXX$\left( x\right) ,\forall
x=(p_{1},\ldots ,p_{j})\in \Real^{3N}$, $p_{l}\in A$, $l=1,\ldots ,j$ .

The first property does not allow to have $p_{m}=p_{n}$ with $m \neq n$. Therefore,
$p_{i}$, $i=1,\ldots ,j$ are separate points in A$\subset \Real^{3}$. Moreover,
we can translate and set $x_{1}$ $=$ $(0,0,0)$ without changing the value of EXX.
Therefore $\exists\varepsilon _{0} >0$ for solving
SOCCXX($j$) and there is not need of the points, $p\in \Real^{3}$ such that
$\left\|
p\right\| <\varepsilon _{0}$ because they are never going to participate by
the condition 1.

For other coordinates of $x^{\ast }$, the second property provides a basin
or convexity region around it. A Taylor series for a potential function
around of $x^{\ast }\in \Real^{3j}$ for a direction $d\in \Real^{3j}$ with $%
0<\varepsilon \ll 1$ is
\begin{equation*}
\hbox{VXX}\left( x^{\ast }+\varepsilon d\right) = \hbox{VXX}\left( x^{\ast }\right)
+\varepsilon d\cdot \nabla \hbox{VXX}\left( x^{\ast }\right) +\frac{1}{2}%
\varepsilon ^{2}d\cdot \nabla ^{2} \hbox{VXX}\left( x^{\ast }\right) \cdot d+O\left(
\varepsilon ^{3}\right) .
\end{equation*}

By the convexity, $\frac{1}{2}\varepsilon ^{2}d\cdot \nabla ^{2}V\left(
x^{\ast }\right) \cdot d\geq 0$. Therefore
\begin{equation*}
\left| \hbox{VXX}\left( x^{\ast }+\varepsilon d\right) - \hbox{VXX}\left( x^{\ast }\right)
\right| \leq \left| \varepsilon \right| \left\| d\right\| \left\| \nabla
\hbox{VXX}\left( x^{\ast }\right) \right\| ,
\end{equation*}
\begin{equation*}
\frac{\left| \hbox{VXX}\left( x^{\ast }+\varepsilon d\right) - \hbox{VXX}\left( x^{\ast
}\right) \right| }{\left| \hbox{VXX}\left( x^{\ast }\right) \right| }\leq \left|
\varepsilon \delta _{0}\right| .
\end{equation*}

This property allows to select a truncated representation of $x^{\ast }$ for
some $\varepsilon _{1}=\varepsilon \delta _{0}>0$.
Therefore, $x=\left(
p_{1},\ldots ,p_{j}\right) \in \Real^{3j}$ such that $\left\| x-x^{\ast
}\right\| <\varepsilon _{1}$ and $\left\| p_{l}-p_{l}^{\ast }\right\|
<\varepsilon _{0}$ $l=1,\ldots ,j$ are not necessary to consider because
they are never going to improve the potential of $x^{\ast}$. Let $%
\varepsilon _{j}=\min \left\{ \varepsilon _{0},\varepsilon _{1}\right\} $.

Finally, $\Omega ^{o}=\cup _{j=2}^{\infty }C_{j}^{\ast }$ for $\varepsilon
^{\ast }=\min \left\{ \varepsilon _{j}\right\} _{j=2}^{\infty }$ and by the
construction of $\Omega ^{o}$ it follows immediately that SOCDXX($j$) and
SOCCXX($j$) have the same solution $\forall j\in \Natural$, $j\geq 2$.
\end{proof}

\begin{remark}
In the previous proposition, A is a subset of $\Real$. A has the
cardinality of $\Real$ but the proposition states that a discrete set of points of
A is sufficient in order to have the same solution between SOCDXX($j$) and
SOCCXX($j$) where XX is BU or LJ.
\end{remark}

\begin{proposition}~\label{prop:Omega_Local_p2}
The set $\Omega ^{l}$=$\{p_{i{_j}}$ $\in$ $C_{j}^{k}$  $|$  $C_{j}^{k}$ is a local optimal
cluster,
such that $C_{j}^{k}$ = $\{ p_{i{_j}}\}$, ${i{_j}} \in I^{k}$, $|I^{k}|=j$\} under a well potential function (a
potential that fulfill the properties of the
Proposition~\ref{prop:DiscreteLattice_p1}) is numerable.
\end{proposition}

\begin{proof}
Let assume that the $\cup _{j=1,\ldots ,N, |I^{k}|=j} C_{j}^{k},$ is not
numerable. This means $\exists K_{j} = \cup_{|I^{k}|=j} I^{k}$ is not
numerable. From the previous proposition, clusters can be created from the
continuous search depicted in the previous proposition, therefore each one fulfill properties 1)
and 2) and we add the coordinates of each $C_{j}$ founded to some $\Omega$.
Then for $j,\ \exists \varepsilon _{j}$ such that
$\overrightarrow{C}_{j}^{m},\overrightarrow{C}_{j}^{n}\in \Real^{3j}$ and
$\overrightarrow{C}_{j}^{m}\neq \overrightarrow{C}_{j}^{n}$ for $\forall m\neq
n$, $m,n\in K_{j}$. But each $\overrightarrow{C}_{j}^{m}$ can be approximated
$\underline{\overrightarrow{C}}_{j}^{m}\in \Rational^{3j}$ if $\left\|
\underline{\overrightarrow{C}}_{j}^{m} - \overrightarrow{C}_{j}^{m}\right\|
<\varepsilon _{j}$ $\forall m \in K_j$, which imply that $\cup _{m \in
K_{j}}\underline{C}_{j}^{m}\subset \Rational^{3}$ is not numerable!
\end{proof}

\begin{proposition}~\label{prop:Omega_global_p3}
The set
$\Omega ^{o}$=$\{p$ $\in$ $C_{j}^{\ast}$ $|$ $C_{j}^{\ast}$is the global optimal cluster
 $\forall$ $j\in\Natural$\} is numerable.
\end{proposition}

\begin{proof}
$\Omega^{o}$ is the union of finite
set of points, therefore is numerable.
\end{proof}

\begin{proposition}~\label{prop:Omega_basin_p4}
The set $\Omega ^{b}$=$\{p$ $\in$ $C_{j}^{k'}$  $|$  $C_{j}^{k'}$ is a cluster in
a basin for the optimal local clusters $C_{j}^{k}$ of size $j$,
$\forall$ $j$ $\in$ $\Natural$\} is not numerable.
\end{proposition}

\begin{proof}
Given an optimal local cluster $C_{j}^{k}$ by the condition 2
of Proposition~\ref{prop:DiscreteLattice_p1}, $\exists \delta_0 >0$ and
$d\in\Real^{3j}$ such that
$\forall$ 0 $\le$ $\delta \le \delta_0$
$\overrightarrow{C}_{j}^{k\delta}$ = $\overrightarrow{C}_{j}^{k} + \delta d$, then
$\overrightarrow{C}_{j}^{k\delta}$ $\rightarrow$ $C_{j}^{k}$,
$\forall$ 0 $\le$ $\delta \le \delta_0$.
Therefore, $\Omega ^{b}$ is union of non-numerable sets for each
optimal local cluster.
\end{proof}

\begin{proposition}~\label{prop:set_Omega_Omega_lattice_p5}
Exist a set, $\Omega ^{\ast}$ such that $\exists$
$C_{k}^{'}$ $\in$ $\Omega ^{\ast}$,  $C_{k}^{'}$ $\rightarrow$ $C^{\ast}_{k}$ $\in$ $\Omega ^{o}$
$\forall k\ge2$.
\end{proposition}
\begin{proof}
The results follows from $\Omega ^{b} \bigcap \Rational^{3}$ $\neq$ $\emptyset$.
\end{proof}

\begin{remark}
The last proposition states that $\Omega ^{\ast}$=$\Omega ^{b} \bigcap \Rational^{3}$ is
one trivial set where $\exists$ $C_{j}$, such that $C_{j} \rightarrow C^{\ast}_{j}$,
$\forall j \ge 2$.
In order to find
IF, I add each putative optimal LJ cluster, $C^{\ast}_{j}$, $j$=2,$\ldots$,1000.
It was a surprise that taking $C_{13}^{\ast}$ and adjusting the other putative optimal LJ
clusters to it, the IF structure show up naturally.
\end{remark}

The next proposition states that is not possible to find a function from
$\Natural$ to $\Omega$ capable to give all optimal clusters. Hereafter,
$\Omega$ is a numerable set and could be $\Omega^{l}$ or  $\Omega^{\ast}$.

\begin{proposition}~\label{prop:CantorDiagonal_p6}
$\nexists s:\Natural \rightarrow \Omega$, $%
s(j)=C_{j}=\left\{ x_{i_{1}},\ldots ,x_{i_{j}}\right\} $, $x_{i_{k}}\in
\Omega$ $\forall k=1,\ldots ,j$ such that $s\left( j\right)
=C_{j}^{\ast },\forall j\in \Natural$.
\end{proposition}

\begin{proof}
The proof is based in building a Cantor's Diagonal schema. Suppose that such
selection function, $s$, exists for some order in $\Omega$, which is
numerable. Then changing the first particle in $\Omega$ that belong
to the $C_{2}^{\ast }$ for any other different and far way from this one,
the new order $\Omega _{2}$ is an order where $s$ can not give
$C_{2}^{\ast}$. This procedure is repeated for $C_{k}^{\ast}$, $k=3,\ldots
,\infty $ giving $\Omega _{k},$ $k=3,\ldots ,\infty $ where $s$ can not
give $C_{k}^{\ast}$. The set of points that belong to the diagonal differs
from all the enumerations $\Omega _{k},$ $k=2,\ldots \infty $ which
are all the possible enumerations of $\Omega$!
\end{proof}

\begin{proposition}~\label{NP}
It is not possible to find an algorithm with polynomial complexity
to solve SOCD($j$), $\forall j$ $\in$ $\Natural$.
\end{proposition}

\begin{proof}
If such algorithm exist, it means that it is possible to find $M\in \Natural$,
$M>0$, $T(j) \leq O\left( j^{M}\right) \forall j\in \Natural$
where $T$ is the time to
take this algorithm to find $C_{j}^{\ast}$. But this means that such
algorithm is the function $s$ of the previous proposition!
\end{proof}

\begin{remark}
The last proposition states that it is not possible to build a selection
function in computational time for finding all the optimal
clusters from $j\geq2$, $j\in\Natural$. In particular, it states that
the complexity of finding all the optimal clusters from $j=2,\ldots ,\infty$
cannot be derived from some arbitrary inhered order of $\Omega$ (numerable).

One of the reason of the success of the methods for SOCCXX($j$) and SOCDXX($j$)
is the combination of different lattices, which are subsets of $\Omega^{\ast}$. Moreover,
from the cardinality point of view, $\Omega ^{\ast}$ is a smaller set than
$\Real^{3}$, and it seems to be the right search space to explore the
complexity of the NP problem SOCDXX($j$), $\forall j\geq 2.$
\end{remark}

The Proposition~\ref{prop:DiscreteLattice_p1} permits to build a discrete set of
points after the solution of the SOCDXX($j$), $\forall j\geq 2$ and also
 proofs that exist a set where SOCC and
SOCD have the same solution, therefore SOCC
is not efficient way for SOC.

It was not easy to build a set of points as
a discrete lattice from basin regions for solving
SOCDLJ($j$), $\forall j\geq 2$ for the putative optimal clusters,
i.e., a set of points in $\Real^3$ with a regular structure.
However, combining IC and FC with an appropriate separation was the surprising answer.
Section~\ref{sc:results} presents numerical experiment of the propositions of this section.
Particularly,  for SOCDLJ($j$) a lattice and a set,
IF9483 and MIF1739, are presented
with the property, $\exists C_{j} \rightarrow C_{j}^{\ast }$, $j=2,\ldots ,1000$
in the sense that ELJ($C^{\ast}_{j}$) are
the putative optimal potential LJ values from~\cite{http:CCD} or better ones.

\subsection{Symmetry reduces the Complexity of Potential's Evaluation}~\label{ssc:simmandPot}
For SOC, the symmetry inhered from a lattice $\Omega^{\ast}$ can reduce
the number of operations to evaluate a potential
function. A simple example, taking $\Omega^{\ast}$= IF and $C_{13}$ as a
centered icosahedron inside of a ball of ratio,
$r^{\ast}=\sqrt[6]{2}$.
Here, without lost of generality the points of $C_{13}=\left\{
p_{1},\ldots ,p_{13}\right\} $ are:  \newline
  $p_{1}$ =  (0.000000000000, 0.000000000000, 0.000000000000),  \newline
  $p_{2} $ =  (0.000000000000, 1.081838288553, 0.000000000000), \newline
  $p_{3} $ =  (0.967625581547, 0.483812790773, 0.000000000000),  \newline
  $p_{4} $ =  (0.299012748890, 0.483812790773, -0.920266614664),\newline
  $p_{5} $ =  (-0.782825539663, 0.483812790773, -0.568756046574), \newline
  $p_{6} $ =  (5,-0.782825539663, 0.483812790773, 0.568756046574),\newline
  $p_{7} $ =  (0.299012748890, 0.483812790773, 0.920266614664), \newline
  $p_{8} $ =  (0.782825539663, -0.483812790773, -0.568756046574),\newline
  $p_{9} $ =  (-0.299012748890, -0.483812790773, -0.920266614664), \newline
  $p_{10} $ =  (-0.967625581547, -0.483812790773, 0.000000000000),\newline
  $p_{11} $ =  (-0.299012748890, -0.483812790773, 0.920266614664), \newline
  $p_{12} $ =  (0.782825539663, -0.483812790773, 0.568756046574), and \newline
  $p_{13} $ =  (0.000000000000, -1.081838288553, 0.000000000000).

Then by the symmetry on these points,
EXX$\left( C_{13}\right)$ =12\hbox{VXX}$_{1,2}$+30\hbox{VXX}$_{2,3}$+
30\hbox{VXX}$_{2,8}$+ 6\hbox{VXX}$_{2,13}$

which requires five points \{$p_{1}$, $p_{2}$, $p_{3}$, $p_{8}$,
$p_{13}$\}  and
the four factors $\hbox{VXX}_{1,2}$, $\hbox{VXX}_{2,3}$, $\hbox{VXX}_{2,8}$, and
$\hbox{VXX}_{2,13}$ for any
potential function. But without symmetry, EXX$\left( C_{13}\right) =$
$\sum_{i=0}^{n-1}\sum_{j=i+1}^{13}\hbox{VXX}{i,j}$ needs thirteen points and
$13(13+1)/2=91$ factors $\hbox{VXX}_{i,j}$. In particular for this cluster
ELJ($C_{13}$) = -44.326801 = ELJ($C^{\ast}_{13}$).

\section{Methods for IF}~\label{sc:methods}
There are several references that explains how to build IC and
FC~\cite{jgo:Leary1997, ps:Maier1992, ape:Solovyov2003, jpca:Xiang2004A}.
I build an IF for the Lennard-Jones Potential using the
propositions of the previous section.
The first approach was to use Proposition~\ref{prop:DiscreteLattice_p1} to build a set
from the $C_{j}^{\ast }$, $j=2,\ldots ,1000$ by adding in growing order the
points of each $C_{j}^{\ast }$ but after few numerical experiments, a fixed
combination of an IC and an FC together with an step ratio, $r^{\ast
}=1.08183839$, between shells makes possible to build an IF such that $\exists
C_{j}$ $\rightarrow C_{j}^{\ast }$ using a
minimization procedure based on the CGM.
The possibility to find a lattice was predicted
by Proposition~\ref{prop:set_Omega_Omega_lattice_p5}.
The value $r^{\ast }$ correspond to the icosahedron
described in section~\ref{ssc:simmandPot}.
In addition, this is the only cluster where SOCDLJ(13) does not need a relaxation, moreover,
ELJ($C_{13})=ELJ(C_{13}^{\ast}$). Note that the particular order of the
sequence of points is very important to reproduce the putative optimal LJ clusters,
therefore
tables~\ref{tb:idMinLattice_0}-\ref{tb:idMinLattice_15} give all the
coordinates of MIF1739.

\begin{figure}
\centerline{
\psfig{figure=\IMAGESPATH/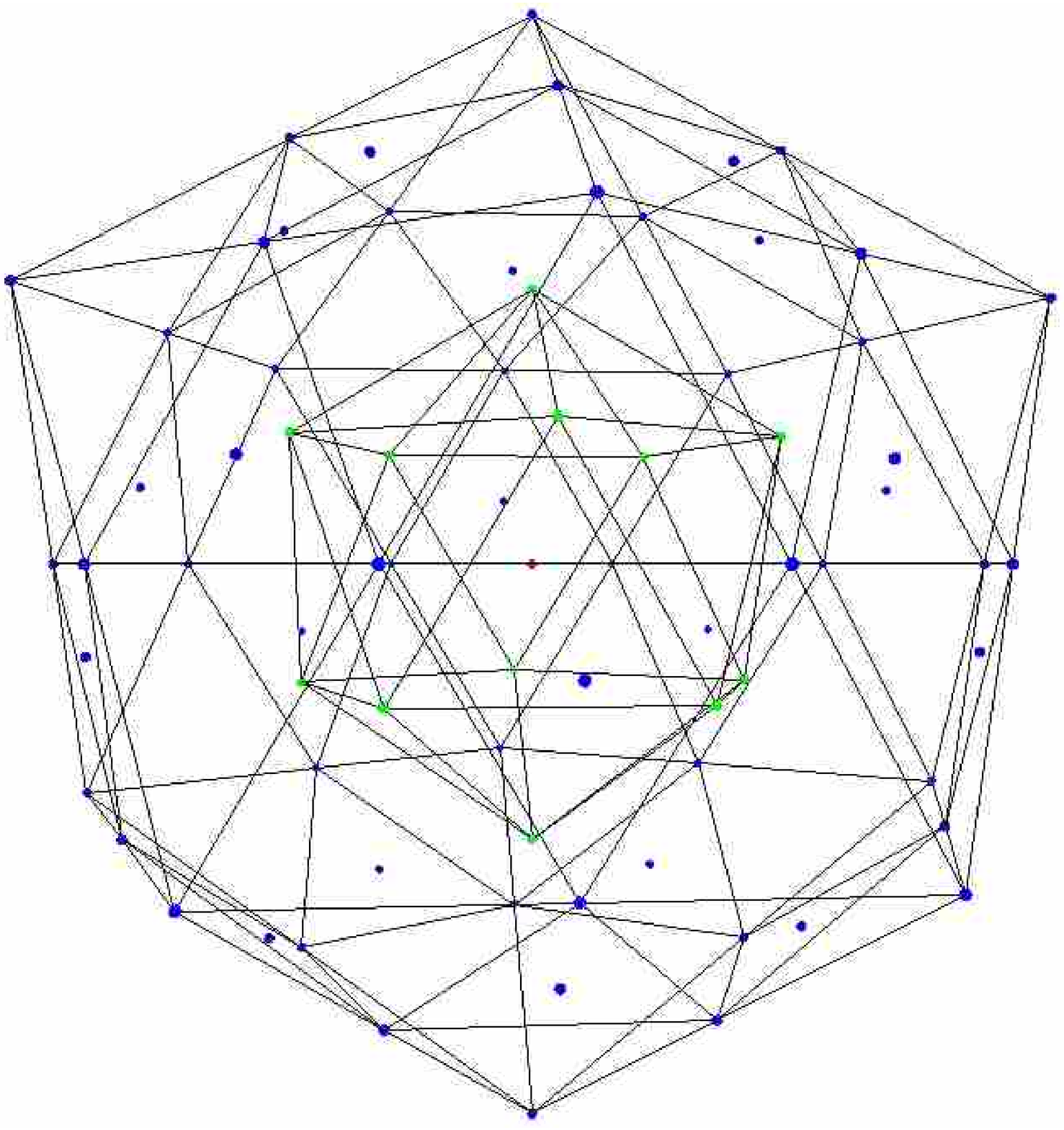,height=40mm}
\psfig{figure=\IMAGESPATH/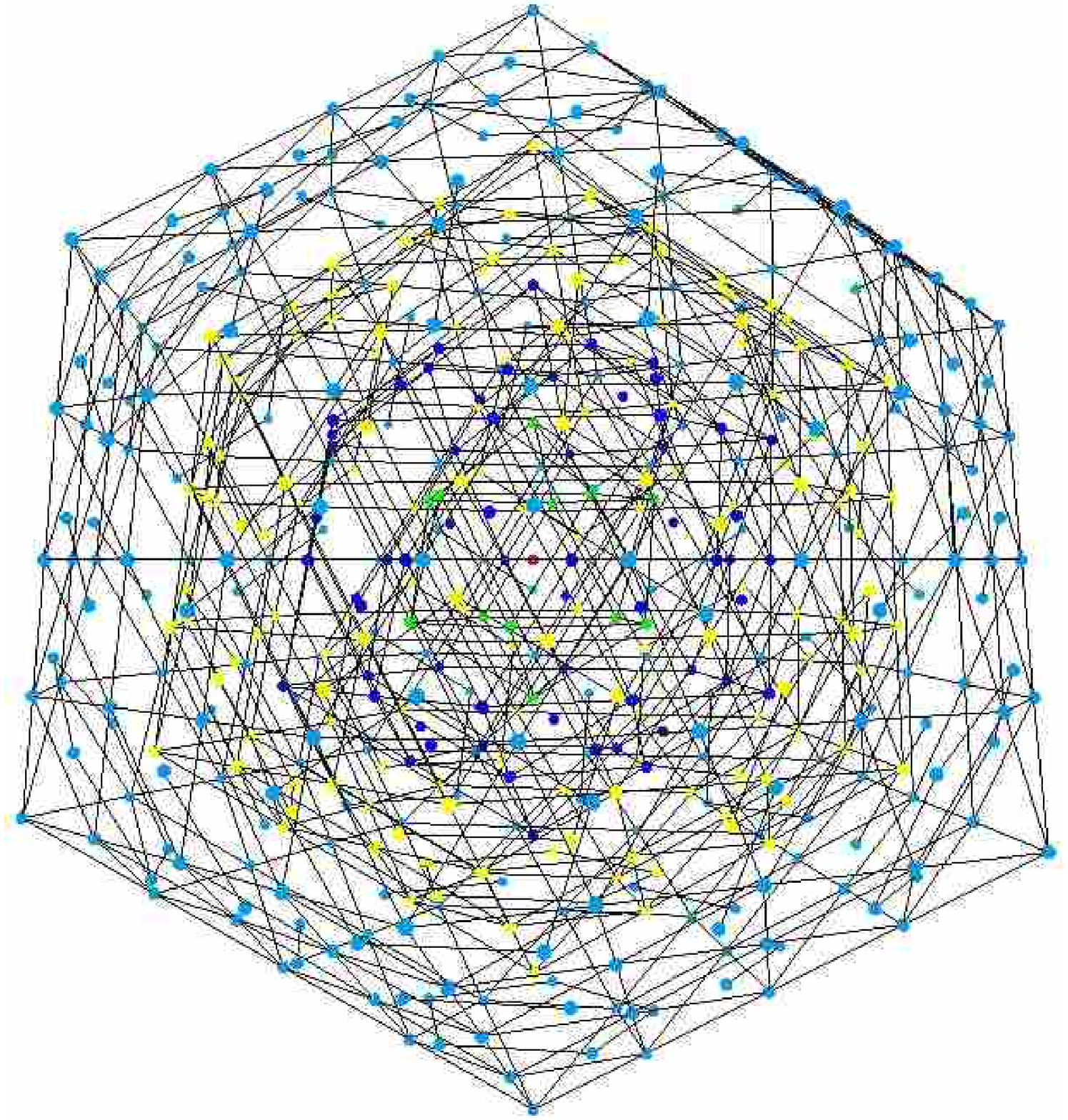,height=40mm}
\psfig{figure=\IMAGESPATH/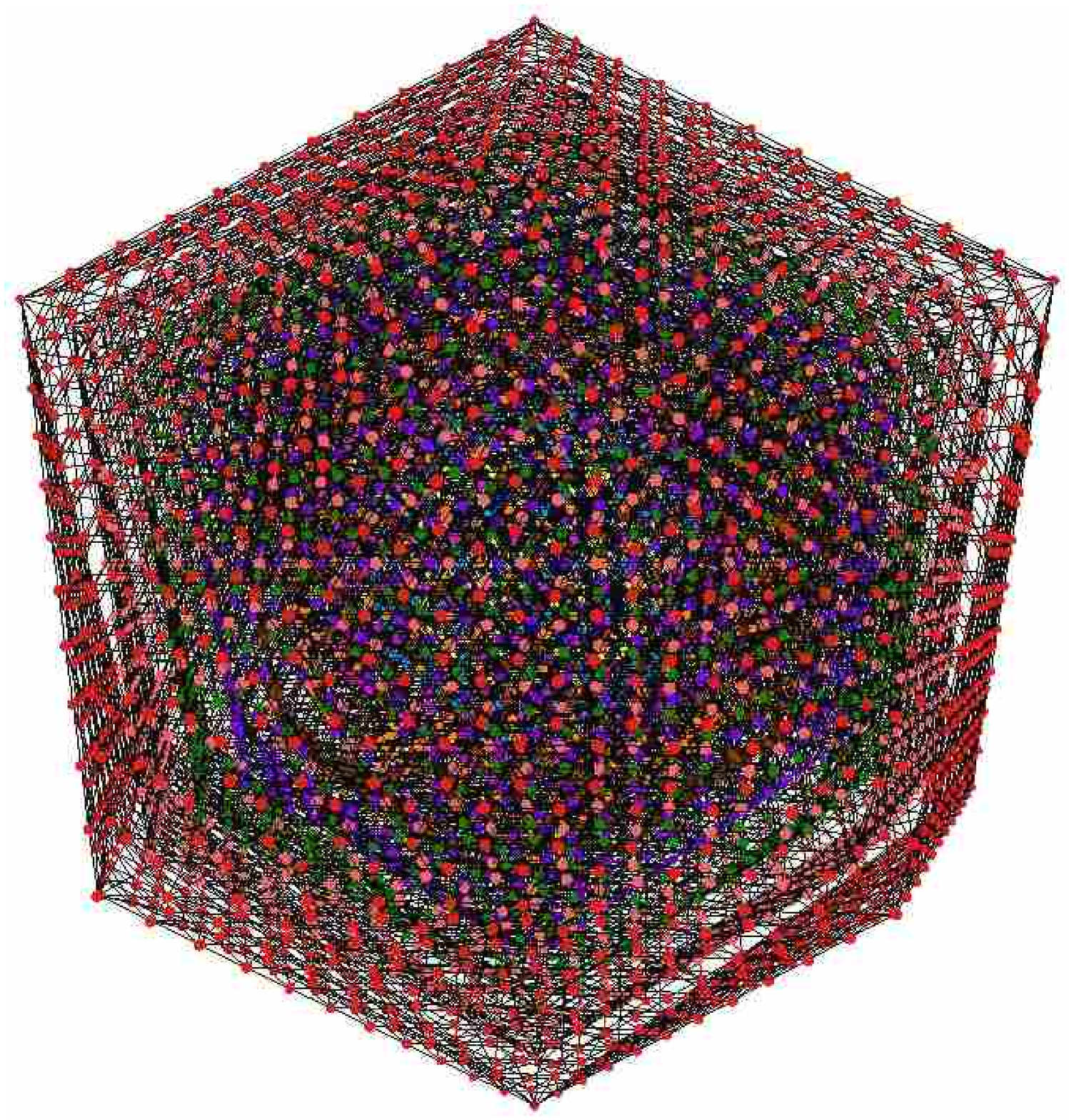,height=40mm}
}
\centerline{ \makebox[1.5in][c]{a)}\makebox[1.5in][c]{b)}\makebox[1.5in][c]{c)} }
\caption{a) IF75, b) IF509, and c) IF9483, where IF =
IC$\bigcup$FC.}~\label{fig:latticeicfc9483}
\end{figure}

\begin{figure}
\centerline{
\psfig{figure=\IMAGESPATH/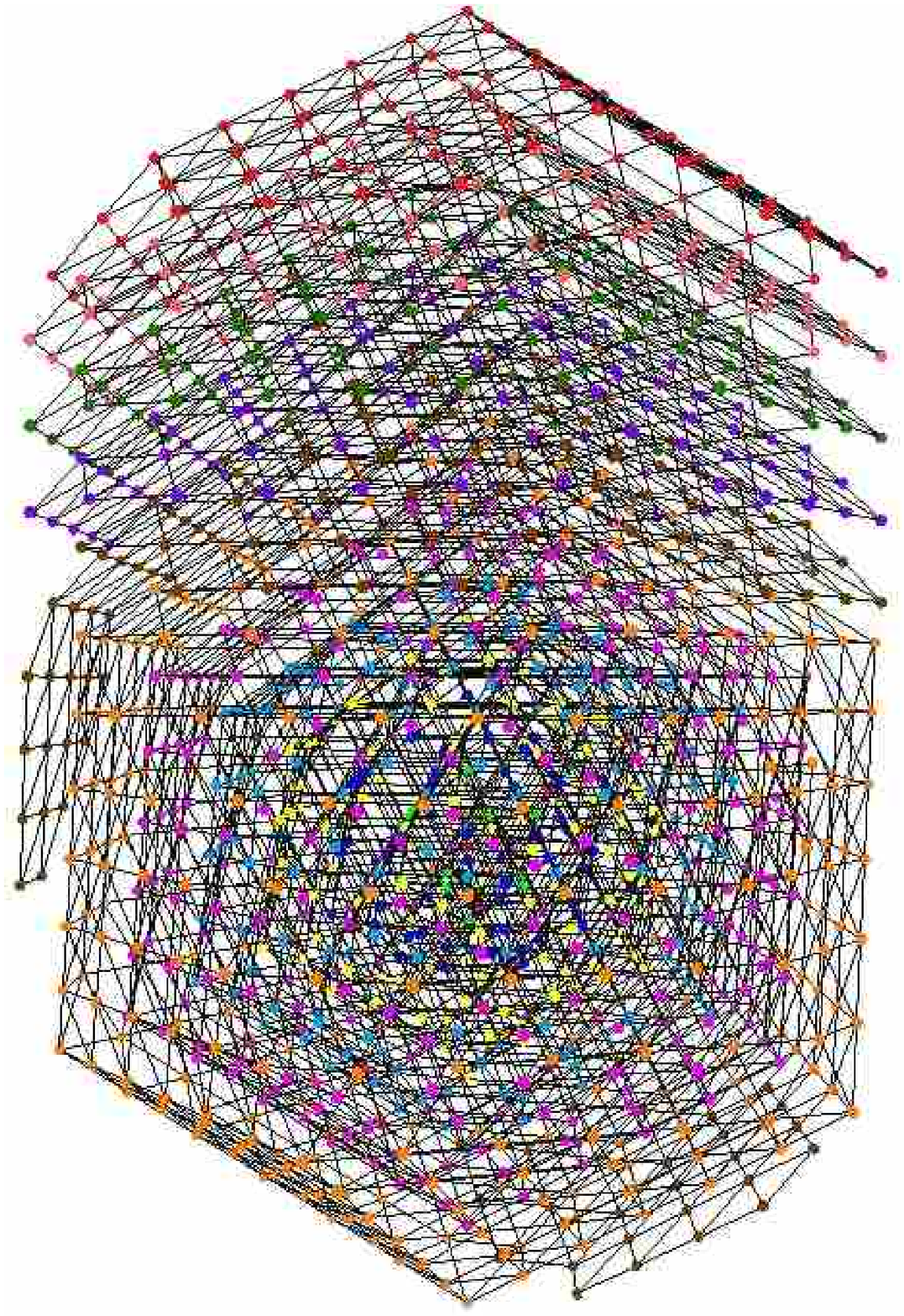, height=50mm}
\psfig{figure=\IMAGESPATH/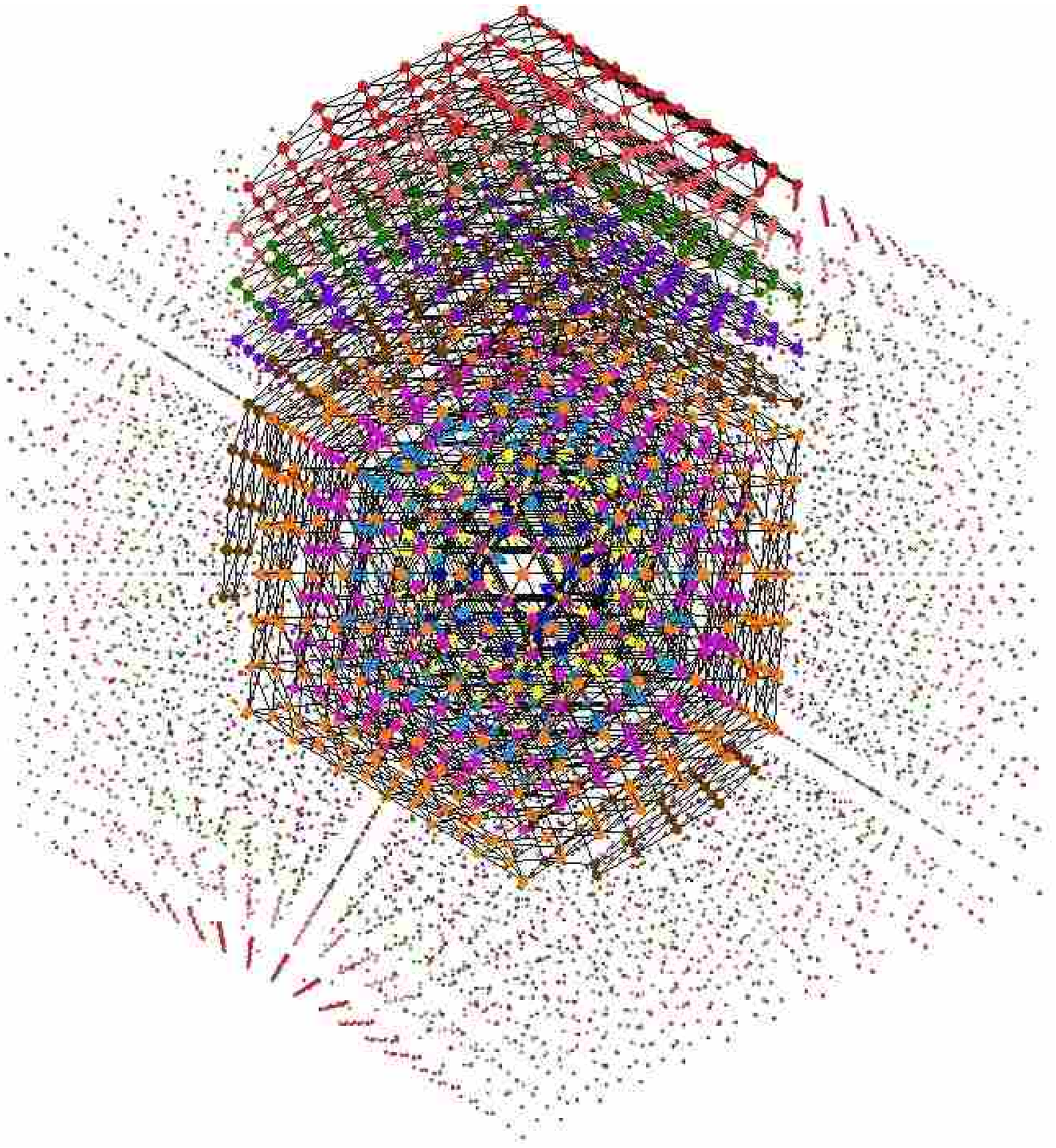, height=50mm}
} \centerline{ \makebox[1.7in][c]{ a) }\makebox[1.6in][c]{ b) } } \caption{a)
MIF1739 and b) MIF1739 inside of IF9483. MIF1739 and IF9483 contain $C_n$, such
that $C^*_n$ = $\min$($C_n$) $n=2,\ldots,1000$.}~\label{fig:min_lattice1739}
\end{figure}

Give the result in IF9483 is lengthy but with MIF1739 is a short way to
present it. Meanwhile MIF1739 contains only 1739 points, the complete IF with
the same property needs 9443 points. The number of points of the IF9483 comes
from sum of the magic numbers of the particles of the complete shells IC and FC for
the shells 0 to 11. Figure~\ref{fig:latticeicfc9483} depicts IF75, IF509, and
IF9483. Figure~\ref{fig:min_lattice1739} depicts MIF1739 alone and inside of an
IF9483.

The construction of the MIF1739 is done by the
following algorithm:
\begin{TightList}
\item [{\bf 1:~~}] $C_{1000}^{^{\prime }}$ is a rotation of $C_{1000}$ to set as many
particles as possible of the last shell over the semi-axes $Y^+$.
\item [{\bf 2:~~}] for j=999, 2
\item [{\bf 3:~~}]\hskip 0.5cm $C_{j}^{\prime }=\min $Adj$\left( C_{j+1}^{^{\prime }}\text{,
}C_{j}\right) $ over all rotations of $C_{j}$ based on the symmetry of the
centered icosahedron in
\item [{\bf ~~}]\hspace{0.5cm}IF.
\item [{\bf 4:~~}] end for
\item [{\bf 5:~~}] MIF=$\cup _{j=2,\ldots ,1000}C_{j}^{\prime}$
\end{TightList}

\begin{remark}
In the step 3, for the clusters that are not centered IC or FC as the
$C_{38}^{\ast }$, $C_{98}^{\ast }$ and the Ino's decahedrons the rotation are
five and they are around of the axes $Y$. In addition, these clusters exist on infinity
positions of an infinite IF, therefore they were manually translated to closed
position toward the center of IF and over the semi-axes $Y^+$.
\end{remark}

The tables~\ref{tb:On_off_0}-\ref{tb:On_off_32} allow to build all the
$C_{j}^{\ast }$ from MIF1739. The algorithm is
\begin{TightList}
\item [{\bf 1:~~}]  $C_{1000}$=\{$p_{i}\in$ MIF1739 $|$ $i$ is in the column On in
tables~\ref{tb:On_off_0}-\ref{tb:On_off_32} for $C_{1000}$\}.

\item [{\bf 2:~~}] $C_{1000}^{\ast}=$ $\min
$($C_{1000}$).
\item [{\bf 3:~~}]  for j=999 to 2
\item [{\bf 4:~~}] \hskip 0.5cm  $C_{j}$=$C_{j+1}\setminus
$\{$p_{i}\in $MIF1739 $|$ $i$ is in the column Off in
tables~\ref{tb:On_off_0}-\ref{tb:On_off_32} for $C_{j}\} \cup $ \{$p_{i}\in
$MIF $|$ $i$ is in the column
\item [{\bf ~~}]\hspace{0.5cm}On in tables~\ref{tb:On_off_0}-\ref{tb:On_off_32} for
$C_{j}\} \cup $ \{$p_{i}\in $MIF $|$ $i$ is in the column On
in tables~\ref{tb:On_off_0}-\ref{tb:On_off_32} for $C_{j}$\}
\item [{\bf 5:~~}]\hspace{0.5cm}$C_{j}^{\ast }=$ $\min$($C_{j}$).
\item [{\bf 6:~~}] end for.
\end{TightList}

The tables~\ref{tb:Changes_0}-\ref{tb:Changes_9} give the type of the
putative optimal LJ clusters in IF as: 1=IC, 2=Ino's decahedron (ID), 3=
truncated octahedron (TO), and 5=FC; the initial, optimal and
difference of LJ,  and the value of
Adj$\left(C_{j+1}^{^{\prime }}\text{, }C_{j}\right)$.

The classification of cluster is done automatically by identified the particles
of a cluster with the type of particles of MIF1739, type is IC  when all
the particles of a cluster are only IC around the center of IF,
type is ID if there is a particle in
the cluster close to the center of mass of the cluster, such that it is on the semi-axes
$Y^+$, type is
TO when all particles are IC and they are inside of a tetrahedron formed by
three internal axis of the IF, type is FC when at least one particle
of the cluster is FC.

\subsection{Methods for search in IF}
The classification of the algorithms for SOC has many
different approaches~\cite{arcp:Wille1999}. Here tree classes are depicted on
a scale from comparisons versus properties (necessary and sufficient conditions
of a problem):
\begin{description}

\item[Exhaustive Algorithm] It explores a search space of a problem verifying
that the global optimum is founded. Here, for the comparisons an objective
function is used to provide the way to determine the optimum. For small discrete and
continuous problems, the algorithm's complexity is not an issue. There are many
global optimization methods that work fine for low dimension problems. By
example, the classical Grid Method divides the search space in small boxes.
Therefore, it can locate the global minimum by an exhaustive search.
Generally, the complexity grows rapidly, exponentially and for the NP
problems, there is not hope that exist a polynomial complexity algorithm.

\item[Scout Algorithm] It is a fact widely accepted that using previous
knowledge and natural (Physical, Chemical, Thermodynamical, Biological,
Medical, and so on) understanding of the process and phenomena involved in a
problem will help to design an efficient method. Here, some authors argue about
how much and what type of knowledge could be used. Other authors apply the
rule: "Achievements kill doubt". From a practical point of view, this category
contains algorithms that can use all or a part of whatever is available. Most
of the methods for clusters optimization belong to this category. A method in
this category could find a novel solution without a guarantee of
optimality. Most of the justifications for algorithm's efficiency are done by
numerical experiments on a set of problems (benchmark). This type of analysis
depends on the researcher and his/her particular computers and working
conditions. Therefore, a claim that SOC can be done in polynomial time O($n^3$)
is a very strong statement. If this could be extended and proved then the NP
problems will be class P! Exploring IF is like a travel in one axes of the IF
towards axes Y$^+$. MIF1739 is also the minimum region to explore without
to many repetitions caused by the icosahedral symmetry.

\item[Wizard Algorithm] There are problems where necessary and sufficient
optimal conditions can be established for the solution. Generally
these methods  are efficient and they do not need to do exhaustive comparisons.
By example, an optimization problem with a convex function can be efficiently
solved by the Conjugate Gradient method or by the large family of Newton and
Quasi-Newton methods.
\end{description}

In \cite{aml:Barron1997} the Peeling Method was presented. This method is
similar to the algorithm 4.1 of Maier {\it et al}.~\cite{ps:Maier1992}. It
executes a greedy strategy to set Off the particles on the outer shell of a
cluster and to set On particles inside of a lattice that are neighbor of the
previous ones. In this way, an small change is done and a minimization routine
computes the potential of this cluster. The Modified Peeling Method has
three basic operations: forward, backward, itself. Moreover, the basic idea is to
adjust a cluster by turning on a neighbor particle and turning Off a particle
in the outer shell of a cluster or the centered particle. The complexity of
each operation is $O($n$^{3})$ for a forward ($C_{n}$ to $C_{n+1}$) and
backward ($C_{n}$  to$C_{n-1}$), and $O($n$^{4})$ for itself. These operations
are quite similar to the proposed pivot algorithm~\cite{jgo:Xue1994}, reverse
greedy operator by Leary~\cite{jgo:Leary1997}, the ''final repair'' step of
Hartke~\cite{jcc:Hartke1999}, Fusion Process by adding one particle to C$_{n}$
of Solov'yov {\it et al}.~\cite{ape:Solovyov2003}, and the greedy search
method~\cite{jpca:Xiang2004B} but the novelty is that in right search space
this Modified Peeling Method is capable to find new solutions or reproduce the
existent ones.

Given $C_{n}=\left\{p_{i_{1}},p_{i_{2}},...,p_{i_{n}}\right\}$,
 $p_{i_{j}}\in$ IF, $\forall j=1,\ldots,n$, computes the sets:
\begin{list}{}{}
    \addtolength{\topsep}{-\baselineskip}
    \addtolength{\itemsep}{-\baselineskip}
    \setlength{\parskip}{\baselineskip}
\item $K_C$ = \{ $i_k$ $|$ $i_{k}$ such that $\exists p_{i_{k}} \in C_{n}$,
$\exists p_{i_{l}} \in \hbox{\ IF\ } \setminus C_{n}$, where $p_{i_{k}}$ and
$p_{i_{l}}$ are neighbors or $i_l=0$ (centered particle of  $C_{n}$)\}.
\item $K_{\hbox{IF}}$ = \{$i_l$ $|$ $i_{l}$ such that $\exists p_{i_{k}} \in
C_{n}$, $ p_{i_{l}} \in \hbox{\ IF\ } \setminus C_{n}$, where $p_{i_{k}}$ and
$p_{i_{l}}$ are neighbors\}.
\end{list}

The Modified Peeling Method executes this three operations:
\begin{TightList}
\item [{\bf forward:~~}]\hspace{0.3cm}$C^\ast_{n+1}$ = $\min_{\hbox{set On\   } j  \in K_{\hbox{IF}} }$
\{ $p_{j}$ \} $\bigcup$ $C_{n}$ \}.
\item [{\bf backward:~~}] $C^\ast_{n-1}$ = $\min_{\hbox{set Off \ } k \in K_C}$ $C_{n}
\setminus \{ p_{k}$ \}.
\item [{\bf itself:~~}]\hspace{0.8cm}$C^\ast_{n}$ = $\min_{ \hbox{set Off\ } k \in K_C \hbox{, set On\
} j \in K_{\hbox{IF}} }$ $C_{n} \setminus \{ p_{k}$ \} $\bigcup$ \{ $p_{j}$ \}.
\end{TightList}

\begin{remark}
These operations do not guarantee global optimality.
\end{remark}

For SOCD, it seems that a wise algorithm only could exist after founded the
optimal clusters. One of the reason is that the global optimality is an open
question for clusters with more than five particles~\cite{arcp:Wille1999}. The
efficient method after founded the optimal cluster is a telephone directory
method. This type of method uses a table with the data needed. The only
operation is retrieving an entry of the table by an index, i.e., given the
indexed table $(t[1],\ldots,t[K])$, $K\in\Natural$, $K>0$ and $j\in[1,K]$ then
retrieve $t[j]$.

\begin{proposition}~\label{prop:TelephoneIFOPT_p7}
 The telephone directory method in $\Omega ^{o}$
 (see Proposition~\ref{prop:Omega_Local_p2}) has complexity $O(1)$.
\end{proposition}
\begin{proof}
Here, the only operation is to select the points of $\Omega ^{o}$ from a table
with the collection of the indexes of each optimal cluster. Therefore the
complexity is only one operation.
\end{proof}

\begin{remark}
$\Omega ^{o}$ is not symmetric because the PES of small clusters.
If the function $s$ of the Proposition~\ref{prop:CantorDiagonal_p6} exists, its
complexity would be one!
\end{remark}

\begin{proposition}~\label{prop:TelephoneIF_p8}
The telephone directory method on $\Omega^{\ast}$, and particularly on IF
 has complexity $O(n^3)$.
\end{proposition}
 \begin{proof}
This limit comes from the complexity of the relaxation multiplying by the
complexity of the evaluation of a pair potential function for $n$ particles. It
is well know that the a minimization method as the CGM converges in at most the
number of variables of the problem, which in this case is $3n$ and the worst
case for evaluation of the potential is $O(n^2)$.
\end{proof}

For space consideration, a table of the optimal clusters for a telephone
directory method is not given. However, this table can be obtained from
the tables \ref{tb:On_off_0}-\ref{tb:On_off_32}. The algorithm for the
telephone directory method on this table is:
\begin{TightList}
\item [{\bf 1:~~}] $C_{n}$=\{ $p_{i}\in$ MIF1739 $|$ $i$ is in the column On in the
telephone table of optimal clusters \}.
\item [{\bf 2:~~}] $C_{n}^{\ast}=$ $\min $($C_{n}$).
\end{TightList}

\begin{remark}
The telephone directory method with MIF1739 or
IF9483 has polynomial complexity $O(n^3)$ for SOCD($n$). An estimation of
complexity less than $O(n^3)$ by other authors is probably biased by the particular data
and environment of their numerical experiments. It is possible that this small
difference was influenced by the data over the number of iterative steps of the
relaxation or minimization procedure on a very good initial population of clusters,
which is highly possible if an algorithm gets initial random clusters
from IC, ID, and FC lattices.
\end{remark}

The results presented in the next Section show that optimal clusters are not
always around on the same localization in IF. Therefore exploring IF by
Exhaustive Algorithms is hard and it has exponential complexity caused by the combinations of
possible particles.

\begin{proposition}~\label{prop:Adj_p9}
The function Adj is not bounded.
\end{proposition}
 \begin{proof}
Let assume that $\exists M>0$ such that Adj$(C_{n}$, $C_{m})$ $<M$ $\forall n,m
\geq 2$. This means that for $m \gg M$ because the number of adjust is bounded,
the complexity of founding SOCD($m$) from SOCD($m-1$) or from SOCD($m+1$) is
$O(m^2)$ which imply the existence of the function $s$ of
Proposition~\ref{prop:CantorDiagonal_p6}.
\end{proof}
\begin{remark}
Also, if Adj is bounded, the complexity of a telephone directory method in an
sufficient large appropriate lattice IF or set $\Omega^{l}$
(see Proposition~\ref{prop:Omega_Local_p2})
 is greater than $O(m^2)$, the complexity  of SOCD($m$) for large clusters, which is also
impossible, moreover then SOCD is not NP!
\end{remark}

Adding previous knowledge to build an scout algorithm is seemed to be the way
to address SOCD, and the complexity can be decreased by symmetry. However, it
easy to see that exists a cluster where symmetry cannot reduce the worst  case
of the complete pair potential's evaluation for some clusters, which is $O(n^2)$.
Finally, by the Proposition~\ref{prop:TelephoneIF_p8}, $O(n^3)$  is a lower limit
of the complexity of SOCD($n$) in IF.

\section{Results}~\label{sc:results}
MIF1739 is a discrete lattice for LJ where $\exists$ $C_{n}$, such that
$ C_{n}^{\ast }$ = $\min(C_{n})$ under LJ and $C_{n}^{\ast}$ is the putative optimal cluster for
$n=2,\ldots ,1000$. I took the data from CCD~\cite{http:CCD} to verify that my results.
Data for $C_{n}^{\ast }$ $n=148,\ldots ,309$ came from Romero,
Barr\'{o}n, and G\'{o}mez \cite{cpc:Romero1999A} and data for
$C_{n}^{\ast }$, $n=310,\ldots ,1000$ came from the recently results of
Shao {\it et al}.~\cite{jpca:Xiang2004A, jpca:Xiang2004B}.
I was able to repeat and even improve some of the putative optimal LJ
clusters. From this comparison new putative optimal LJ clusters are $
C_{537}^{\ast }$, $C_{542}^{\ast }$, $C_{543}^{\ast }$, $C_{546}^{\ast }$,
$C_{547}^{\ast }$, $C_{548}^{\ast }$, $C_{664}^{\ast }$, and
$C_{813}^{\ast }$. The
particular interest are the $C_{542}^{\ast }$, $C_{543}^{\ast }$, $%
C_{546}^{\ast }$, $C_{547}^{\ast }$, and $C_{548}^{\ast }$ because to my
knowledge they are first optimal LJ clusters type IC with central vacancy (CV)
reported in the shell 310-561. The prediction of CV was stated in~\cite{cp:Shao2004}
 for the next shell,
562-923. Table~\ref{tb:newClusters} summarizes these
new results.

\begin{table}
\centerline{
\begin{tabular}[h]{ r | r r r l} \hline
\multicolumn{1}{c}{n} & \multicolumn{1}{c}{$E^{\text{new}\ast }$} &
\multicolumn{1}{c}{$E^{\text{old}\ast }$ } &
\multicolumn{1}{c}{$E^{\text{new}\ast }-E^{\text{old}\ast}$} &
\multicolumn{1}{c}{T} \\ \hline
537 & -3659.52825 & -3659.70629 & -0.17804 & 1 \\ \cline{2-5}
542 & -3698.95403 & -3699.22727 & -0.27324 & 1/CV \\ \cline{2-5}
543 & -3706.94784 & -3708.21090 & -1.26306 & 1/CV \\ \cline{2-5}
546 & -3730.50408 & -3730.69222 & -0.18814 & 1/CV \\ \cline{2-5}
547 & -3738.38788 & -3738.68065 & -0.29277 & 1/CV \\ \cline{2-5}
548 & -3746.37071 & -3747.67942 & -1.30871 & 1/CV \\ \cline{2-5}
664 & -4596.1978 & -4596.1971 & -0.0007 & 2 \\ \cline{2-5}
813 & -5712.2517 & -5712.2506 & -0.0011 & 1 \\ \hline
\end{tabular}
}
\caption{Novel $C^*_n$.}~\label{tb:newClusters}
\end{table}

\begin{figure}
\centerline{
\psfig{figure=\IMAGESPATH/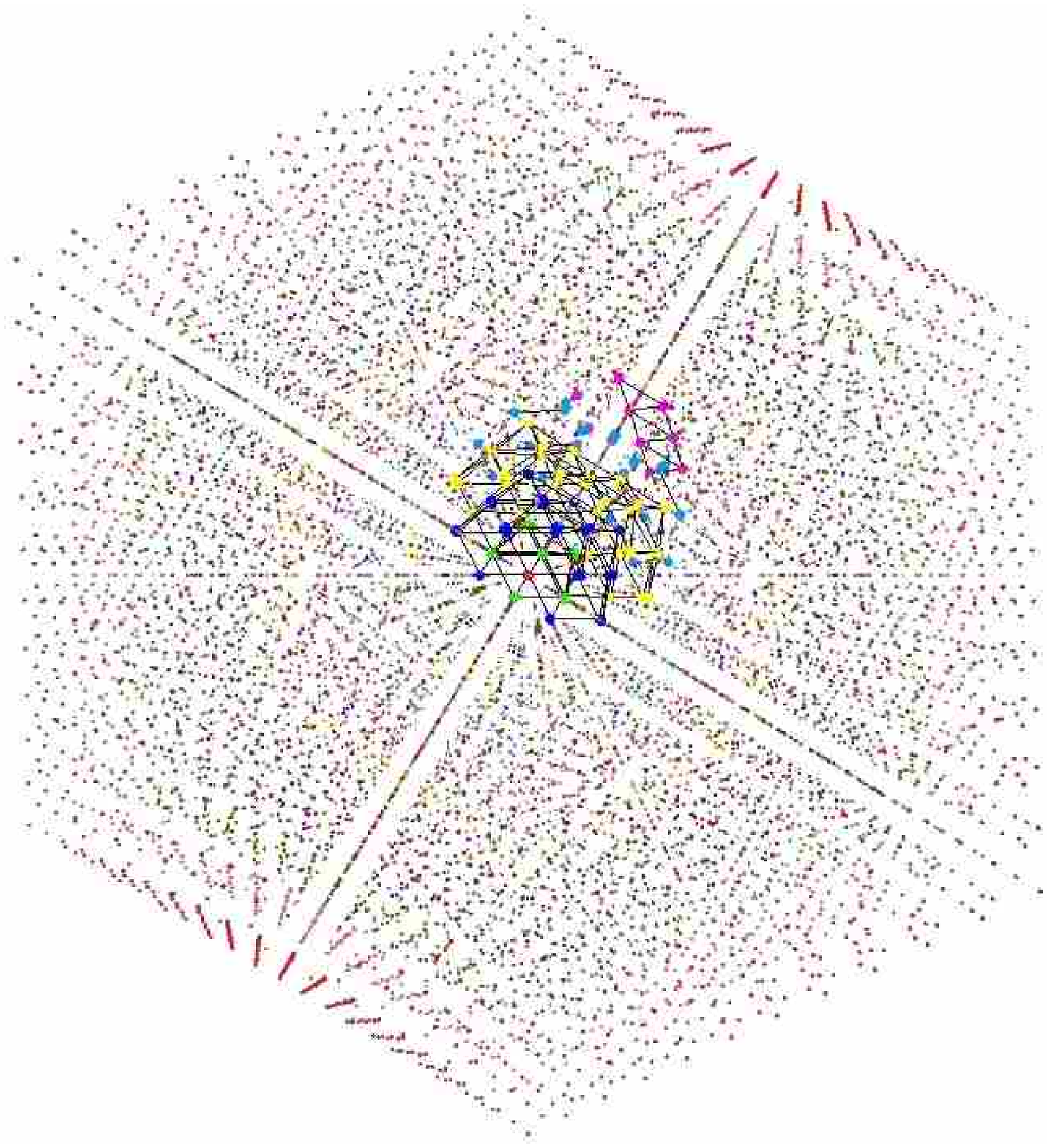, height=40mm}
\psfig{figure=\IMAGESPATH/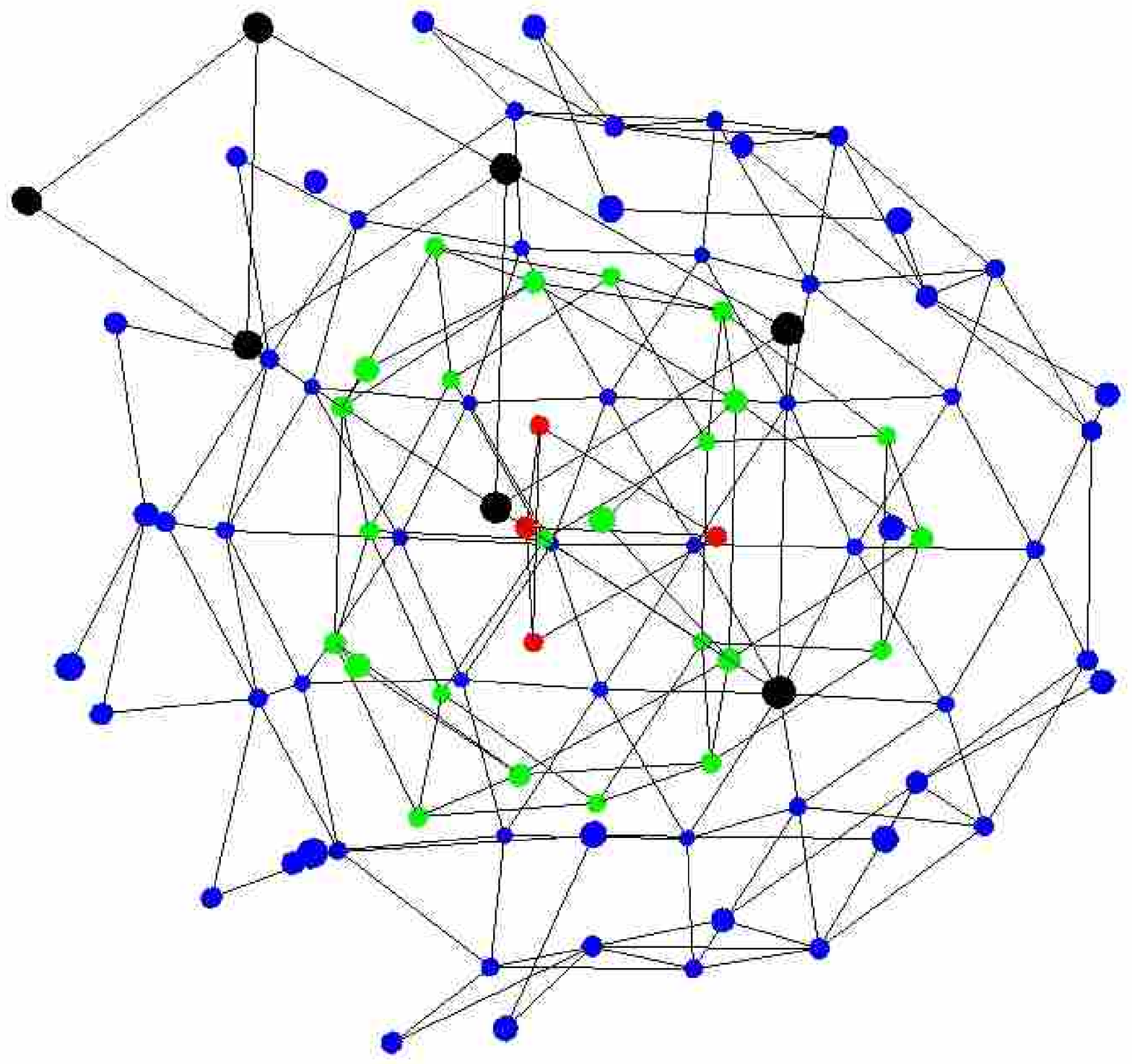, height=40mm}}
\centerline{ \makebox[1.7in][c]{ a) }\makebox[1.7in][c]{ b) } }
\centerline{
\psfig{figure=\IMAGESPATH/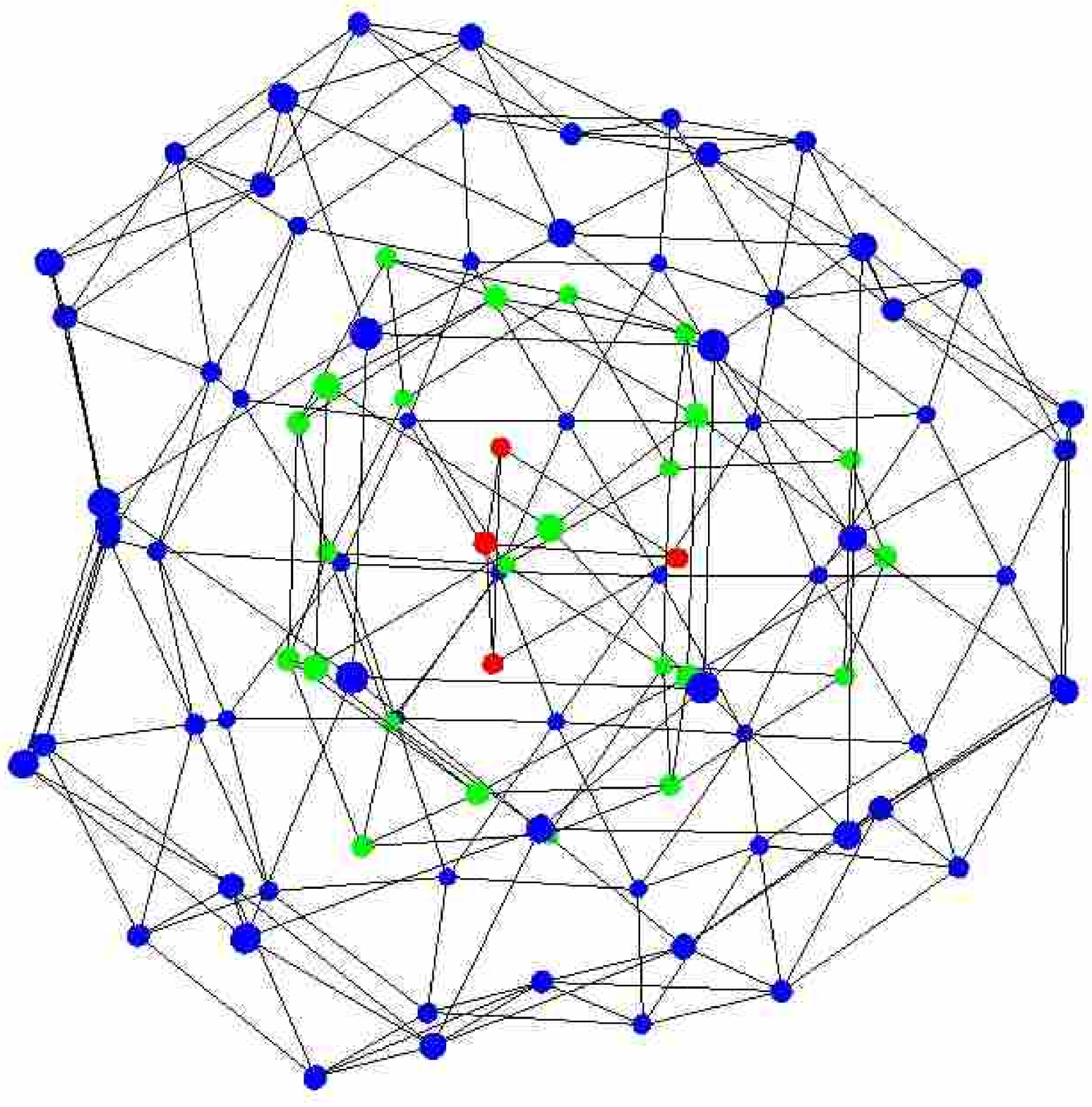, height=40mm}
\psfig{figure=\IMAGESPATH/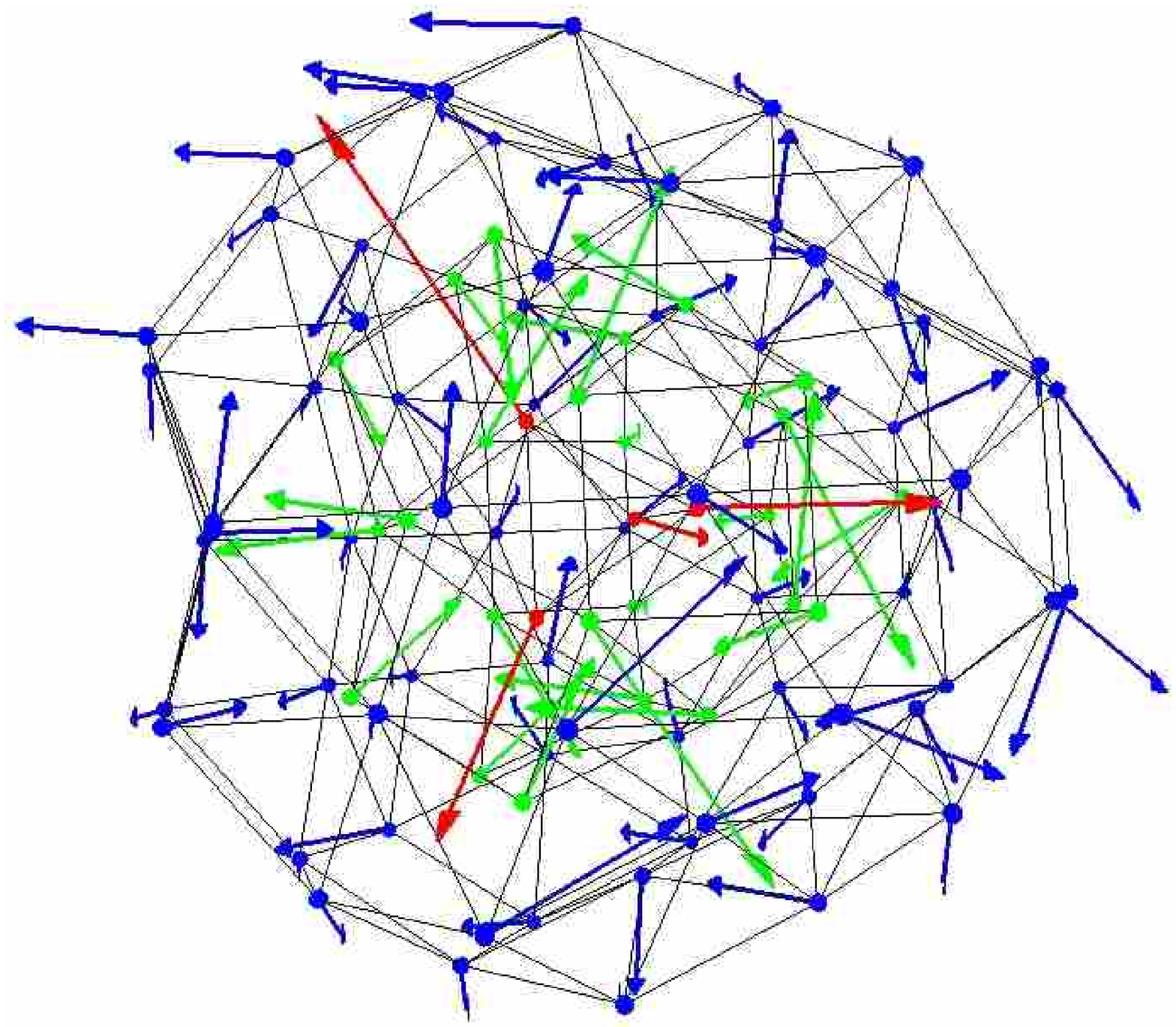, height=40mm}
} \centerline{ \makebox[1.7in][c]{ c) }\makebox[1.7in][c]{ d) } } \caption{ a)
$C_{98}$ inside of IF9483, b) $C_{98}$, c) $C^*_{98}$, and d) $C^*_{98}$ with
gradient.}~\label{fig:LJ98}
\end{figure}

\begin{figure}
\centerline{
\psfig{figure=\IMAGESPATH/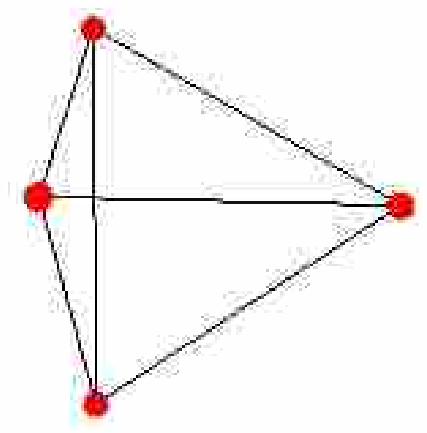, height=40mm}
\psfig{figure=\IMAGESPATH/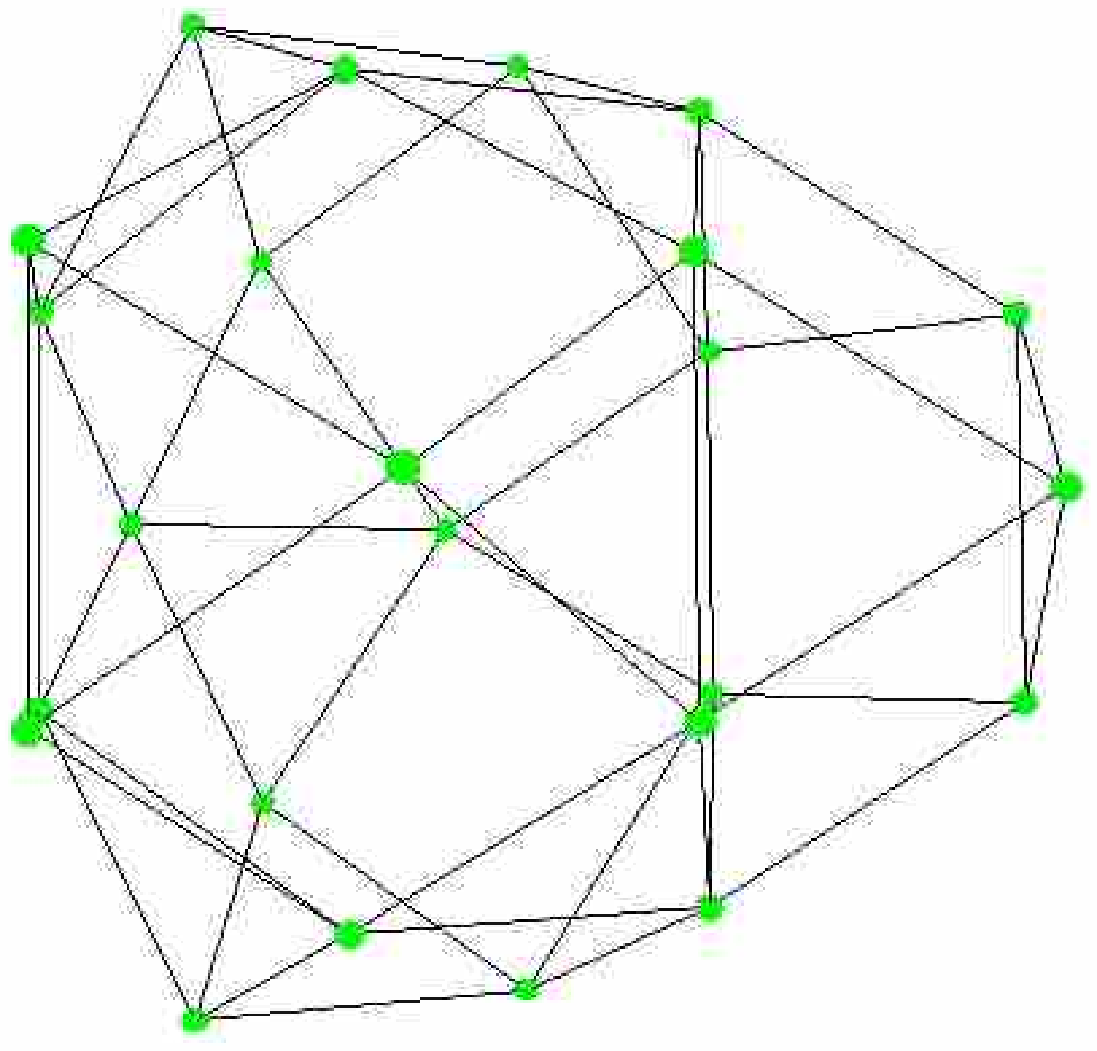, height=40mm}
\psfig{figure=\IMAGESPATH/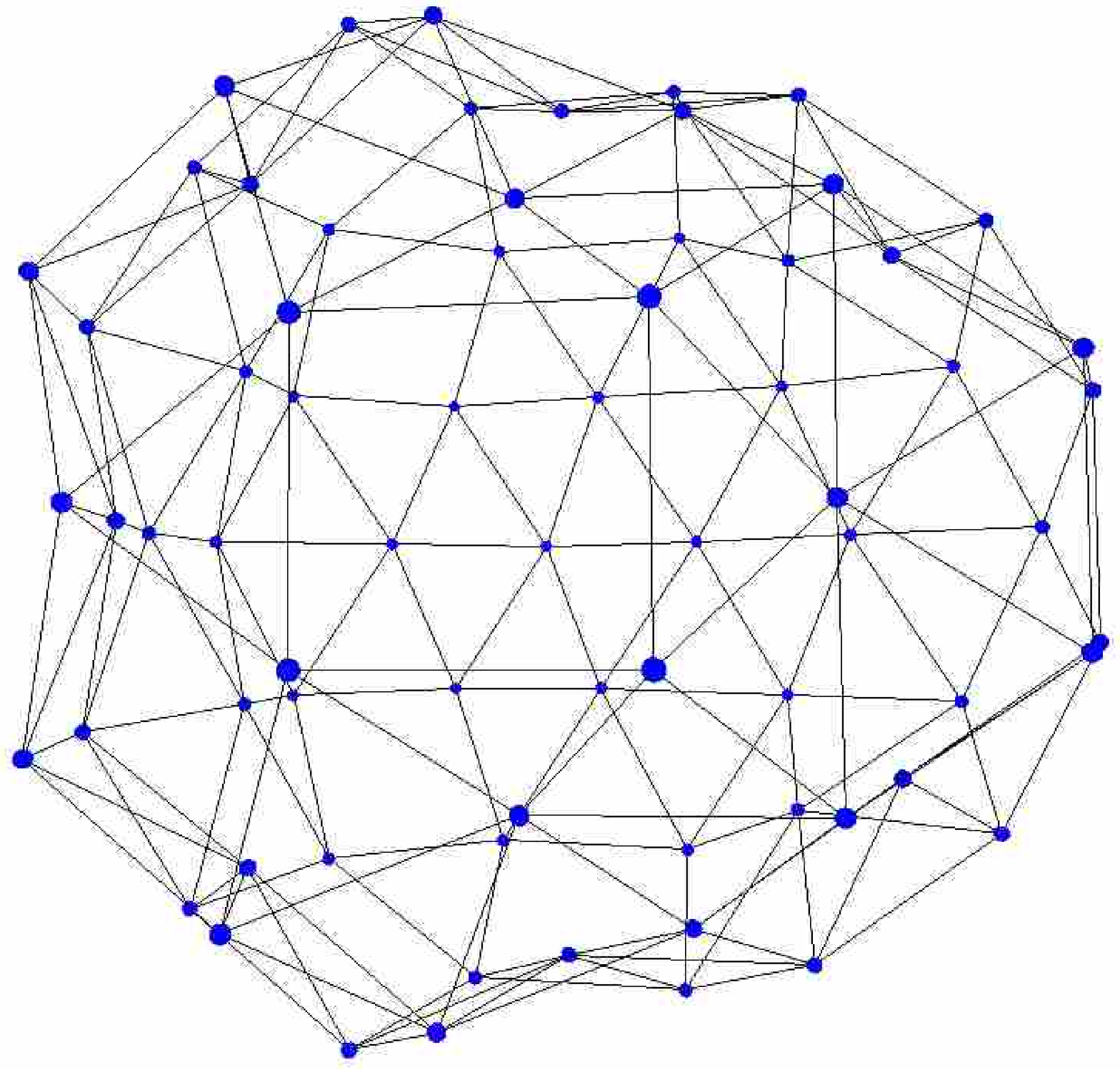, height=40mm}
} \centerline{ \makebox[1.7in][c]{a)}\makebox[1.7in][c]{b)}
\makebox[1.7in][c]{c)} } \caption{Novel views of tetrahedral shells of
$C^*_{98}$ a) interior shell with 4 particles, b) medium shell with 24, and c)
outer shell with 70 particles.}~\label{fig:LJ98_Capas}
\end{figure}

What is easy to accept from our classification is that a cluster type ID is
contained on any of the twelve axis of symmetry of IF (here
the clusters type ID are on the semi-axes Y$^+$), a cluster type
TO is in a centered tetrahedron formed by the center and three axis of symmetry
forming four triangular faces. One controversial result of this automatic
classification is the cluster's type for $C_{98}^{\ast }$, moreover, this is the
only cluster in MIF1739 where the adjacency
matrix~\cite{ap:Hoare1983} is not the same between $C_{98}$ and
$C_{98}^{\ast }$, however, $C_{98}^{\ast }=\min $($C_{98})$. The right
type of $C_{98}^{\ast }$ is tetrahedral (depicted by Leary~\cite{pre:Leary1999}).
The different type reported here comes from the distance between
$C_{98}$ and $C_{98}^{\ast}$
but it is a fact that the basin around $C_{98}^{\ast }$ attracts this $C_{98}$
of type FC. Figures \ref{fig:LJ98} and \ref{fig:LJ98_Capas} depict
novel views of $C_{98}$ and $C_{98}^{\ast}$.

\begin{figure}
\centerline{
\psfig{figure=\IMAGESPATH/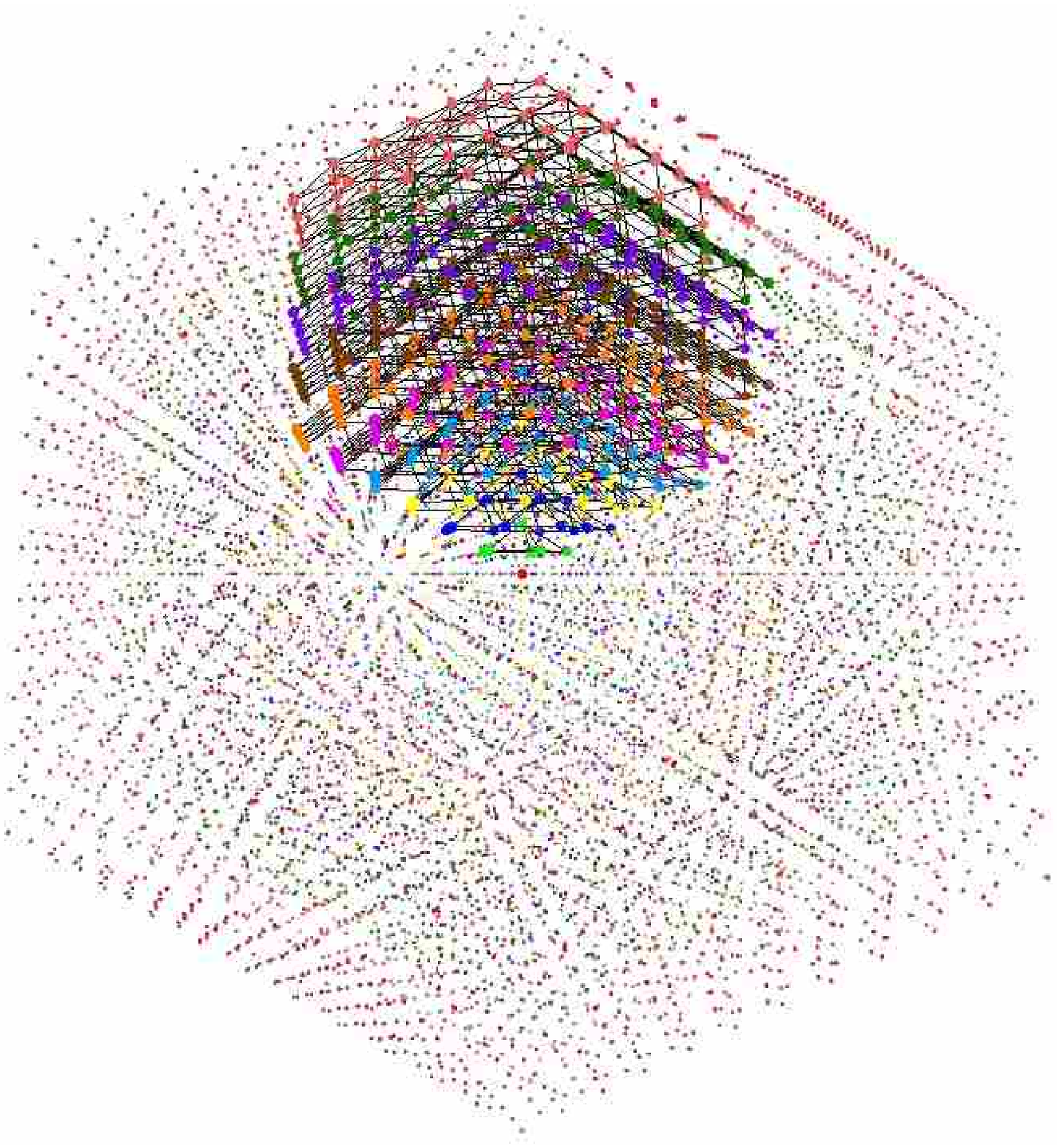, height=50mm}
\psfig{figure=\IMAGESPATH/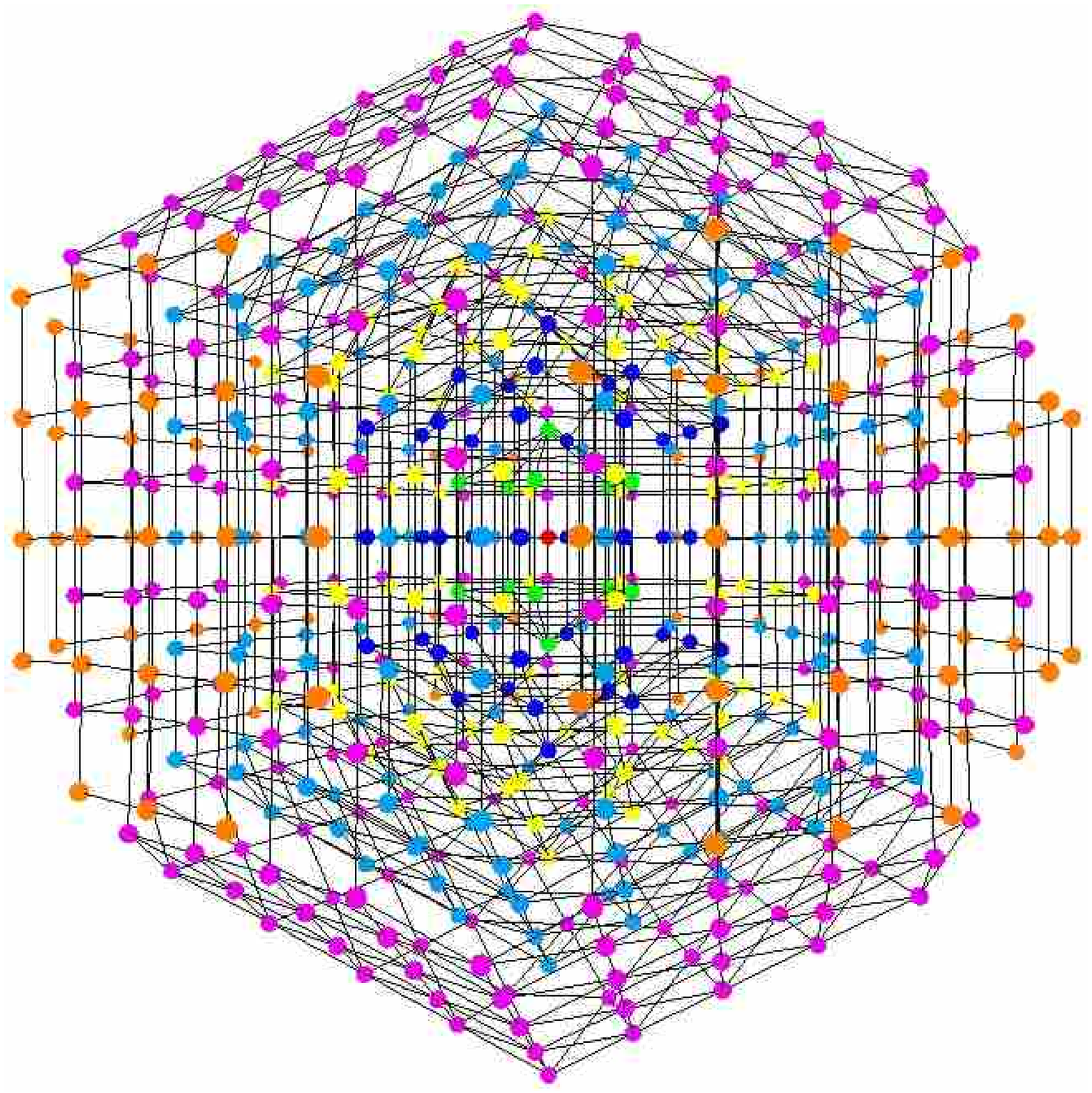, height=50mm}
}
\centerline{ \makebox[2.2in][c]{ a) }\makebox[2.2in][c]{ b) } }
\caption{a) $C_{664}$ inside of IF9483 and b) $C^*_{664}$.}~\label{fig:LJ664}
\end{figure}

Figure \ref{fig:LJ664} a) depicts $C_{664}$ inside of IF9483 and b) depicts
$C_{664}^{\ast }$ as ID.

\begin{figure}
\centerline{
\psfig{figure=\IMAGESPATH/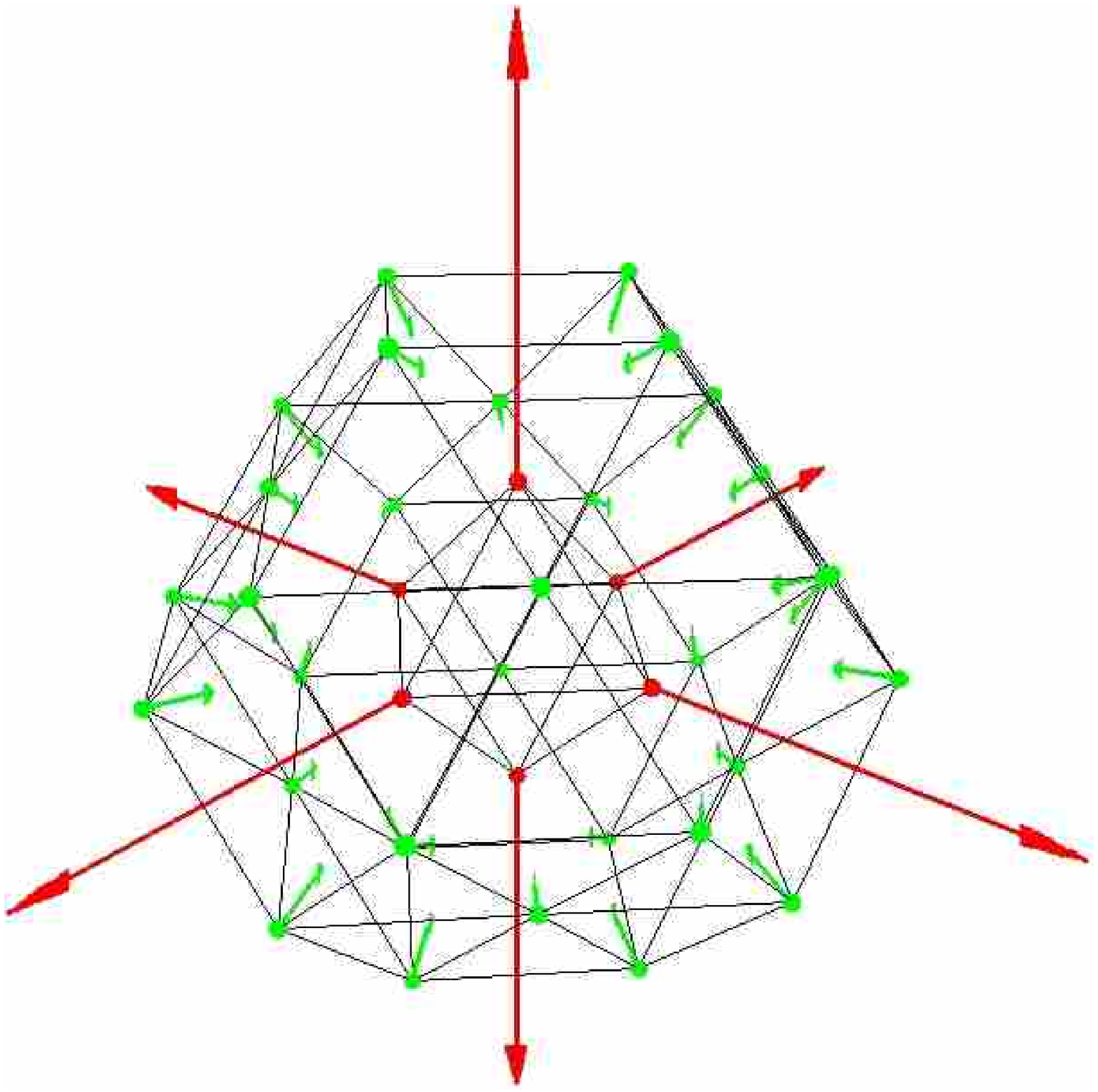, height=50mm}
\psfig{figure=\IMAGESPATH/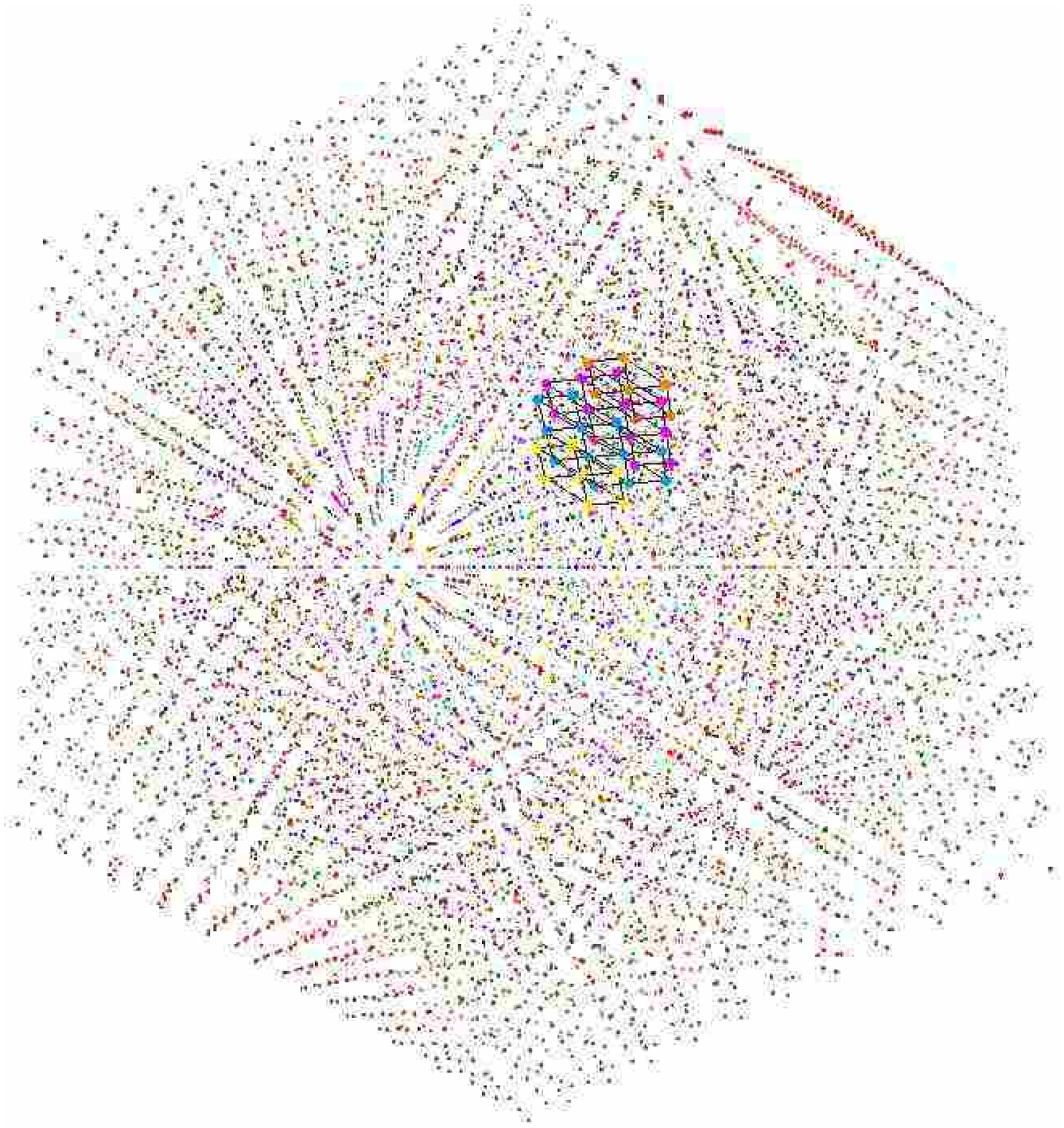, height=50mm}
} \centerline{ \makebox[2.2in][c]{ a) }\makebox[2.2in][c]{ b) } } \caption{a)
$C^*_{38}$ with gradient and b) $C_{38}$ inside of IF9483.}~\label{fig:LJ38}
\end{figure}

Wales and Doye~\cite{jpca:Wales1997} have been used basin for search of optimal
LJ clusters and get through the barrier of the PES caused be the deformation of
the particles on the surface. Figure \ref{fig:LJ38} b) gives another perspective
of this. Because, $C_{37}^{\ast }$ and C$_{39}^{\ast }$ are close to the center
of IF, $C_{38}^{\ast }$ \ is a jump from the center of IF to the first
truncated octahedron in one of the tetrahedron formed by three axes of symmetry
of IF. From table \ref{tb:On_off_32}, the indices of the particles for these
three clusters show that they do not share any particle. This is the unique
case in the results where there is not intersection between three consecutive
clusters. Figure \ref{fig:LJ38} a) depicts $C_{38}^{\ast }$ with gradient,
where $\left\| \nabla \hbox{VXX}\left( x^{\ast }\right) \right\| \thicksim
$3.7$\cdot $10$^{-6}$. The gradient looks big by the normalization. The
gradient of $C_{38}^{\ast }$ shows that this cluster is stable and elastic in
the sense that it can be deformed by $\alpha \nabla \hbox{VXX}\left( x^{\ast
}\right) $ for some small values of $\alpha $ and with a minimization routine, the deformed
$C_{38}$ will return to $C_{38}^{\ast }$. The gradient of the thirty-two
particles in the outer shell point toward the center and the six particles of
the inner octahedron shell point towards the square faces of the outer shell
making this cluster stable to twisting deformations.

\begin{figure}
\centerline{
\psfig{figure=\IMAGESPATH/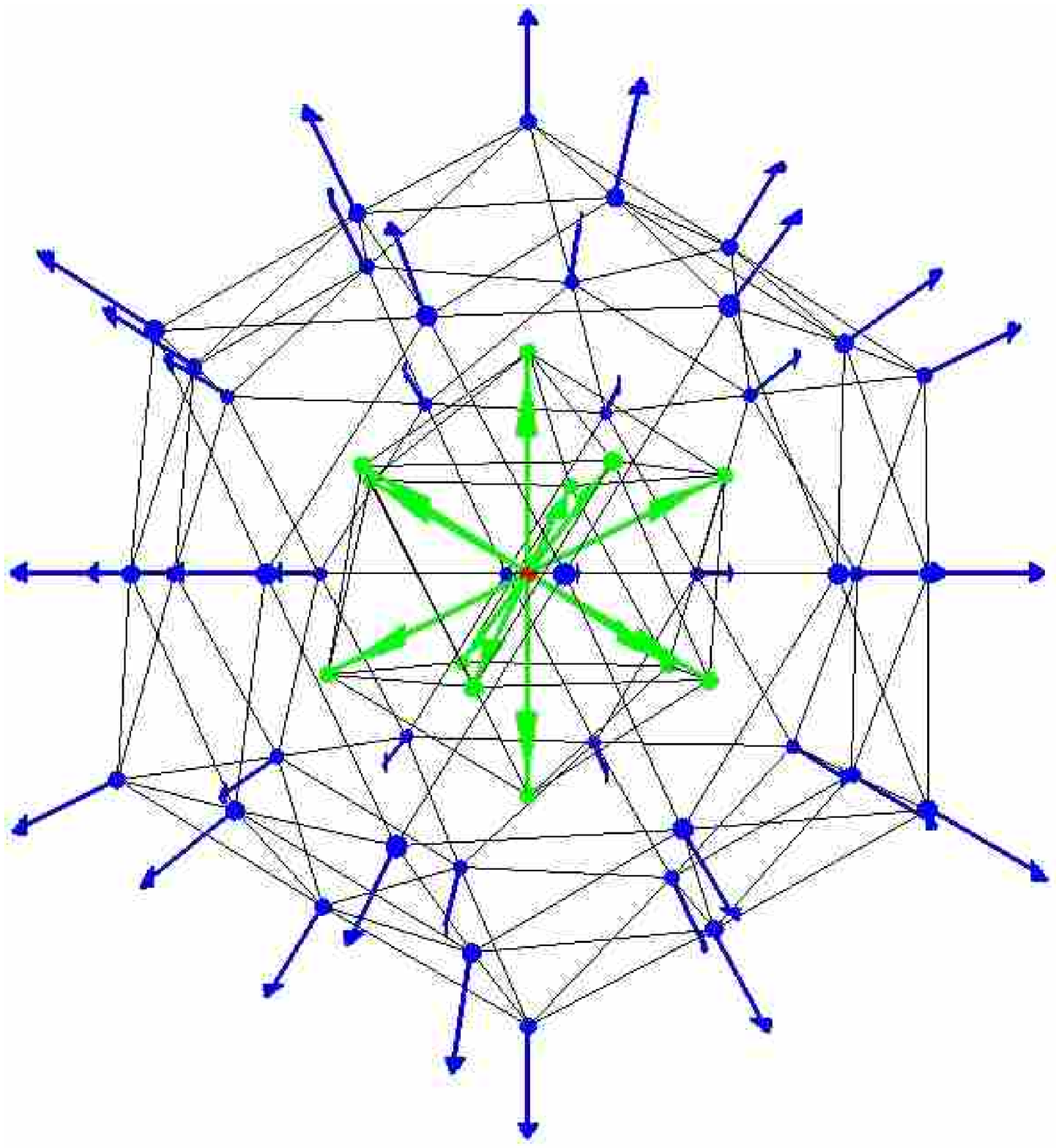, height=50mm}
\psfig{figure=\IMAGESPATH/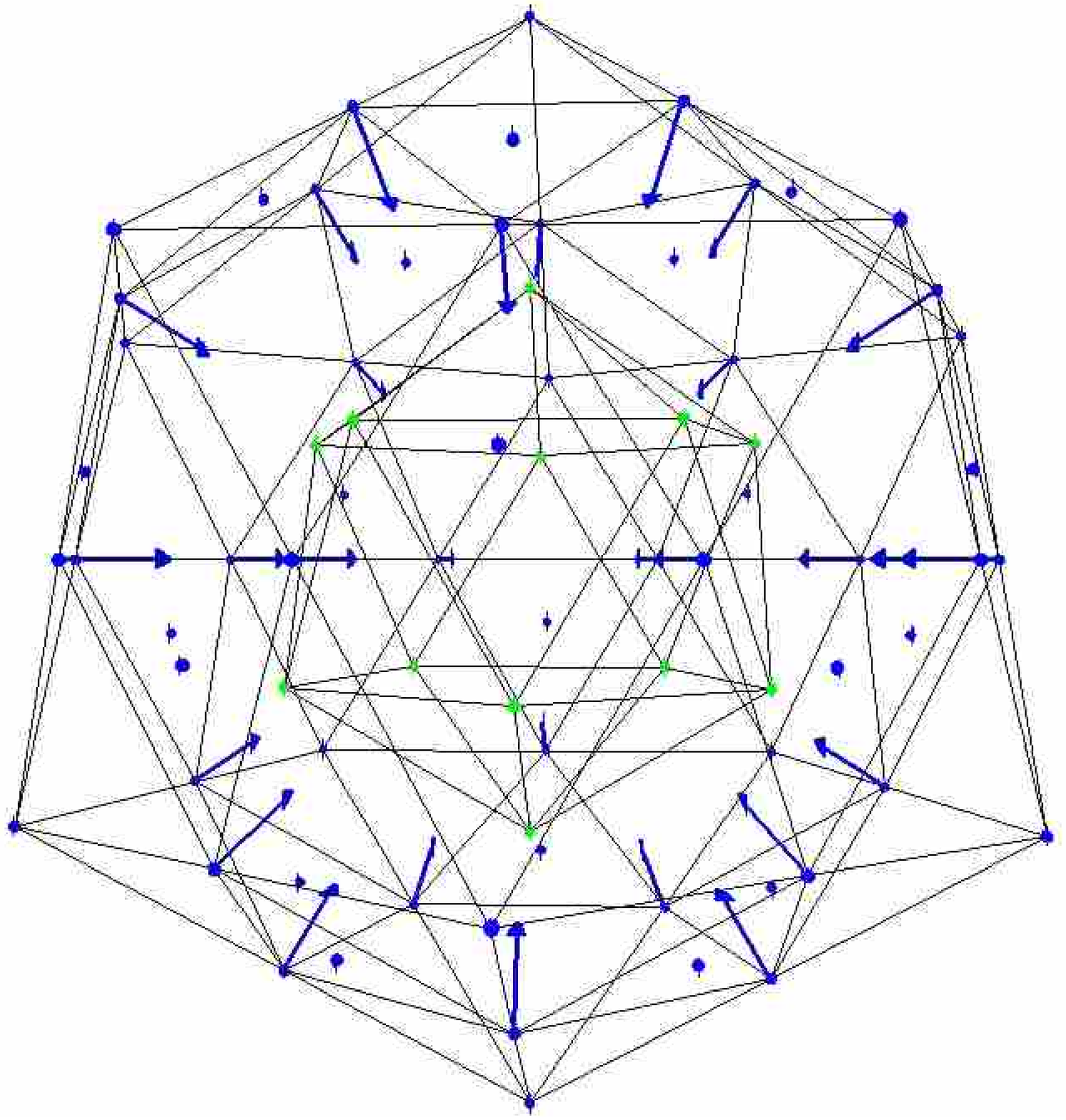, height=50mm}
}
\centerline{ \makebox[2.2in][c]{ a) }\makebox[2.2in][c]{ b) } } \caption{a)
$C^*_{55}$=IC$55^*$ and b) IF75 with gradient.}~\label{fig:ic55icfc75}
\end{figure}

\begin{figure}
\centerline{
\psfig{figure=\IMAGESPATH/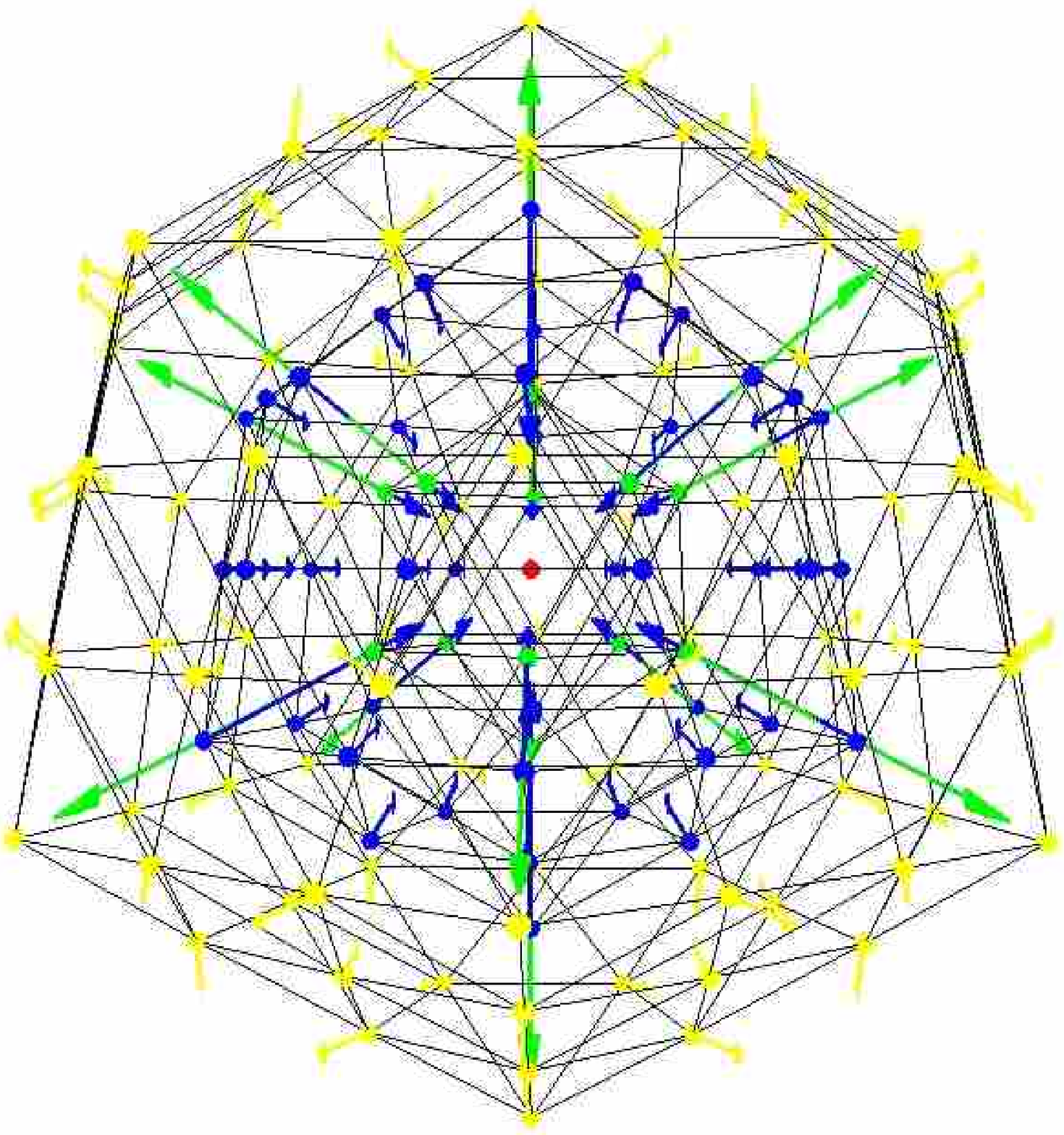, height=60mm}
\psfig{figure=\IMAGESPATH/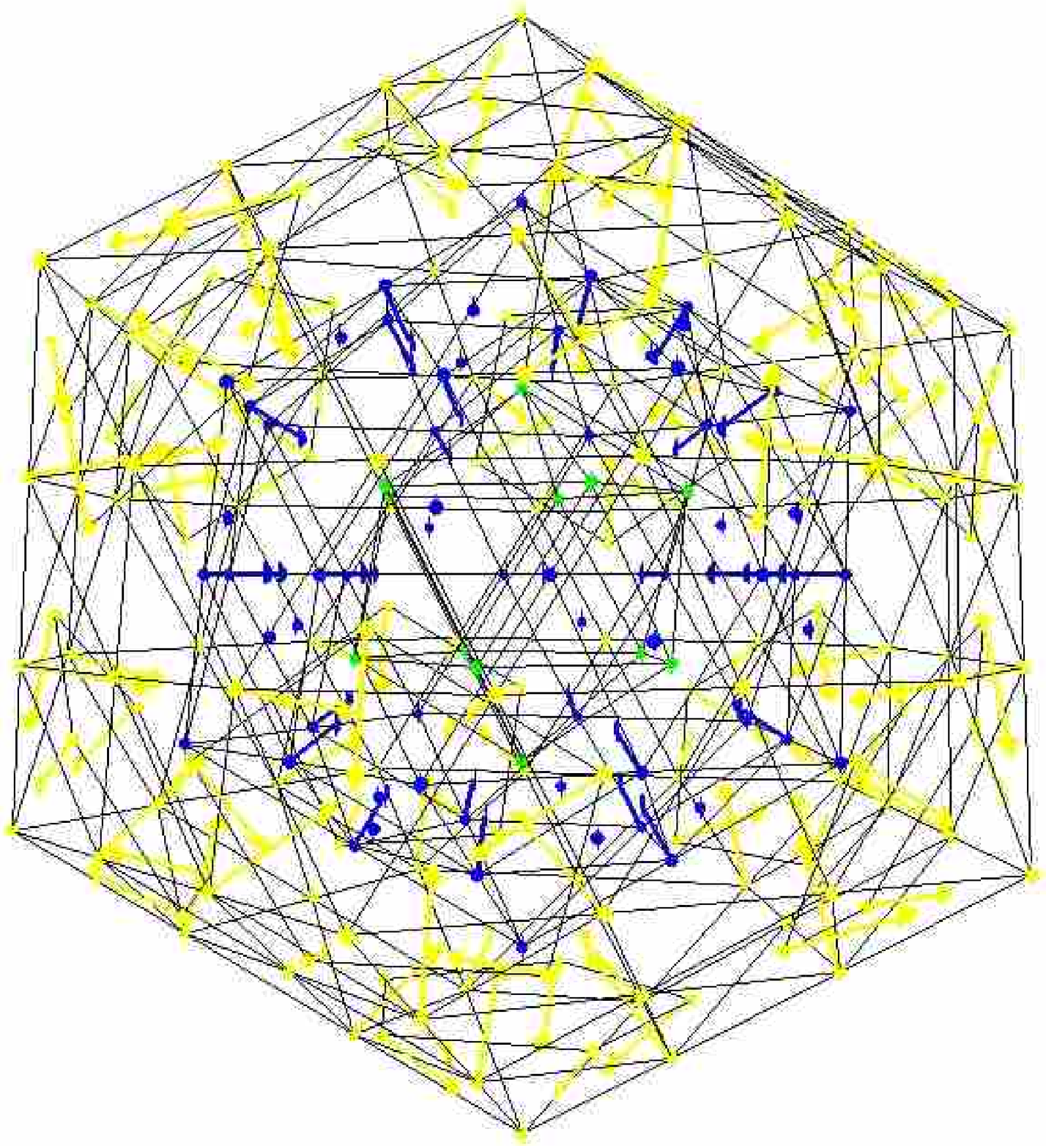, height=60mm}
} \centerline{ \makebox[2.4in][c]{ a) }\makebox[2.4in][c]{ b) } } \caption{a)
$C^*_147$=IC$147^*$ and b) IF227 with gradient.}~\label{fig:ic147icfc227}
\end{figure}

Figure \ref{fig:ic55icfc75} a) depicts $C_{55}^{\ast }$ with gradient. Here the
gradient shows that the particles on the top of the outer shell can move to
close positions around the semi-axes $Y^+$, in similar way for the bottom. The
gradient of the particles in the inner icosahedron points in diametrical
directions and if this effect is combined with an incomplete outer shell is
possible to have great displacement of particles from their positions on IF
and, also, a possible twisting and shrinking on the top or bottom cap of a
cluster. Similar results can be seen from the gradient of $C_{147}^{\ast }$
depicted in figure \ref{fig:ic147icfc227} a).

The graphical representation of the gradient helped to design IF.
Our selection of the shell's step size is to have the
corresponding component of the gradient for the particles
IC and FC pointed towards $(0,0,0)$. Figures \ref{fig:ic55icfc75} b) and
\ref{fig:ic147icfc227} b) depicted IF75 and IF227 with their gradient.


\include{\TABLESPATH/TABLA_idMinLattice}

\include{\TABLESPATH/tabla_DiffMinLatice}

\include{\TABLESPATH/TABLA_on_off}

\section{Conclusions and future work}~\label{sc:conclusions and future work}
All the details of the construction of the IF9483 and MIF1739 are not given here but in a
later article. Our first option was to build only
$\Omega^o$  from Propositions\ref{prop:DiscreteLattice_p1} and~\ref{prop:Omega_global_p3}
but it was not as regular
and symmetrical as IF. IF was the lattice that close the gap of
Northby's statement~\cite{jcp:Northby1987} in a regular, elegant and symmetrical lattice but hot.
It seems that Proposition~\ref{prop:DiscreteLattice_p1} is focus in build arbitrary sets
from arbitrary points, that can be obtained from the separable points of
$\Real^{3}$ under BU or LJ. However,  IF came naturally from using the $C_{13}^{\ast}$
as the seed. Of course, there could be other sets or lattices without such
motifs but IF joints in one framework the well know IC and FC and all the
putative optimal LJ clusters. Also, in the sets  $\Omega^l$ and $\Omega^o$
of the Propositions~\ref{prop:Omega_Local_p2} and ~\ref{prop:Omega_global_p3} are not
necessary a minimization procedure for any cluster. This gives a
witness property for the optimality verification in the neighborhood
of a cluster in polynomial time.
Depicting the components of the gradient of the potential on each particle
helps to understand the deformation of the PES for
small clusters, but it need further study.
Propositions~\ref{prop:DiscreteLattice_p1},~\ref{prop:Omega_Local_p2},
and~\ref{prop:Omega_global_p3} permit see SOCDXX as the Traveling Salesman
Problem in $\Omega^l$ or $\Omega^o$
and let to have a framework to analyze the complexity of
SOCDXX where XX is BU or LJ. An open question is: Are there similar results for
Kihara or Morse Potentials? Here the obstacle is that there is not strong
rejection when $r\rightarrow 0$ and this does not allow to have separate
points. Moreover: Is $C_{13}^{\ast }$ the global minimum? In my opinion it will be
possible to answer this by a combination of an exhaustive search and by the
symmetrical properties of IF and such proposition could
be proved by using a computer to explore in wise fashion the local optimal
clusters of 13 particles in IF. Maybe in similar way to the proof of the
Four-Color Theorem. Other conjecture open by this work is: Are all the optimal
cluster under LJ or for equivalent potentials in IF? The answer of this
conjecture does not change the results of my propositions, it is clear that if
there are new cases, these can be added without problem but IF will possibly
lose the beautiful minimum combination of IC and FC. In the case, of huge
clusters it is possible that other lattices could be added to IF with other
geometrical motifs.

Finally, this novel formulation brings
new perspectives for NP's complexity and will be extended
in the future to other potential functions.

\section*{Acknowledgement}
This work was supported by Elsa Guillermina Barr\'{o}n and Emma Romero. It is
dedicated to them and to Roland Glowinski, Ioannis A. Kakadiaris, Alberto
Santamar\'ia, David Romero, Susana G\'{o}mez, and Hector Lorenzo Ju\'arez.
This work started long ago, but I would not have finished it without the help,
support, and friendship of Alberto Santamar\'ia. Also, thanks to following
institutions University of Houston, IIMAS-UNAM, UAM-I, and CIMAT.

\bibliographystyle{abbrv}
\bibliography{\BIBPATH/LJ_01}

\end{document}

%% file: definitions.tex
\newcommand{\Real}{\mathbb R}
\newcommand{\Natural}{\mathbb N}
\newcommand{\Rational}{\mathbb Q}

\newcommand{\A}{\mathcal{A}}
\newenvironment{TightList}{
\begin{list}{}{\itemsep 0in\parsep 0in\itemindent 0in\labelsep 0in\rightmargin 0in\leftmargin
\labelwidth\topsep 0in\parskip 0in\partopsep 0in}}{\end{list}}
\def\IMAGESPATH{.}

\def\BIBPATH{.}


%% file: TABLA_idMinLattice.tex
\begin{table}
\centerline{\small
\begin{tabular}[t]{r r | r r r}
\multicolumn{1}{c}{$\bf \#$} & \multicolumn{1}{c}{$\bf i_p$} & \multicolumn{1}{c}{$\bf X$} & \multicolumn{1}{c}{$\bf Y$} & \multicolumn{1}{c}{$\bf Z$}  \\ \hline \hline
1 & 1 & $0.00000000$ & $0.00000000$ & $0.00000000$ \\ \cline{2-5}
2 & 2 & $0.00000000$ & $1.08183839$ & $0.00000000$ \\ \cline{2-5}
3 & 3 & $0.96762567$ & $0.48381284$ & $0.00000000$ \\ \cline{2-5}
4 & 4 & $0.29901278$ & $0.48381284$ & $-0.92026670$ \\ \cline{2-5}
5 & 5 & $-0.78282561$ & $0.48381284$ & $-0.56875610$ \\ \cline{2-5}
6 & 6 & $-0.78282561$ & $0.48381284$ & $0.56875610$ \\ \cline{2-5}
7 & 7 & $0.29901278$ & $0.48381284$ & $0.92026670$ \\
\end{tabular}
\ \ \ \ \ \ 
\begin{tabular}[t]{r r | r r r}
\multicolumn{1}{c}{$\bf \#$} & \multicolumn{1}{c}{$\bf i_p$} & \multicolumn{1}{c}{$\bf X$} & \multicolumn{1}{c}{$\bf Y$} & \multicolumn{1}{c}{$\bf Z$}  \\ \hline \hline
56 & 56 & $0.84442563$ & $1.36630937$ & $-0.61351113$ \\ \cline{2-5}
57 & 57 & $1.36630937$ & $0.32254189$ & $-0.99268187$ \\ \cline{2-5}
58 & 58 & $1.68885127$ & $-0.32254189$ & $0.00000000$ \\ \cline{2-5}
59 & 59 & $1.04376748$ & $-1.36630937$ & $0.00000000$ \\ \cline{2-5}
60 & 60 & $-0.32254189$ & $1.36630937$ & $-0.99268187$ \\ \cline{2-5}
61 & 61 & $-0.52188374$ & $0.32254189$ & $-1.60619300$ \\ \cline{2-5}
62 & 62 & $0.52188374$ & $-0.32254189$ & $-1.60619300$ \\
\end{tabular}
}
\caption{MIF1739 contains $C_n$, such that $C^*_n$ = $\min$($C_n$) $n=2,\ldots,1000$.}~\label{tb:idMinLattice_0}
\end{table}

\begin{table}
\centerline{\small
\begin{tabular}[t]{r r | r r r}
\multicolumn{1}{c}{$\bf \#$} & \multicolumn{1}{c}{$\bf i_p$} & \multicolumn{1}{c}{$\bf X$} & \multicolumn{1}{c}{$\bf Y$} & \multicolumn{1}{c}{$\bf Z$}  \\ \hline \hline
111 & 116 & $-1.86466400$ & $0.48381284$ & $2.05777890$ \\ \cline{2-5}
112 & 117 & $-0.78282561$ & $0.48381284$ & $2.40928950$ \\ \cline{2-5}
113 & 118 & $0.29901278$ & $0.48381284$ & $2.76080010$ \\ \cline{2-5}
114 & 119 & $1.38085117$ & $0.48381284$ & $2.40928950$ \\ \cline{2-5}
115 & 120 & $2.04946406$ & $0.48381284$ & $1.48902280$ \\ \cline{2-5}
116 & 121 & $2.71807696$ & $0.48381284$ & $0.56875610$ \\ \cline{2-5}
117 & 122 & $2.53327690$ & $-0.48381284$ & $-1.13751220$ \\
\end{tabular}
\ \ \ \ \ \ 
\begin{tabular}[t]{r r | r r r}
\multicolumn{1}{c}{$\bf \#$} & \multicolumn{1}{c}{$\bf i_p$} & \multicolumn{1}{c}{$\bf X$} & \multicolumn{1}{c}{$\bf Y$} & \multicolumn{1}{c}{$\bf Z$}  \\ \hline \hline
166 & 171 & $2.15891033$ & $-0.22453832$ & $-1.56854017$ \\ \cline{2-5}
167 & 172 & $2.36183488$ & $0.83798841$ & $-0.94400263$ \\ \cline{2-5}
168 & 173 & $1.62764697$ & $0.83798841$ & $-1.95452560$ \\ \cline{2-5}
169 & 174 & $2.46563538$ & $-0.83798841$ & $-0.62453754$ \\ \cline{2-5}
170 & 175 & $2.66855993$ & $0.22453832$ & $0.00000000$ \\ \cline{2-5}
171 & 176 & $2.46563538$ & $-0.83798841$ & $0.62453754$ \\ \cline{2-5}
172 & 177 & $1.85218529$ & $-1.83057152$ & $-0.62453754$ \\
\end{tabular}
}
\caption{MIF1739 contains $C_n$, such that $C^*_n$ = $\min$($C_n$) $n=2,\ldots,1000$ (cont.)}~\label{tb:idMinLattice_1}
\end{table}

\begin{table}
\centerline{\small
\begin{tabular}[t]{r r | r r r}
\multicolumn{1}{c}{$\bf \#$} & \multicolumn{1}{c}{$\bf i_p$} & \multicolumn{1}{c}{$\bf X$} & \multicolumn{1}{c}{$\bf Y$} & \multicolumn{1}{c}{$\bf Z$}  \\ \hline \hline
221 & 238 & $-1.56565123$ & $3.13130245$ & $-1.13751220$ \\ \cline{2-5}
222 & 239 & $-1.56565123$ & $3.13130245$ & $0.00000000$ \\ \cline{2-5}
223 & 240 & $-1.56565123$ & $3.13130245$ & $1.13751220$ \\ \cline{2-5}
224 & 241 & $-0.48381284$ & $3.13130245$ & $1.48902280$ \\ \cline{2-5}
225 & 242 & $0.59802555$ & $3.13130245$ & $1.84053340$ \\ \cline{2-5}
226 & 243 & $1.26663845$ & $3.13130245$ & $0.92026670$ \\ \cline{2-5}
227 & 244 & $2.90287702$ & $2.53327690$ & $0.00000000$ \\
\end{tabular}
\ \ \ \ \ \ 
\begin{tabular}[t]{r r | r r r}
\multicolumn{1}{c}{$\bf \#$} & \multicolumn{1}{c}{$\bf i_p$} & \multicolumn{1}{c}{$\bf X$} & \multicolumn{1}{c}{$\bf Y$} & \multicolumn{1}{c}{$\bf Z$}  \\ \hline \hline
276 & 293 & $-0.48381284$ & $0.96762567$ & $3.32955620$ \\ \cline{2-5}
277 & 294 & $0.59802555$ & $0.96762567$ & $3.68106680$ \\ \cline{2-5}
278 & 295 & $1.67986394$ & $0.96762567$ & $3.32955620$ \\ \cline{2-5}
279 & 296 & $2.34847684$ & $0.96762567$ & $2.40928950$ \\ \cline{2-5}
280 & 297 & $3.01708973$ & $0.96762567$ & $1.48902280$ \\ \cline{2-5}
281 & 298 & $3.68570263$ & $0.96762567$ & $0.56875610$ \\ \cline{2-5}
282 & 299 & $3.50090257$ & $0.00000000$ & $-1.13751220$ \\
\end{tabular}
}
\caption{MIF1739 contains $C_n$, such that $C^*_n$ = $\min$($C_n$) $n=2,\ldots,1000$ (cont.)}~\label{tb:idMinLattice_2}
\end{table}

\begin{table}
\centerline{\small
\begin{tabular}[t]{r r | r r r}
\multicolumn{1}{c}{$\bf \#$} & \multicolumn{1}{c}{$\bf i_p$} & \multicolumn{1}{c}{$\bf X$} & \multicolumn{1}{c}{$\bf Y$} & \multicolumn{1}{c}{$\bf Z$}  \\ \hline \hline
331 & 348 & $-3.20188979$ & $-1.93525134$ & $0.92026670$ \\ \cline{2-5}
332 & 349 & $-2.53327690$ & $-1.93525134$ & $1.84053340$ \\ \cline{2-5}
333 & 350 & $-1.86466400$ & $-1.93525134$ & $2.76080010$ \\ \cline{2-5}
334 & 351 & $-1.19605111$ & $-1.93525134$ & $3.68106680$ \\ \cline{2-5}
335 & 352 & $-0.11421272$ & $-1.93525134$ & $3.32955620$ \\ \cline{2-5}
336 & 353 & $0.96762567$ & $-1.93525134$ & $2.97804560$ \\ \cline{2-5}
337 & 354 & $2.04946406$ & $-1.93525134$ & $2.62653500$ \\
\end{tabular}
\ \ \ \ \ \ 
\begin{tabular}[t]{r r | r r r}
\multicolumn{1}{c}{$\bf \#$} & \multicolumn{1}{c}{$\bf i_p$} & \multicolumn{1}{c}{$\bf X$} & \multicolumn{1}{c}{$\bf Y$} & \multicolumn{1}{c}{$\bf Z$}  \\ \hline \hline
386 & 416 & $-1.97744435$ & $2.30531085$ & $-2.14059242$ \\ \cline{2-5}
387 & 417 & $-0.81793346$ & $2.30531085$ & $-2.51734034$ \\ \cline{2-5}
388 & 418 & $-1.13841439$ & $2.94627270$ & $-1.53100147$ \\ \cline{2-5}
389 & 419 & $0.02109650$ & $2.94627270$ & $-1.90774939$ \\ \cline{2-5}
390 & 438 & $-0.96882835$ & $3.58723456$ & $0.00000000$ \\ \cline{2-5}
391 & 439 & $-2.64688828$ & $2.30531085$ & $-1.21918190$ \\ \cline{2-5}
392 & 440 & $-2.64688828$ & $2.30531085$ & $1.21918190$ \\
\end{tabular}
}
\caption{MIF1739 contains $C_n$, such that $C^*_n$ = $\min$($C_n$) $n=2,\ldots,1000$ (cont.)}~\label{tb:idMinLattice_3}
\end{table}

\begin{table}
\centerline{\small
\begin{tabular}[t]{r r | r r r}
\multicolumn{1}{c}{$\bf \#$} & \multicolumn{1}{c}{$\bf i_p$} & \multicolumn{1}{c}{$\bf X$} & \multicolumn{1}{c}{$\bf Y$} & \multicolumn{1}{c}{$\bf Z$}  \\ \hline \hline
441 & 537 & $-0.18480006$ & $3.61511529$ & $2.40928950$ \\ \cline{2-5}
442 & 538 & $0.89703833$ & $3.61511529$ & $2.76080010$ \\ \cline{2-5}
443 & 539 & $1.56565123$ & $3.61511529$ & $1.84053340$ \\ \cline{2-5}
444 & 540 & $2.23426412$ & $3.61511529$ & $0.92026670$ \\ \cline{2-5}
445 & 541 & $3.87050269$ & $3.01708973$ & $0.00000000$ \\ \cline{2-5}
446 & 542 & $3.20188979$ & $3.01708973$ & $-0.92026670$ \\ \cline{2-5}
447 & 543 & $2.53327690$ & $3.01708973$ & $-1.84053340$ \\
\end{tabular}
\ \ \ \ \ \ 
\begin{tabular}[t]{r r | r r r}
\multicolumn{1}{c}{$\bf \#$} & \multicolumn{1}{c}{$\bf i_p$} & \multicolumn{1}{c}{$\bf X$} & \multicolumn{1}{c}{$\bf Y$} & \multicolumn{1}{c}{$\bf Z$}  \\ \hline \hline
496 & 592 & $-0.18480006$ & $1.45143851$ & $-4.24982290$ \\ \cline{2-5}
497 & 593 & $-1.26663845$ & $1.45143851$ & $-3.89831230$ \\ \cline{2-5}
498 & 594 & $-2.34847684$ & $1.45143851$ & $-3.54680170$ \\ \cline{2-5}
499 & 595 & $-3.43031523$ & $1.45143851$ & $-3.19529110$ \\ \cline{2-5}
500 & 596 & $-4.09892812$ & $1.45143851$ & $-2.27502440$ \\ \cline{2-5}
501 & 597 & $-4.09892812$ & $1.45143851$ & $-1.13751220$ \\ \cline{2-5}
502 & 598 & $-4.09892812$ & $1.45143851$ & $0.00000000$ \\
\end{tabular}
}
\caption{MIF1739 contains $C_n$, such that $C^*_n$ = $\min$($C_n$) $n=2,\ldots,1000$ (cont.)}~\label{tb:idMinLattice_4}
\end{table}
\clearpage

\begin{table}
\centerline{\small
\begin{tabular}[t]{r r | r r r}
\multicolumn{1}{c}{$\bf \#$} & \multicolumn{1}{c}{$\bf i_p$} & \multicolumn{1}{c}{$\bf X$} & \multicolumn{1}{c}{$\bf Y$} & \multicolumn{1}{c}{$\bf Z$}  \\ \hline \hline
551 & 647 & $-4.46852824$ & $-0.48381284$ & $0.00000000$ \\ \cline{2-5}
552 & 648 & $-4.46852824$ & $-0.48381284$ & $1.13751220$ \\ \cline{2-5}
553 & 649 & $-3.79991535$ & $-0.48381284$ & $2.05777890$ \\ \cline{2-5}
554 & 650 & $-3.13130245$ & $-0.48381284$ & $2.97804560$ \\ \cline{2-5}
555 & 651 & $-2.46268956$ & $-0.48381284$ & $3.89831230$ \\ \cline{2-5}
556 & 652 & $-1.38085117$ & $-0.48381284$ & $4.24982290$ \\ \cline{2-5}
557 & 653 & $-0.29901278$ & $-0.48381284$ & $4.60133350$ \\
\end{tabular}
\ \ \ \ \ \ 
\begin{tabular}[t]{r r | r r r}
\multicolumn{1}{c}{$\bf \#$} & \multicolumn{1}{c}{$\bf i_p$} & \multicolumn{1}{c}{$\bf X$} & \multicolumn{1}{c}{$\bf Y$} & \multicolumn{1}{c}{$\bf Z$}  \\ \hline \hline
606 & 702 & $-0.41322549$ & $-2.41906418$ & $4.24982290$ \\ \cline{2-5}
607 & 703 & $0.66861290$ & $-2.41906418$ & $3.89831230$ \\ \cline{2-5}
608 & 704 & $1.75045129$ & $-2.41906418$ & $3.54680170$ \\ \cline{2-5}
609 & 705 & $2.83228968$ & $-2.41906418$ & $3.19529110$ \\ \cline{2-5}
610 & 706 & $3.91412807$ & $-2.41906418$ & $2.84378050$ \\ \cline{2-5}
611 & 707 & $3.91412807$ & $-2.41906418$ & $1.70626830$ \\ \cline{2-5}
612 & 708 & $3.91412807$ & $-2.41906418$ & $0.56875610$ \\
\end{tabular}
}
\caption{MIF1739 contains $C_n$, such that $C^*_n$ = $\min$($C_n$) $n=2,\ldots,1000$ (cont.)}~\label{tb:idMinLattice_5}
\end{table}

\begin{table}
\centerline{\small
\begin{tabular}[t]{r r | r r r}
\multicolumn{1}{c}{$\bf \#$} & \multicolumn{1}{c}{$\bf i_p$} & \multicolumn{1}{c}{$\bf X$} & \multicolumn{1}{c}{$\bf Y$} & \multicolumn{1}{c}{$\bf Z$}  \\ \hline \hline
661 & 757 & $-0.29901278$ & $-4.81116640$ & $-0.92026670$ \\ \cline{2-5}
662 & 758 & $-0.96762567$ & $-4.81116640$ & $0.00000000$ \\ \cline{2-5}
663 & 759 & $-0.29901278$ & $-4.81116640$ & $0.92026670$ \\ \cline{2-5}
664 & 760 & $0.78282561$ & $-4.81116640$ & $0.56875610$ \\ \cline{2-5}
665 & 761 & $0.00000000$ & $-5.40919195$ & $0.00000000$ \\ \cline{2-5}
666 & 767 & $1.40459143$ & $3.41577344$ & $-2.50615402$ \\ \cline{2-5}
667 & 768 & $2.11106408$ & $3.41577344$ & $-1.53377783$ \\
\end{tabular}
\ \ \ \ \ \ 
\begin{tabular}[t]{r r | r r r}
\multicolumn{1}{c}{$\bf \#$} & \multicolumn{1}{c}{$\bf i_p$} & \multicolumn{1}{c}{$\bf X$} & \multicolumn{1}{c}{$\bf Y$} & \multicolumn{1}{c}{$\bf Z$}  \\ \hline \hline
716 & 962 & $0.00000000$ & $6.49103034$ & $0.00000000$ \\ \cline{2-5}
717 & 963 & $0.96762567$ & $5.89300479$ & $0.00000000$ \\ \cline{2-5}
718 & 964 & $0.29901278$ & $5.89300479$ & $-0.92026670$ \\ \cline{2-5}
719 & 965 & $-0.78282561$ & $5.89300479$ & $-0.56875610$ \\ \cline{2-5}
720 & 966 & $-0.78282561$ & $5.89300479$ & $0.56875610$ \\ \cline{2-5}
721 & 967 & $0.29901278$ & $5.89300479$ & $0.92026670$ \\ \cline{2-5}
722 & 968 & $1.93525134$ & $5.29497923$ & $0.00000000$ \\
\end{tabular}
}
\caption{MIF1739 contains $C_n$, such that $C^*_n$ = $\min$($C_n$) $n=2,\ldots,1000$ (cont.)}~\label{tb:idMinLattice_6}
\end{table}

\begin{table}
\centerline{\small
\begin{tabular}[t]{r r | r r r}
\multicolumn{1}{c}{$\bf \#$} & \multicolumn{1}{c}{$\bf i_p$} & \multicolumn{1}{c}{$\bf X$} & \multicolumn{1}{c}{$\bf Y$} & \multicolumn{1}{c}{$\bf Z$}  \\ \hline \hline
771 & 1017 & $2.16367678$ & $3.50090257$ & $-3.68106680$ \\ \cline{2-5}
772 & 1018 & $1.49506388$ & $3.50090257$ & $-4.60133350$ \\ \cline{2-5}
773 & 1019 & $0.41322549$ & $3.50090257$ & $-4.24982290$ \\ \cline{2-5}
774 & 1020 & $-0.66861290$ & $3.50090257$ & $-3.89831230$ \\ \cline{2-5}
775 & 1021 & $-1.75045129$ & $3.50090257$ & $-3.54680170$ \\ \cline{2-5}
776 & 1022 & $-2.83228968$ & $3.50090257$ & $-3.19529110$ \\ \cline{2-5}
777 & 1023 & $-3.91412807$ & $3.50090257$ & $-2.84378050$ \\
\end{tabular}
\ \ \ \ \ \ 
\begin{tabular}[t]{r r | r r r}
\multicolumn{1}{c}{$\bf \#$} & \multicolumn{1}{c}{$\bf i_p$} & \multicolumn{1}{c}{$\bf X$} & \multicolumn{1}{c}{$\bf Y$} & \multicolumn{1}{c}{$\bf Z$}  \\ \hline \hline
826 & 1072 & $2.94650239$ & $1.93525134$ & $-4.24982290$ \\ \cline{2-5}
827 & 1073 & $2.27788950$ & $1.93525134$ & $-5.17008960$ \\ \cline{2-5}
828 & 1074 & $1.19605111$ & $1.93525134$ & $-5.52160021$ \\ \cline{2-5}
829 & 1075 & $0.11421272$ & $1.93525134$ & $-5.17008960$ \\ \cline{2-5}
830 & 1076 & $-0.96762567$ & $1.93525134$ & $-4.81857900$ \\ \cline{2-5}
831 & 1077 & $-2.04946406$ & $1.93525134$ & $-4.46706840$ \\ \cline{2-5}
832 & 1078 & $-3.13130245$ & $1.93525134$ & $-4.11555780$ \\
\end{tabular}
}
\caption{MIF1739 contains $C_n$, such that $C^*_n$ = $\min$($C_n$) $n=2,\ldots,1000$ (cont.)}~\label{tb:idMinLattice_7}
\end{table}

\begin{table}
\centerline{\small
\begin{tabular}[t]{r r | r r r}
\multicolumn{1}{c}{$\bf \#$} & \multicolumn{1}{c}{$\bf i_p$} & \multicolumn{1}{c}{$\bf X$} & \multicolumn{1}{c}{$\bf Y$} & \multicolumn{1}{c}{$\bf Z$}  \\ \hline \hline
881 & 1127 & $5.43615392$ & $0.96762567$ & $0.00000000$ \\ \cline{2-5}
882 & 1128 & $5.25135386$ & $0.00000000$ & $-1.70626830$ \\ \cline{2-5}
883 & 1129 & $4.58274096$ & $0.00000000$ & $-2.62653500$ \\ \cline{2-5}
884 & 1130 & $3.91412807$ & $0.00000000$ & $-3.54680170$ \\ \cline{2-5}
885 & 1131 & $3.24551517$ & $0.00000000$ & $-4.46706840$ \\ \cline{2-5}
886 & 1132 & $2.16367678$ & $0.00000000$ & $-4.81857900$ \\ \cline{2-5}
887 & 1133 & $1.08183839$ & $0.00000000$ & $-5.17008960$ \\
\end{tabular}
\ \ \ \ \ \ 
\begin{tabular}[t]{r r | r r r}
\multicolumn{1}{c}{$\bf \#$} & \multicolumn{1}{c}{$\bf i_p$} & \multicolumn{1}{c}{$\bf X$} & \multicolumn{1}{c}{$\bf Y$} & \multicolumn{1}{c}{$\bf Z$}  \\ \hline \hline
936 & 1182 & $3.72932801$ & $-0.96762567$ & $4.11555780$ \\ \cline{2-5}
937 & 1183 & $4.39794090$ & $-0.96762567$ & $3.19529110$ \\ \cline{2-5}
938 & 1184 & $5.06655380$ & $-0.96762567$ & $2.27502440$ \\ \cline{2-5}
939 & 1185 & $5.06655380$ & $-0.96762567$ & $1.13751220$ \\ \cline{2-5}
940 & 1186 & $5.06655380$ & $-0.96762567$ & $0.00000000$ \\ \cline{2-5}
941 & 1187 & $5.06655380$ & $-0.96762567$ & $-1.13751220$ \\ \cline{2-5}
942 & 1188 & $4.88175374$ & $-1.93525134$ & $-2.84378050$ \\
\end{tabular}
}
\caption{MIF1739 contains $C_n$, such that $C^*_n$ = $\min$($C_n$) $n=2,\ldots,1000$ (cont.)}~\label{tb:idMinLattice_8}
\end{table}

\begin{table}
\centerline{\small
\begin{tabular}[t]{r r | r r r}
\multicolumn{1}{c}{$\bf \#$} & \multicolumn{1}{c}{$\bf i_p$} & \multicolumn{1}{c}{$\bf X$} & \multicolumn{1}{c}{$\bf Y$} & \multicolumn{1}{c}{$\bf Z$}  \\ \hline \hline
991 & 1237 & $-0.71223827$ & $-2.90287702$ & $5.17008960$ \\ \cline{2-5}
992 & 1238 & $0.36960012$ & $-2.90287702$ & $4.81857900$ \\ \cline{2-5}
993 & 1239 & $1.45143851$ & $-2.90287702$ & $4.46706840$ \\ \cline{2-5}
994 & 1240 & $2.53327690$ & $-2.90287702$ & $4.11555780$ \\ \cline{2-5}
995 & 1241 & $3.61511529$ & $-2.90287702$ & $3.76404720$ \\ \cline{2-5}
996 & 1242 & $4.69695368$ & $-2.90287702$ & $3.41253660$ \\ \cline{2-5}
997 & 1243 & $4.69695368$ & $-2.90287702$ & $2.27502440$ \\
\end{tabular}
\ \ \ \ \ \ 
\begin{tabular}[t]{r r | r r r}
\multicolumn{1}{c}{$\bf \#$} & \multicolumn{1}{c}{$\bf i_p$} & \multicolumn{1}{c}{$\bf X$} & \multicolumn{1}{c}{$\bf Y$} & \multicolumn{1}{c}{$\bf Z$}  \\ \hline \hline
1046 & 1292 & $3.13130245$ & $-4.09892812$ & $-1.13751220$ \\ \cline{2-5}
1047 & 1293 & $2.34847684$ & $-4.69695368$ & $-1.70626830$ \\ \cline{2-5}
1048 & 1294 & $1.26663845$ & $-4.69695368$ & $-2.05777890$ \\ \cline{2-5}
1049 & 1295 & $0.18480006$ & $-4.69695368$ & $-2.40928950$ \\ \cline{2-5}
1050 & 1296 & $-0.89703833$ & $-4.69695368$ & $-2.76080010$ \\ \cline{2-5}
1051 & 1297 & $-1.56565123$ & $-4.69695368$ & $-1.84053340$ \\ \cline{2-5}
1052 & 1298 & $-2.23426412$ & $-4.69695368$ & $-0.92026670$ \\
\end{tabular}
}
\caption{MIF1739 contains $C_n$, such that $C^*_n$ = $\min$($C_n$) $n=2,\ldots,1000$ (cont.)}~\label{tb:idMinLattice_9}
\end{table}
\clearpage

\begin{table}
\centerline{\small
\begin{tabular}[t]{r r | r r r}
\multicolumn{1}{c}{$\bf \#$} & \multicolumn{1}{c}{$\bf i_p$} & \multicolumn{1}{c}{$\bf X$} & \multicolumn{1}{c}{$\bf Y$} & \multicolumn{1}{c}{$\bf Z$}  \\ \hline \hline
1101 & 1632 & $0.59802555$ & $6.37681762$ & $-1.84053340$ \\ \cline{2-5}
1102 & 1633 & $-0.48381284$ & $6.37681762$ & $-1.48902280$ \\ \cline{2-5}
1103 & 1634 & $-1.56565123$ & $6.37681762$ & $-1.13751220$ \\ \cline{2-5}
1104 & 1635 & $-1.56565123$ & $6.37681762$ & $0.00000000$ \\ \cline{2-5}
1105 & 1636 & $-1.56565123$ & $6.37681762$ & $1.13751220$ \\ \cline{2-5}
1106 & 1637 & $-0.48381284$ & $6.37681762$ & $1.48902280$ \\ \cline{2-5}
1107 & 1638 & $0.59802555$ & $6.37681762$ & $1.84053340$ \\
\end{tabular}
\ \ \ \ \ \ 
\begin{tabular}[t]{r r | r r r}
\multicolumn{1}{c}{$\bf \#$} & \multicolumn{1}{c}{$\bf i_p$} & \multicolumn{1}{c}{$\bf X$} & \multicolumn{1}{c}{$\bf Y$} & \multicolumn{1}{c}{$\bf Z$}  \\ \hline \hline
1156 & 1687 & $-3.91412807$ & $4.58274096$ & $-0.56875610$ \\ \cline{2-5}
1157 & 1688 & $-3.91412807$ & $4.58274096$ & $0.56875610$ \\ \cline{2-5}
1158 & 1689 & $-3.91412807$ & $4.58274096$ & $1.70626830$ \\ \cline{2-5}
1159 & 1690 & $-3.91412807$ & $4.58274096$ & $2.84378050$ \\ \cline{2-5}
1160 & 1691 & $-2.83228968$ & $4.58274096$ & $3.19529110$ \\ \cline{2-5}
1161 & 1692 & $-1.75045129$ & $4.58274096$ & $3.54680170$ \\ \cline{2-5}
1162 & 1693 & $-0.66861290$ & $4.58274096$ & $3.89831230$ \\
\end{tabular}
}
\caption{MIF1739 contains $C_n$, such that $C^*_n$ = $\min$($C_n$) $n=2,\ldots,1000$ (cont.)}~\label{tb:idMinLattice_10}
\end{table}

\begin{table}
\centerline{\small
\begin{tabular}[t]{r r | r r r}
\multicolumn{1}{c}{$\bf \#$} & \multicolumn{1}{c}{$\bf i_p$} & \multicolumn{1}{c}{$\bf X$} & \multicolumn{1}{c}{$\bf Y$} & \multicolumn{1}{c}{$\bf Z$}  \\ \hline \hline
1211 & 1783 & $-5.66457935$ & $2.41906418$ & $1.13751220$ \\ \cline{2-5}
1212 & 1784 & $-5.66457935$ & $2.41906418$ & $2.27502440$ \\ \cline{2-5}
1213 & 1815 & $-5.84937941$ & $1.45143851$ & $-1.70626830$ \\ \cline{2-5}
1214 & 1816 & $-5.84937941$ & $1.45143851$ & $-0.56875610$ \\ \cline{2-5}
1215 & 1817 & $-5.84937941$ & $1.45143851$ & $0.56875610$ \\ \cline{2-5}
1216 & 1818 & $-5.84937941$ & $1.45143851$ & $1.70626830$ \\ \cline{2-5}
1217 & 1840 & $2.46268956$ & $0.48381284$ & $-5.73884570$ \\
\end{tabular}
\ \ \ \ \ \ 
\begin{tabular}[t]{r r | r r r}
\multicolumn{1}{c}{$\bf \#$} & \multicolumn{1}{c}{$\bf i_p$} & \multicolumn{1}{c}{$\bf X$} & \multicolumn{1}{c}{$\bf Y$} & \multicolumn{1}{c}{$\bf Z$}  \\ \hline \hline
1266 & 2086 & $1.26663845$ & $-5.77879207$ & $-2.05777890$ \\ \cline{2-5}
1267 & 2087 & $0.18480006$ & $-5.77879207$ & $-2.40928950$ \\ \cline{2-5}
1268 & 2098 & $2.34847684$ & $-5.77879207$ & $0.56875610$ \\ \cline{2-5}
1269 & 2099 & $2.34847684$ & $-5.77879207$ & $-0.56875610$ \\ \cline{2-5}
1270 & 2100 & $1.56565123$ & $-6.37681762$ & $-1.13751220$ \\ \cline{2-5}
1271 & 2101 & $0.48381284$ & $-6.37681762$ & $-1.48902280$ \\ \cline{2-5}
1272 & 2109 & $1.56565123$ & $-6.37681762$ & $0.00000000$ \\
\end{tabular}
}
\caption{MIF1739 contains $C_n$, such that $C^*_n$ = $\min$($C_n$) $n=2,\ldots,1000$ (cont.)}~\label{tb:idMinLattice_11}
\end{table}

\begin{table}
\centerline{\small
\begin{tabular}[t]{r r | r r r}
\multicolumn{1}{c}{$\bf \#$} & \multicolumn{1}{c}{$\bf i_p$} & \multicolumn{1}{c}{$\bf X$} & \multicolumn{1}{c}{$\bf Y$} & \multicolumn{1}{c}{$\bf Z$}  \\ \hline \hline
1321 & 2584 & $1.86466400$ & $6.26260490$ & $2.76080010$ \\ \cline{2-5}
1322 & 2585 & $2.53327690$ & $6.26260490$ & $1.84053340$ \\ \cline{2-5}
1323 & 2586 & $3.20188979$ & $6.26260490$ & $0.92026670$ \\ \cline{2-5}
1324 & 2587 & $4.83812836$ & $5.66457935$ & $0.00000000$ \\ \cline{2-5}
1325 & 2588 & $4.16951547$ & $5.66457935$ & $-0.92026670$ \\ \cline{2-5}
1326 & 2589 & $3.50090257$ & $5.66457935$ & $-1.84053340$ \\ \cline{2-5}
1327 & 2590 & $2.83228968$ & $5.66457935$ & $-2.76080010$ \\
\end{tabular}
\ \ \ \ \ \ 
\begin{tabular}[t]{r r | r r r}
\multicolumn{1}{c}{$\bf \#$} & \multicolumn{1}{c}{$\bf i_p$} & \multicolumn{1}{c}{$\bf X$} & \multicolumn{1}{c}{$\bf Y$} & \multicolumn{1}{c}{$\bf Z$}  \\ \hline \hline
1376 & 2658 & $-5.47977929$ & $4.46852824$ & $-1.70626830$ \\ \cline{2-5}
1377 & 2659 & $-5.47977929$ & $4.46852824$ & $-0.56875610$ \\ \cline{2-5}
1378 & 2660 & $-5.47977929$ & $4.46852824$ & $0.56875610$ \\ \cline{2-5}
1379 & 2661 & $-5.47977929$ & $4.46852824$ & $1.70626830$ \\ \cline{2-5}
1380 & 2662 & $-5.47977929$ & $4.46852824$ & $2.84378050$ \\ \cline{2-5}
1381 & 2664 & $-4.39794090$ & $4.46852824$ & $4.33280330$ \\ \cline{2-5}
1382 & 2665 & $-3.31610251$ & $4.46852824$ & $4.68431390$ \\
\end{tabular}
}
\caption{MIF1739 contains $C_n$, such that $C^*_n$ = $\min$($C_n$) $n=2,\ldots,1000$ (cont.)}~\label{tb:idMinLattice_12}
\end{table}

\begin{table}
\centerline{\small
\begin{tabular}[t]{r r | r r r}
\multicolumn{1}{c}{$\bf \#$} & \multicolumn{1}{c}{$\bf i_p$} & \multicolumn{1}{c}{$\bf X$} & \multicolumn{1}{c}{$\bf Y$} & \multicolumn{1}{c}{$\bf Z$}  \\ \hline \hline
1431 & 3777 & $-3.13130245$ & $7.34444329$ & $-2.27502440$ \\ \cline{2-5}
1432 & 3778 & $-3.13130245$ & $7.34444329$ & $-1.13751220$ \\ \cline{2-5}
1433 & 3779 & $-3.13130245$ & $7.34444329$ & $0.00000000$ \\ \cline{2-5}
1434 & 3780 & $-3.13130245$ & $7.34444329$ & $1.13751220$ \\ \cline{2-5}
1435 & 3781 & $-3.13130245$ & $7.34444329$ & $2.27502440$ \\ \cline{2-5}
1436 & 3782 & $-2.04946406$ & $7.34444329$ & $2.62653500$ \\ \cline{2-5}
1437 & 3783 & $-0.96762567$ & $7.34444329$ & $2.97804560$ \\
\end{tabular}
\ \ \ \ \ \ 
\begin{tabular}[t]{r r | r r r}
\multicolumn{1}{c}{$\bf \#$} & \multicolumn{1}{c}{$\bf i_p$} & \multicolumn{1}{c}{$\bf X$} & \multicolumn{1}{c}{$\bf Y$} & \multicolumn{1}{c}{$\bf Z$}  \\ \hline \hline
1486 & 3835 & $-1.45143851$ & $6.14839219$ & $4.46706840$ \\ \cline{2-5}
1487 & 3836 & $-0.36960012$ & $6.14839219$ & $4.81857900$ \\ \cline{2-5}
1488 & 3837 & $0.71223827$ & $6.14839219$ & $5.17008960$ \\ \cline{2-5}
1489 & 3838 & $1.79407666$ & $6.14839219$ & $5.52160021$ \\ \cline{2-5}
1490 & 3839 & $2.46268956$ & $6.14839219$ & $4.60133350$ \\ \cline{2-5}
1491 & 3840 & $3.13130245$ & $6.14839219$ & $3.68106680$ \\ \cline{2-5}
1492 & 3841 & $3.79991535$ & $6.14839219$ & $2.76080010$ \\
\end{tabular}
}
\caption{MIF1739 contains $C_n$, such that $C^*_n$ = $\min$($C_n$) $n=2,\ldots,1000$ (cont.)}~\label{tb:idMinLattice_13}
\end{table}

\begin{table}
\centerline{\small
\begin{tabular}[t]{r r | r r r}
\multicolumn{1}{c}{$\bf \#$} & \multicolumn{1}{c}{$\bf i_p$} & \multicolumn{1}{c}{$\bf X$} & \multicolumn{1}{c}{$\bf Y$} & \multicolumn{1}{c}{$\bf Z$}  \\ \hline \hline
1541 & 5300 & $2.23426412$ & $9.02430724$ & $0.92026670$ \\ \cline{2-5}
1542 & 5301 & $3.87050269$ & $8.42628168$ & $0.00000000$ \\ \cline{2-5}
1543 & 5302 & $3.20188979$ & $8.42628168$ & $-0.92026670$ \\ \cline{2-5}
1544 & 5303 & $2.53327690$ & $8.42628168$ & $-1.84053340$ \\ \cline{2-5}
1545 & 5304 & $1.86466400$ & $8.42628168$ & $-2.76080010$ \\ \cline{2-5}
1546 & 5305 & $1.19605111$ & $8.42628168$ & $-3.68106680$ \\ \cline{2-5}
1547 & 5306 & $0.11421272$ & $8.42628168$ & $-3.32955620$ \\
\end{tabular}
\ \ \ \ \ \ 
\begin{tabular}[t]{r r | r r r}
\multicolumn{1}{c}{$\bf \#$} & \multicolumn{1}{c}{$\bf i_p$} & \multicolumn{1}{c}{$\bf X$} & \multicolumn{1}{c}{$\bf Y$} & \multicolumn{1}{c}{$\bf Z$}  \\ \hline \hline
1596 & 5357 & $-3.61511529$ & $7.23023058$ & $-3.76404720$ \\ \cline{2-5}
1597 & 5359 & $-4.69695368$ & $7.23023058$ & $-2.27502440$ \\ \cline{2-5}
1598 & 5360 & $-4.69695368$ & $7.23023058$ & $-1.13751220$ \\ \cline{2-5}
1599 & 5361 & $-4.69695368$ & $7.23023058$ & $0.00000000$ \\ \cline{2-5}
1600 & 5362 & $-4.69695368$ & $7.23023058$ & $1.13751220$ \\ \cline{2-5}
1601 & 5363 & $-4.69695368$ & $7.23023058$ & $2.27502440$ \\ \cline{2-5}
1602 & 5364 & $-4.69695368$ & $7.23023058$ & $3.41253660$ \\
\end{tabular}
}
\caption{MIF1739 contains $C_n$, such that $C^*_n$ = $\min$($C_n$) $n=2,\ldots,1000$ (cont.)}~\label{tb:idMinLattice_14}
\end{table}
\clearpage

\begin{table}
\centerline{\small
\begin{tabular}[t]{r r | r r r}
\multicolumn{1}{c}{$\bf \#$} & \multicolumn{1}{c}{$\bf i_p$} & \multicolumn{1}{c}{$\bf X$} & \multicolumn{1}{c}{$\bf Y$} & \multicolumn{1}{c}{$\bf Z$}  \\ \hline \hline
1651 & 7193 & $-1.26663845$ & $10.10614563$ & $-2.05777890$ \\ \cline{2-5}
1652 & 7194 & $-2.34847684$ & $10.10614563$ & $-1.70626830$ \\ \cline{2-5}
1653 & 7195 & $-2.34847684$ & $10.10614563$ & $-0.56875610$ \\ \cline{2-5}
1654 & 7196 & $-2.34847684$ & $10.10614563$ & $0.56875610$ \\ \cline{2-5}
1655 & 7197 & $-2.34847684$ & $10.10614563$ & $1.70626830$ \\ \cline{2-5}
1656 & 7198 & $-1.26663845$ & $10.10614563$ & $2.05777890$ \\ \cline{2-5}
1657 & 7199 & $-0.18480006$ & $10.10614563$ & $2.40928950$ \\
\end{tabular}
\ \ \ \ \ \ 
\begin{tabular}[t]{r r | r r r}
\multicolumn{1}{c}{$\bf \#$} & \multicolumn{1}{c}{$\bf i_p$} & \multicolumn{1}{c}{$\bf X$} & \multicolumn{1}{c}{$\bf Y$} & \multicolumn{1}{c}{$\bf Z$}  \\ \hline \hline
1706 & 7250 & $4.46852824$ & $8.31206897$ & $-1.84053340$ \\ \cline{2-5}
1707 & 7251 & $3.79991535$ & $8.31206897$ & $-2.76080010$ \\ \cline{2-5}
1708 & 7252 & $3.13130245$ & $8.31206897$ & $-3.68106680$ \\ \cline{2-5}
1709 & 7256 & $-0.36960012$ & $8.31206897$ & $-4.81857900$ \\ \cline{2-5}
1710 & 7257 & $-1.45143851$ & $8.31206897$ & $-4.46706840$ \\ \cline{2-5}
1711 & 7258 & $-2.53327690$ & $8.31206897$ & $-4.11555780$ \\ \cline{2-5}
1712 & 7259 & $-3.61511529$ & $8.31206897$ & $-3.76404720$ \\
\end{tabular}
}
\caption{MIF1739 contains $C_n$, such that $C^*_n$ = $\min$($C_n$) $n=2,\ldots,1000$ (cont.)}~\label{tb:idMinLattice_15}
\end{table}

%% file: tabla_DiffMinLatice.tex
\begin{table}
\centerline{\small
\begin{tabular}{c | l c c c r}
\multicolumn{1}{c}{$n$} & \multicolumn{1}{c}{T} & \multicolumn{1}{c}{ - E$^{MIF}$} & \multicolumn{1}{c}{ - E$^*$}   &  \multicolumn{1}{c}{-(E$^*$ -E$^{MIF}$) } & \multicolumn{1}{c}{Adj} \\ \hline \hline
 1000  & 1 & $6993.629435$ & $7128.821829$ & $135.192394$ & 0 \\ \cline{2-6}
 999  & 1 & $6985.902614$ & $7120.842638$ & $134.940023$ & 1 \\ \cline{2-6}
 998  & 1 & $6978.241464$ & $7113.103883$ & $134.862419$ & 5 \\ \cline{2-6}
 997  & 1 & $6971.087388$ & $7105.735481$ & $134.648093$ & 1 \\ \cline{2-6}
 996  & 1 & $6963.933352$ & $7098.367102$ & $134.433750$ & 1 \\ \cline{2-6}
 995  & 1 & $6955.926596$ & $7090.208621$ & $134.282025$ & 1 \\ \cline{2-6}
 994  & 1 & $6948.023102$ & $7082.249979$ & $134.226877$ & 3 \\
\end{tabular}
\ \ 
\begin{tabular}{c | l c c c r}
\multicolumn{1}{c}{$n$} & \multicolumn{1}{c}{T} & \multicolumn{1}{c}{ - E$^{MIF}$} & \multicolumn{1}{c}{ - E$^*$}   &  \multicolumn{1}{c}{-(E$^*$ -E$^{MIF}$) } & \multicolumn{1}{c}{Adj} \\ \hline \hline
 946  & 1 & $6593.942750$ & $6721.248620$ & $127.305870$ & 1 \\ \cline{2-6}
 945  & 1 & $6586.835928$ & $6713.940121$ & $127.104193$ & 1 \\ \cline{2-6}
 944  & 1 & $6579.729107$ & $6706.631600$ & $126.902493$ & 1 \\ \cline{2-6}
 943  & 1 & $6571.915736$ & $6698.737233$ & $126.821497$ & 3 \\ \cline{2-6}
 942  & 1 & $6564.808928$ & $6691.428659$ & $126.619731$ & 1 \\ \cline{2-6}
 941  & 1 & $6557.196696$ & $6683.553791$ & $126.357095$ & 5 \\ \cline{2-6}
 940  & 1 & $6549.888846$ & $6676.224783$ & $126.335937$ & 7 \\
\end{tabular}
}
\caption{Results from MIF1739: type, initial, minimum, and difference of potential, and Adj($C_{n+1}$,$C_{n}$).}~\label{tb:Changes_0}
\end{table}

\begin{table}
\centerline{\small
\begin{tabular}{c | l c c c r}
\multicolumn{1}{c}{$n$} & \multicolumn{1}{c}{T} & \multicolumn{1}{c}{ - E$^{MIF}$} & \multicolumn{1}{c}{ - E$^*$}   &  \multicolumn{1}{c}{-(E$^*$ -E$^{MIF}$) } & \multicolumn{1}{c}{Adj} \\ \hline \hline
 892  & 1 & $6196.870994$ & $6315.533826$ & $118.662832$ & 1 \\ \cline{2-6}
 891  & 1 & $6189.681742$ & $6308.162614$ & $118.480872$ & 3 \\ \cline{2-6}
 890  & 1 & $6182.615707$ & $6300.903543$ & $118.287835$ & 1 \\ \cline{2-6}
 889  & 1 & $6173.769778$ & $6291.887710$ & $118.117931$ & 1 \\ \cline{2-6}
 888  & 1 & $6165.973161$ & $6283.878319$ & $117.905158$ & 3 \\ \cline{2-6}
 887  & 1 & $6158.248389$ & $6275.969721$ & $117.721332$ & 1 \\ \cline{2-6}
 886  & 1 & $6151.059154$ & $6268.598027$ & $117.538873$ & 3 \\
\end{tabular}
\ \ 
\begin{tabular}{c | l c c c r}
\multicolumn{1}{c}{$n$} & \multicolumn{1}{c}{T} & \multicolumn{1}{c}{ - E$^{MIF}$} & \multicolumn{1}{c}{ - E$^*$}   &  \multicolumn{1}{c}{-(E$^*$ -E$^{MIF}$) } & \multicolumn{1}{c}{Adj} \\ \hline \hline
 838  & 1 & $5789.342198$ & $5901.636273$ & $112.294075$ & 7 \\ \cline{2-6}
 837  & 1 & $5782.276279$ & $5894.377564$ & $112.101285$ & 1 \\ \cline{2-6}
 836  & 1 & $5774.735034$ & $5886.706637$ & $111.971603$ & 5 \\ \cline{2-6}
 835  & 1 & $5767.669481$ & $5879.448440$ & $111.778959$ & 1 \\ \cline{2-6}
 834  & 1 & $5760.603572$ & $5872.189749$ & $111.586176$ & 1 \\ \cline{2-6}
 833  & 1 & $5752.877886$ & $5864.257422$ & $111.379536$ & 1 \\ \cline{2-6}
 832  & 1 & $5745.136150$ & $5856.444416$ & $111.308266$ & 5 \\
\end{tabular}
}
\caption{Results from MIF1739: type, initial, minimum, and difference of potential, and Adj($C_{n+1}$,$C_{n}$) (cont.).}~\label{tb:Changes_1}
\end{table}

\begin{table}
\centerline{\small
\begin{tabular}{c | l c c c r}
\multicolumn{1}{c}{$n$} & \multicolumn{1}{c}{T} & \multicolumn{1}{c}{ - E$^{MIF}$} & \multicolumn{1}{c}{ - E$^*$}   &  \multicolumn{1}{c}{-(E$^*$ -E$^{MIF}$) } & \multicolumn{1}{c}{Adj} \\ \hline \hline
 784  & 1 & $5389.030236$ & $5494.160631$ & $105.130395$ & 23 \\ \cline{2-6}
 783  & 1 & $5381.965653$ & $5486.904329$ & $104.938676$ & 1 \\ \cline{2-6}
 782  & 1 & $5374.901110$ & $5479.648067$ & $104.746957$ & 1 \\ \cline{2-6}
 781  & 1 & $5367.835486$ & $5472.390187$ & $104.554701$ & 1 \\ \cline{2-6}
 780  & 1 & $5360.293437$ & $5464.707535$ & $104.414099$ & 5 \\ \cline{2-6}
 779  & 1 & $5353.228903$ & $5457.451291$ & $104.222387$ & 1 \\ \cline{2-6}
 778  & 1 & $5346.163522$ & $5450.193696$ & $104.030174$ & 1 \\
\end{tabular}
\ \ 
\begin{tabular}{c | l c c c r}
\multicolumn{1}{c}{$n$} & \multicolumn{1}{c}{T} & \multicolumn{1}{c}{ - E$^{MIF}$} & \multicolumn{1}{c}{ - E$^*$}   &  \multicolumn{1}{c}{-(E$^*$ -E$^{MIF}$) } & \multicolumn{1}{c}{Adj} \\ \hline \hline
 730  & 1 & $4991.452265$ & $5088.472966$ & $97.020701$ & 9 \\ \cline{2-6}
 729  & 1 & $4984.281993$ & $5081.042333$ & $96.760340$ & 5 \\ \cline{2-6}
 728  & 1 & $4976.531159$ & $5073.284623$ & $96.753464$ & 13 \\ \cline{2-6}
 727  & 1 & $4969.351473$ & $5065.896282$ & $96.544809$ & 15 \\ \cline{2-6}
 726  & 1 & $4960.988873$ & $5057.800861$ & $96.811988$ & 31 \\ \cline{2-6}
 725  & 1 & $4953.925068$ & $5050.546387$ & $96.621319$ & 1 \\ \cline{2-6}
 724  & 1 & $4946.860866$ & $5043.291232$ & $96.430366$ & 27 \\
\end{tabular}
}
\caption{Results from MIF1739: type, initial, minimum, and difference of potential, and Adj($C_{n+1}$,$C_{n}$) (cont.).}~\label{tb:Changes_2}
\end{table}

\begin{table}
\centerline{\small
\begin{tabular}{c | l c c c r}
\multicolumn{1}{c}{$n$} & \multicolumn{1}{c}{T} & \multicolumn{1}{c}{ - E$^{MIF}$} & \multicolumn{1}{c}{ - E$^*$}   &  \multicolumn{1}{c}{-(E$^*$ -E$^{MIF}$) } & \multicolumn{1}{c}{Adj} \\ \hline \hline
 676  & 1 & $4605.622417$ & $4686.556156$ & $80.933739$ & 1 \\ \cline{2-6}
 675  & 1 & $4598.558789$ & $4679.303997$ & $80.745207$ & 1 \\ \cline{2-6}
 674  & 1 & $4591.495168$ & $4672.051833$ & $80.556665$ & 1 \\ \cline{2-6}
 673  & 1 & $4575.585597$ & $4664.692986$ & $89.107389$ & 1 \\ \cline{2-6}
 672  & 1 & $4567.862777$ & $4656.743036$ & $88.880259$ & 1 \\ \cline{2-6}
 671  & 1 & $4568.967545$ & $4649.022423$ & $80.054878$ & 5 \\ \cline{2-6}
 670  & 1 & $4561.817349$ & $4641.668692$ & $79.851343$ & 7 \\
\end{tabular}
\ \ 
\begin{tabular}{c | l c c c r}
\multicolumn{1}{c}{$n$} & \multicolumn{1}{c}{T} & \multicolumn{1}{c}{ - E$^{MIF}$} & \multicolumn{1}{c}{ - E$^*$}   &  \multicolumn{1}{c}{-(E$^*$ -E$^{MIF}$) } & \multicolumn{1}{c}{Adj} \\ \hline \hline
 622  & 1 & $4212.818660$ & $4285.785595$ & $72.966935$ & 5 \\ \cline{2-6}
 621  & 1 & $4205.668824$ & $4278.428797$ & $72.759973$ & 1 \\ \cline{2-6}
 620  & 1 & $4189.767363$ & $4270.926392$ & $81.159029$ & 1 \\ \cline{2-6}
 619  & 1 & $4182.045625$ & $4262.967594$ & $80.921969$ & 1 \\ \cline{2-6}
 618  & 1 & $4183.148787$ & $4255.392520$ & $72.243733$ & 5 \\ \cline{2-6}
 617  & 1 & $4175.999673$ & $4248.036524$ & $72.036850$ & 1 \\ \cline{2-6}
 616  & 1 & $4160.098779$ & $4240.525779$ & $80.426999$ & 1 \\
\end{tabular}
}
\caption{Results from MIF1739: type, initial, minimum, and difference of potential, and Adj($C_{n+1}$,$C_{n}$) (cont.).}~\label{tb:Changes_3}
\end{table}

\begin{table}
\centerline{\small
\begin{tabular}{c | l c c c r}
\multicolumn{1}{c}{$n$} & \multicolumn{1}{c}{T} & \multicolumn{1}{c}{ - E$^{MIF}$} & \multicolumn{1}{c}{ - E$^*$}   &  \multicolumn{1}{c}{-(E$^*$ -E$^{MIF}$) } & \multicolumn{1}{c}{Adj} \\ \hline \hline
 568  & 1 & $3823.709861$ & $3889.051094$ & $65.341234$ & 1 \\ \cline{2-6}
 567  & 1 & $3816.647847$ & $3881.798364$ & $65.150517$ & 1 \\ \cline{2-6}
 566  & 1 & $3809.585951$ & $3874.545764$ & $64.959813$ & 1 \\ \cline{2-6}
 565  & 1 & $3802.524173$ & $3867.293295$ & $64.769122$ & 1 \\ \cline{2-6}
 564  & 5 & $3795.116425$ & $3860.238707$ & $65.122283$ & 13 \\ \cline{2-6}
 563  & 5 & $3788.439649$ & $3853.287758$ & $64.848109$ & 1 \\ \cline{2-6}
 562  & 1 & $3782.730387$ & $3847.376685$ & $64.646298$ & 3 \\
\end{tabular}
\ \ 
\begin{tabular}{c | l c c c r}
\multicolumn{1}{c}{$n$} & \multicolumn{1}{c}{T} & \multicolumn{1}{c}{ - E$^{MIF}$} & \multicolumn{1}{c}{ - E$^*$}   &  \multicolumn{1}{c}{-(E$^*$ -E$^{MIF}$) } & \multicolumn{1}{c}{Adj} \\ \hline \hline
 514  & 1 & $3429.039452$ & $3487.179880$ & $58.140429$ & 1 \\ \cline{2-6}
 513  & 1 & $3421.978326$ & $3479.926793$ & $57.948467$ & 1 \\ \cline{2-6}
 512  & 1 & $3414.449890$ & $3472.314244$ & $57.864354$ & 5 \\ \cline{2-6}
 511  & 1 & $3407.388678$ & $3465.061048$ & $57.672370$ & 1 \\ \cline{2-6}
 510  & 1 & $3400.327585$ & $3457.807978$ & $57.480394$ & 1 \\ \cline{2-6}
 509  & 1 & $3392.799163$ & $3450.193949$ & $57.394787$ & 5 \\ \cline{2-6}
 508  & 1 & $3385.737983$ & $3442.940773$ & $57.202790$ & 1 \\
\end{tabular}
}
\caption{Results from MIF1739: type, initial, minimum, and difference of potential, and Adj($C_{n+1}$,$C_{n}$) (cont.).}~\label{tb:Changes_4}
\end{table}
\clearpage

\begin{table}
\centerline{\small
\begin{tabular}{c | l c c c r}
\multicolumn{1}{c}{$n$} & \multicolumn{1}{c}{T} & \multicolumn{1}{c}{ - E$^{MIF}$} & \multicolumn{1}{c}{ - E$^*$}   &  \multicolumn{1}{c}{-(E$^*$ -E$^{MIF}$) } & \multicolumn{1}{c}{Adj} \\ \hline \hline
 460  & 1 & $3037.549004$ & $3089.218142$ & $51.669138$ & 1 \\ \cline{2-6}
 459  & 1 & $3030.492575$ & $3081.971267$ & $51.478692$ & 1 \\ \cline{2-6}
 458  & 1 & $3023.434904$ & $3074.722501$ & $51.287596$ & 1 \\ \cline{2-6}
 457  & 1 & $3016.377352$ & $3067.473860$ & $51.096507$ & 1 \\ \cline{2-6}
 456  & 1 & $3009.317001$ & $3060.221313$ & $50.904312$ & 1 \\ \cline{2-6}
 455  & 1 & $3001.785917$ & $3052.572975$ & $50.787059$ & 5 \\ \cline{2-6}
 454  & 1 & $2994.728397$ & $3045.324373$ & $50.595976$ & 1 \\
\end{tabular}
\ \ 
\begin{tabular}{c | l c c c r}
\multicolumn{1}{c}{$n$} & \multicolumn{1}{c}{T} & \multicolumn{1}{c}{ - E$^{MIF}$} & \multicolumn{1}{c}{ - E$^*$}   &  \multicolumn{1}{c}{-(E$^*$ -E$^{MIF}$) } & \multicolumn{1}{c}{Adj} \\ \hline \hline
 406  & 1 & $2652.605944$ & $2696.958552$ & $44.352608$ & 1 \\ \cline{2-6}
 405  & 1 & $2645.550990$ & $2689.713734$ & $44.162745$ & 1 \\ \cline{2-6}
 404  & 1 & $2637.837823$ & $2681.814373$ & $43.976550$ & 1 \\ \cline{2-6}
 403  & 1 & $2630.124666$ & $2674.029018$ & $43.904352$ & 5 \\ \cline{2-6}
 402  & 1 & $2623.069773$ & $2666.784339$ & $43.714566$ & 1 \\ \cline{2-6}
 401  & 1 & $2615.929390$ & $2659.459309$ & $43.529918$ & 1 \\ \cline{2-6}
 400  & 1 & $2608.201551$ & $2651.675335$ & $43.473783$ & 3 \\
\end{tabular}
}
\caption{Results from MIF1739: type, initial, minimum, and difference of potential, and Adj($C_{n+1}$,$C_{n}$) (cont.).}~\label{tb:Changes_5}
\end{table}

\begin{table}
\centerline{\small
\begin{tabular}{c | l c c c r}
\multicolumn{1}{c}{$n$} & \multicolumn{1}{c}{T} & \multicolumn{1}{c}{ - E$^{MIF}$} & \multicolumn{1}{c}{ - E$^*$}   &  \multicolumn{1}{c}{-(E$^*$ -E$^{MIF}$) } & \multicolumn{1}{c}{Adj} \\ \hline \hline
 352  & 1 & $2270.626129$ & $2307.648892$ & $37.022763$ & 1 \\ \cline{2-6}
 351  & 1 & $2263.572396$ & $2300.415404$ & $36.843008$ & 1 \\ \cline{2-6}
 350  & 1 & $2256.518688$ & $2293.181867$ & $36.663179$ & 1 \\ \cline{2-6}
 349  & 1 & $2249.465430$ & $2285.948844$ & $36.483414$ & 1 \\ \cline{2-6}
 348  & 1 & $2242.416025$ & $2278.712819$ & $36.296794$ & 1 \\ \cline{2-6}
 347  & 1 & $2234.704330$ & $2270.792335$ & $36.088006$ & 1 \\ \cline{2-6}
 346  & 1 & $2226.994520$ & $2262.986228$ & $35.991708$ & 5 \\
\end{tabular}
\ \ 
\begin{tabular}{c | l c c c r}
\multicolumn{1}{c}{$n$} & \multicolumn{1}{c}{T} & \multicolumn{1}{c}{ - E$^{MIF}$} & \multicolumn{1}{c}{ - E$^*$}   &  \multicolumn{1}{c}{-(E$^*$ -E$^{MIF}$) } & \multicolumn{1}{c}{Adj} \\ \hline \hline
 298  & 1 & $1898.393646$ & $1927.638785$ & $29.245140$ & 1 \\ \cline{2-6}
 297  & 1 & $1891.343530$ & $1920.405571$ & $29.062041$ & 1 \\ \cline{2-6}
 296  & 1 & $1882.550887$ & $1911.461967$ & $28.911080$ & 1 \\ \cline{2-6}
 295  & 1 & $1874.752345$ & $1903.525774$ & $28.773429$ & 1 \\ \cline{2-6}
 294  & 1 & $1867.054052$ & $1895.687896$ & $28.633843$ & 1 \\ \cline{2-6}
 293  & 1 & $1859.902111$ & $1888.427400$ & $28.525288$ & 3 \\ \cline{2-6}
 292  & 1 & $1852.852066$ & $1881.193243$ & $28.341177$ & 1 \\
\end{tabular}
}
\caption{Results from MIF1739: type, initial, minimum, and difference of potential, and Adj($C_{n+1}$,$C_{n}$) (cont.).}~\label{tb:Changes_6}
\end{table}

\begin{table}
\centerline{\small
\begin{tabular}{c | l c c c r}
\multicolumn{1}{c}{$n$} & \multicolumn{1}{c}{T} & \multicolumn{1}{c}{ - E$^{MIF}$} & \multicolumn{1}{c}{ - E$^*$}   &  \multicolumn{1}{c}{-(E$^*$ -E$^{MIF}$) } & \multicolumn{1}{c}{Adj} \\ \hline \hline
 244  & 1 & $1512.956517$ & $1536.882829$ & $23.926312$ & 13 \\ \cline{2-6}
 243  & 1 & $1505.908421$ & $1529.712151$ & $23.803729$ & 1 \\ \cline{2-6}
 242  & 1 & $1499.076971$ & $1522.723264$ & $23.646293$ & 7 \\ \cline{2-6}
 241  & 1 & $1492.030024$ & $1515.554577$ & $23.524553$ & 1 \\ \cline{2-6}
 240  & 1 & $1485.202483$ & $1508.562918$ & $23.360435$ & 3 \\ \cline{2-6}
 239  & 1 & $1478.156685$ & $1501.396062$ & $23.239377$ & 1 \\ \cline{2-6}
 238  & 2 & $1458.401342$ & $1494.438248$ & $36.036906$ & 295 \\
\end{tabular}
\ \ 
\begin{tabular}{c | l c c c r}
\multicolumn{1}{c}{$n$} & \multicolumn{1}{c}{T} & \multicolumn{1}{c}{ - E$^{MIF}$} & \multicolumn{1}{c}{ - E$^*$}   &  \multicolumn{1}{c}{-(E$^*$ -E$^{MIF}$) } & \multicolumn{1}{c}{Adj} \\ \hline \hline
 190  & 2 & $1133.097035$ & $1161.301939$ & $28.204904$ & 1 \\ \cline{2-6}
 189  & 2 & $1125.783551$ & $1153.637019$ & $27.853468$ & 5 \\ \cline{2-6}
 188  & 2 & $1118.572796$ & $1146.439107$ & $27.866311$ & 1 \\ \cline{2-6}
 187  & 1 & $1122.743390$ & $1139.455696$ & $16.712305$ & 201 \\ \cline{2-6}
 186  & 1 & $1116.076450$ & $1132.669966$ & $16.593516$ & 1 \\ \cline{2-6}
 185  & 1 & $1109.046091$ & $1125.493794$ & $16.447704$ & 1 \\ \cline{2-6}
 184  & 1 & $1102.422710$ & $1118.767663$ & $16.344953$ & 9 \\
\end{tabular}
}
\caption{Results from MIF1739: type, initial, minimum, and difference of potential, and Adj($C_{n+1}$,$C_{n}$) (cont.).}~\label{tb:Changes_7}
\end{table}

\begin{table}
\centerline{\small
\begin{tabular}{c | l c c c r}
\multicolumn{1}{c}{$n$} & \multicolumn{1}{c}{T} & \multicolumn{1}{c}{ - E$^{MIF}$} & \multicolumn{1}{c}{ - E$^*$}   &  \multicolumn{1}{c}{-(E$^*$ -E$^{MIF}$) } & \multicolumn{1}{c}{Adj} \\ \hline \hline
 136  & 1 & $787.552841$ & $797.453259$ & $9.900418$ & 1 \\ \cline{2-6}
 135  & 1 & $780.538494$ & $790.278120$ & $9.739626$ & 1 \\ \cline{2-6}
 134  & 1 & $771.888084$ & $782.206157$ & $10.318073$ & 13 \\ \cline{2-6}
 133  & 1 & $764.872103$ & $775.023203$ & $10.151101$ & 1 \\ \cline{2-6}
 132  & 1 & $757.624997$ & $768.042203$ & $10.417206$ & 7 \\ \cline{2-6}
 131  & 1 & $752.079580$ & $762.441558$ & $10.361978$ & 1 \\ \cline{2-6}
 130  & 1 & $745.074713$ & $755.271073$ & $10.196360$ & 1 \\
\end{tabular}
\ \ 
\begin{tabular}{c | l c c c r}
\multicolumn{1}{c}{$n$} & \multicolumn{1}{c}{T} & \multicolumn{1}{c}{ - E$^{MIF}$} & \multicolumn{1}{c}{ - E$^*$}   &  \multicolumn{1}{c}{-(E$^*$ -E$^{MIF}$) } & \multicolumn{1}{c}{Adj} \\ \hline \hline
 82  & 1 & $435.141809$ & $440.550425$ & $5.408616$ & 1 \\ \cline{2-6}
 81  & 5 & $418.751871$ & $434.343643$ & $15.591772$ & 49 \\ \cline{2-6}
 80  & 5 & $413.174252$ & $428.083564$ & $14.909313$ & 1 \\ \cline{2-6}
 79  & 5 & $407.599164$ & $421.810897$ & $14.211734$ & 1 \\ \cline{2-6}
 78  & 5 & $400.623898$ & $414.794401$ & $14.170503$ & 1 \\ \cline{2-6}
 77  & 2 & $399.485644$ & $409.083517$ & $9.597873$ & 109 \\ \cline{2-6}
 76  & 2 & $393.350121$ & $402.894866$ & $9.544745$ & 1 \\
\end{tabular}
}
\caption{Results from MIF1739: type, initial, minimum, and difference of potential, and Adj($C_{n+1}$,$C_{n}$) (cont.).}~\label{tb:Changes_8}
\end{table}

\begin{table}
\centerline{\small
\begin{tabular}{c | l c c c r}
\multicolumn{1}{c}{$n$} & \multicolumn{1}{c}{T} & \multicolumn{1}{c}{ - E$^{MIF}$} & \multicolumn{1}{c}{ - E$^*$}   &  \multicolumn{1}{c}{-(E$^*$ -E$^{MIF}$) } & \multicolumn{1}{c}{Adj} \\ \hline \hline
 28  & 5 & $111.743771$ & $117.822402$ & $6.078631$ & 1 \\ \cline{2-6}
 27  & 5 & $107.676832$ & $112.873584$ & $5.196752$ & 1 \\ \cline{2-6}
 26  & 5 & $103.329759$ & $108.315616$ & $4.985858$ & 5 \\ \cline{2-6}
 25  & 5 & $97.407732$ & $102.372663$ & $4.964931$ & 1 \\ \cline{2-6}
 24  & 5 & $93.391233$ & $97.348815$ & $3.957582$ & 1 \\ \cline{2-6}
 23  & 5 & $88.925087$ & $92.844472$ & $3.919385$ & 1 \\ \cline{2-6}
 22  & 5 & $84.331772$ & $86.809782$ & $2.478011$ & 1 \\
\end{tabular}
}
\caption{Results from MIF1739: type, initial, minimum, and difference of potential, and Adj($C_{n+1}$,$C_{n}$) (cont.).}~\label{tb:Changes_9}
\end{table}

%% file: TABLA_on_off.tex
\begin{table}
\centerline{\small
\begin{tabular*}{2.1in}[t]{c | p{0.71in} p{0.71in}}
$\bf n$ &   {\bf On / T} & {\bf Off / T} \\ \hline \hline
1000 &     2    3    4    5    6    7    8    9   10   11   12   13   14   15   16   17   18   19   20   21   22   23   24   25   26   27   28   29   30   31   32   33   34   35   36   37   38   39   40   41   42   43   44   45   46   47   48   49   50   51   52   53   54   55   76   77   78   79   80   81   82   83   84   85   86   87   88   89   90   91   92   93   94   95   96   97   98   99  100  101  102  103  104  105  106  107  108  109  110  111  112  113  114  115  116  117  118  119  120  121  122  123  124  125  126  127  128  129  130  131  132  133  134  135  136  137  138  139  140  141  142  143  144  145  146  147  148  149  150  151  152  153  154  155  156  157  158  159  160  161  162  163  164  165  166  167  228  229  230  231  232  233  234  235  236  237  238  239  240  241  242  243  244  245  246  247  248  249  250  251  252  253  254  255  256  257  258  259  260  261 &  /    0 \\ \hline
\multicolumn{1}{c}{\ }
\end{tabular*}
\ \ \ \ 
\begin{tabular*}{2.1in}[t]{c | p{0.71in} p{0.71in}}
$\bf n$ &   {\bf On / T} & {\bf Off / T} \\ \hline \hline
1000 &   262  263  264  265  266  267  268  269  270  271  272  273  274  275  276  277  278  279  280  281  282  283  284  285  286  287  288  289  290  291  292  293  294  295  296  297  298  299  300  301  302  303  304  305  306  307  308  309  310  311  312  313  314  315  316  317  318  319  320  321  322  323  324  325  326  327  328  329  330  331  332  333  334  335  336  337  338  339  340  341  342  343  344  345  346  347  348  349  350  351  352  353  354  355  356  357  358  359  360  361  362  363  364  365  366  367  368  369  370  371  372  373  374  375  376  377  378  379  380  381  382  383  384  385  386  387  388  389  510  511  512  513  514  515  516  517  518  519  520  521  522  523  524  525  526  527  528  529  530  531  532  533  534  535  536  537  538  539  540 &  \\ \hline
\multicolumn{1}{c}{\ }
\end{tabular*}
\ \ \ \ 
\begin{tabular*}{2.1in}[t]{c | p{0.71in} p{0.71in}}
$\bf n$ &   {\bf On / T} & {\bf Off / T} \\ \hline \hline
1000 &   541  542  543  544  545  546  547  548  549  550  551  552  553  554  555  556  557  558  559  560  561  562  563  564  565  566  567  568  569  570  571  572  573  574  575  576  577  578  579  580  581  582  583  584  585  586  587  588  589  590  591  592  593  594  595  596  597  598  599  600  601  602  603  604  605  606  607  608  609  610  611  612  613  614  615  616  617  618  619  620  621  622  623  624  625  626  627  628  629  630  631  632  633  634  635  636  637  638  639  640  641  642  643  644  645  646  647  648  649  650  651  652  653  654  655  656  657  658  659  660  661  662  663  664  665  666  667  668  669  670  671  672  673  674  675  676  677  678  679  680  681  682  683  684  685  686  687  688  689  690  691  692  693  694  695  696  697  698  699 &  \\ \hline
\multicolumn{1}{c}{\ }
\end{tabular*}
}
\caption{$i_p$ of the particles in MIF1739, to build $C_{n}$ from $C_{n+1}$.}~\label{tb:On_off_0}
\end{table}

\begin{table}
\centerline{\small
\begin{tabular*}{2.1in}[t]{c | p{0.71in} p{0.71in}}
$\bf n$ &   {\bf On / T} & {\bf Off / T} \\ \hline \hline
1000 &   700  701  702  703  704  705  706  707  708  709  710  711  712  713  714  715  716  717  718  719  720  721  722  723  724  725  726  727  728  729  730  731  732  733  734  735  736  737  738  739  740  741  742  743  744  745  746  747  748  749  750  751  752  753  754  755  756  757  758  759  760  761  962  963  964  965  966  967  968  969  970  971  972  973  974  975  976  977  978  979  980  981  982  983  984  985  986  987  988  989  990  991  992  993  994  995  996  997  998  999 1000 1001 1002 1003 1004 1005 1006 1007 1008 1009 1010 1011 1012 1013 1014 1015 1016 1017 1018 1019 1020 1021 1022 1023 1024 1025 1026 1027 1028 1029 1030 1031 1032 1033 1034 1035 1036 1037 1039 &  \\ \hline
\multicolumn{1}{c}{\ }
\end{tabular*}
\ \ \ \ 
\begin{tabular*}{2.1in}[t]{c | p{0.71in} p{0.71in}}
$\bf n$ &   {\bf On / T} & {\bf Off / T} \\ \hline \hline
1000 &  1040 1041 1042 1043 1045 1046 1047 1048 1049 1050 1051 1052 1053 1054 1055 1057 1058 1059 1060 1061 1063 1064 1065 1066 1067 1068 1069 1070 1071 1072 1073 1074 1075 1076 1077 1078 1079 1080 1081 1082 1083 1084 1085 1086 1087 1088 1089 1090 1091 1092 1093 1094 1095 1096 1097 1098 1099 1100 1101 1102 1103 1104 1105 1106 1107 1108 1109 1110 1111 1112 1113 1114 1115 1116 1117 1118 1119 1120 1121 1122 1123 1124 1125 1126 1127 1128 1129 1130 1131 1132 1133 1134 1135 1136 1137 1138 1139 1140 1141 1142 1143 1144 1145 1146 1147 1148 &  \\ \hline
\multicolumn{1}{c}{\ }
\end{tabular*}
\ \ \ \ 
\begin{tabular*}{2.1in}[t]{c | p{0.71in} p{0.71in}}
$\bf n$ &   {\bf On / T} & {\bf Off / T} \\ \hline \hline
1000 &  1149 1150 1151 1152 1153 1154 1155 1156 1157 1158 1159 1160 1161 1162 1163 1164 1165 1166 1167 1168 1169 1170 1171 1172 1173 1174 1175 1176 1177 1178 1179 1180 1181 1182 1183 1184 1185 1186 1187 1188 1189 1190 1191 1192 1193 1194 1195 1196 1197 1198 1199 1200 1201 1202 1203 1204 1205 1206 1207 1208 1209 1210 1211 1212 1213 1214 1215 1216 1217 1219 1220 1221 1222 1223 1225 1226 1227 1228 1229 1231 1232 1233 1234 1235 1237 1238 1239 1240 1241 1243 1244 1245 1246 1247 1248 1249 1250 1251 1252 1253 1254 1255 1256 1257 1258 1259 &  \\ \hline
\multicolumn{1}{c}{\ }
\end{tabular*}
}
\caption{$i_p$ of the particles in MIF1739, to build $C_{n}$ from $C_{n+1}$ (cont.)}~\label{tb:On_off_1}
\end{table}

\begin{table}
\centerline{\small
\begin{tabular*}{2.1in}[t]{c | p{0.71in} p{0.71in}}
$\bf n$ &   {\bf On / T} & {\bf Off / T} \\ \hline \hline
1000 &  1260 1261 1262 1263 1264 1265 1266 1267 1268 1269 1270 1271 1272 1273 1274 1275 1276 1277 1278 1279 1280 1281 1282 1283 1284 1285 1286 1287 1288 1289 1290 1291 1292 1293 1294 1295 1296 1297 1298 1299 1300 1301 1302 1303 1304 1305 1306 1307 1308 1309 1310 1311 1312 1313 1314 1315 1316 1317 1318 1319 1320 1321 1322 1625 1626 1627 1628 1629 1630 1631 1632 1633 1634 1635 1636 1637 1638 1639 1640 1641 1642 1643 1644 1645 1646 1647 1648 1649 1650 1651 1652 1653 1654 1655 1656 1657 1658 1659 1660 1661 1662 1663 1664 1665 1666 1667 &  \\ \hline
\multicolumn{1}{c}{\ }
\end{tabular*}%
}
\caption{$i_p$ of the particles in MIF1739, to build $C_{n}$ from $C_{n+1}$ (cont.)}~\label{tb:On_off_2}
\end{table}

\begin{table}
\centerline{\small
%
}
\caption{$i_p$ of the particles in MIF1739, to build $C_{n}$ from $C_{n+1}$ (cont.)}~\label{tb:On_off_3}
\end{table}

\begin{table}
\centerline{\small
\begin{tabular*}{2.1in}[t]{c | p{0.71in} p{0.71in}}
$\bf n$ &   {\bf On / T} & {\bf Off / T} \\ \hline \hline
883 &  1223 1225 1236 1247 1248 1253 /   17 &  1017 1040 1041 1042 /   18 \\ \cline{2-3}
882 &  /    0 &  1236 /    1 \\ \cline{2-3}
881 &  1236 1323 /    2 &   965  966  967 /    3 \\ \cline{2-3}
880 &  /    0 &  1323 /    1 \\ \cline{2-3}
879 &  /    0 &  1236 /    1 \\ \cline{2-3}
878 &  1230 1323 /    2 &  1067 1068 1097 /    3 \\ \cline{2-3}
877 &  /    0 &  1230 /    1 \\
\end{tabular*}%
}
\caption{$i_p$ of the particles in MIF1739, to build $C_{n}$ from $C_{n+1}$ (cont.)}~\label{tb:On_off_4}
\end{table}
\clearpage

\begin{table}
\centerline{\small
%
}
\caption{$i_p$ of the particles in MIF1739, to build $C_{n}$ from $C_{n+1}$ (cont.)}~\label{tb:On_off_9}
\end{table}
\clearpage

\begin{table}
\centerline{\small
%
}
\caption{$i_p$ of the particles in MIF1739, to build $C_{n}$ from $C_{n+1}$ (cont.)}~\label{tb:On_off_14}
\end{table}
\clearpage

\begin{table}
\centerline{\small
%
}
\caption{$i_p$ of the particles in MIF1739, to build $C_{n}$ from $C_{n+1}$ (cont.)}~\label{tb:On_off_32}
\end{table}